\documentclass[floats,floatfix,showpacs,amssymb,prd,twocolumn,superscriptaddress,nofootinbib,nolongbibliography,reprint]{revtex4-2}

\usepackage{amssymb,amsmath,verbatim,mathtools,needspace,enumitem,etoolbox,graphicx,physics,microtype,afterpage,bigints,gensymb,tabularx,mathtools,siunitx}

\usepackage[dvipsnames, usenames]{xcolor}
\definecolor{linkcolor}{rgb}{0.0,0.3,0.5}
\definecolor{dodgerblue}{HTML}{1E90FF}
\usepackage{hyperref}
\hypersetup{unicode, colorlinks=true, linkcolor=linkcolor, citecolor=linkcolor, filecolor=linkcolor,urlcolor=linkcolor, pdfusetitle}
\usepackage[all]{hypcap}
\usepackage[T1]{fontenc}
\usepackage[utf8]{inputenc}
\usepackage{orcidlink}
\usepackage{makecell}

\usepackage [english]{babel}
\usepackage [autostyle, english = american]{csquotes}
\MakeOuterQuote{"}

\newcommand{\Msol}{\rm \,M_{\odot}}

\def\araa{{Ann.~Rev.~Astron.~Astrophys.}}
\def\apj{{Astrophys.~J.}}
\def\apjl{{Astrophys.~J.~Lett.}}
\def\apjs{{Astrophys.~J.~Supp.}}

\def\aap{{Astron.~Astrophys.}}

\def\jcap{{J. Cosmology Astropart. Phys.}}
\def\jrasc{{J. R. Astron. Soc. Can.}}

\def\mnras{{Mon.~Not.~R.~Astron.~Soc.}}             
\def\na{{New A.}}
\def\nar{{New A. Rev.}}

\def\prc{{Phys.~Rev.~C}}
\def\prd{{Phys.~Rev.~D}}

\def\prl{{Phys.~Rev.~Lett.}}

\def\nat{{Nature}}

\def\physrep{{Phys.~Rep.}}

\def\araa{\rm{ARA\&A}}
\def\apj{\rm{Astrophys.~J.}}
\def\apjl{\rm{Astrophys.~J.~ Lett.}}
\def\apjs{\rm{ApJS}}
\def\aap{\rm{Astron.~Astrophys.}}

\def\jcap{\rm{J. Cosmology Astropart. Phys.}}
\def\mnras{\rm{Mon.~Not.~R.~Astron.~Soc.}}
\def\prd{\rm{Phys.~Rev.~D}}
\def\prl{\rm{Phys.~Rev.~Lett.}}

\def\nat{\rm{Nature}}

\def\physrep{\rm{Phys.~Rep.}}

\renewcommand{\emph}[1]{\textit{#1}}

\newcommand{\umani}{\affiliation{Department of Physics and Astronomy \& Winnipeg Institute for Theoretical Physics, University of Manitoba, Winnipeg, R3T 2N2, Canada}}
\newcommand{\bham}{\affiliation{School of Physics and Astronomy \& Institute for Gravitational Wave Astronomy, University of Birmingham,\vspace{-0.05cm}\\$\;$Birmingham, B15 2TT, UK}}
\newcommand{\qub}{\affiliation{Astrophysics Research Centre, School of Mathematics and Physics, Queens University Belfast,\vspace{-0.05cm}\\$\;$Belfast BT7 1NN, UK}}

\begin{document}

\title{Binary Neutron Star Mergers: Multi-Messenger Systematics \\ and Prospects with Next-Generation Facilities}

\author{Nathan Steinle$\,$\orcidlink{0000-0003-0658-402X}}
\email{nathan.steinle@umanitoba.ca}
\umani
\author{Samar Safi-Harb$\,$\orcidlink{0000-0001-6189-7665}}
\umani
\author{Matt Nicholl$\,$\orcidlink{0000-0002-2555-3192}}
\qub
\author{Isabelle Worssam $\,$\orcidlink{0009-0007-3476-2272}}
\bham
\author{Benjamin P. Gompertz$\,$\orcidlink{0000-0002-5826-0548}}
\bham

\begin{abstract} 
Multi-messenger astronomy was galvanized by the detection of gravitational waves (GWs) from the binary neutron star (BNS) merger GW170817 and electromagnetic (EM) emission from the subsequent kilonova and short gamma ray burst. Maximizing multi-messenger constraints on these systems requires combining models of the progenitors and products of BNS mergers within a single framework.  Motivated by GW170817, we create a combined model that relate the progenitor astrophysics of a BNS population with their GW observability and localizability, kilonova light curves, gamma-ray burst afterglow flux, and kilonova remnant evolution. 
We compute the BNS merger rate by convolving metallicity-dependent star-formation history with population-synthesis predictions, and we sample realistic populations to evaluate their GW and EM observables and joint detection rates. We find that next-generation detectors will typically observe BNS mergers with GW network signal-to-noise ratios of $\sim 10$–$20$, 90th-percentile sky areas of order $\sim 10~\mathrm{deg}^2$, and kilonova $i$-band magnitudes spanning $\sim 23$–$33$.  The variation of the merger rate with respect to the common-envelope efficiency is shown in the GW and EM observables and the resulting multi-messenger detection yield, demonstrating how uncertainties propagate into all stages of joint GW+EM forecasting. Across the models examined, no more than $\sim 4\%$ of BNS mergers are detectable simultaneously by a two–Cosmic-Explorer plus one–Einstein-Telescope network and by both Roman (in a $K$-like band) and Rubin ($i$ and $g$ bands). These results show that assumptions underlying the combination of progenitor evolution and source observables will constitute key multi-messenger modeling systematics for inference of astrophysical, nuclear, and fundamental physics from future datasets. 
\end{abstract}

\maketitle

\section{Introduction}
\label{sec:Intro}

The LIGO-Virgo gravitational-wave (GW) detector network  revolutionized astronomy with the first direct detection of a binary neutron-star (BNS) merger, known as GW170817 \cite{GW170817discovery,GW170817redeux}. Observations of the electromagnetic (EM) counterparts \cite{GW170817mmobs} revealed, within seconds of the GW trigger, a short-duration gamma-ray burst (sGRB) \cite{GW170817grb,Goldstein2017,Savchenko2017}, and, in the ensuing days, the signature of the kilonova AT2017gfo \cite{GW170817kn,Arcavi2017,Coulter2017,Nicholl2017,Lipunov2018,SoaresSantos2017,Cowperthwaite2017,Smartt2017,Tanvir2017,Valenti2017}, confirming the association of sGRBs with BNS mergers and the presence of r-process nucleosynthesis in these mergers \cite{2017Natur.551...80K,2017Sci...358.1570D}. 
The jet that powered the sGRB 
interacted with the surrounding interstellar medium to produce the observable GRB afterglow \cite{DAvanzo2018,Lyman2018,Lamb2019,Balasubramanian2022}. 
Also observed was the kilonova remnant (KNR) \cite{Safi-Harb2019,Troja2020,Ren2022,Hajela2022}, also referred to as a kilonova afterglow, from the expanding kilonova ejecta that interacts with the surrounding interstellar medium. 
If a stable, long-lived, magnetized neutron star formed after the merger it is possible to form a pulsar-wind nebula, but the compact remnant from GW170817 was likely a black hole. \cite{Margalit2017,GW170817postmerger}. 
To date, cosmic rays have not been detected in association with GW170817 \cite{GW170817cosmicrays}, but this is an exciting future direction for astronomy as cosmic rays are expected to be produced in kilonovae of BNS mergers \cite{Rodrigues2019}.

Given the successful detection of GW170817 \cite{Burns2020} and its EM counterparts \cite{Margutti2021,Nicholl2025}, one expects the improved sensitivities of future GW and EM detectors to reveal more BNS mergers with EM counterparts \cite{Chan2018,Colombo2022,Ronchini2022,Li2022,Corsi2024,Li2024,Colombo2025}. 
This will advance many fields of research, from tests of fundamental and nuclear physics to novel probes of dark matter, cosmology, and stellar evolution. 
For example, a grand open problem in binary astrophysics is how to distinguish the contributions of many formation channels to an observed population of stellar-mass compact binary mergers \cite{OShaughnessy2005,deMink2016,Vitale2017,Zevin2017,Taylor2018,Belczynski2018,Bouffanais2019,Zevin2020b,Callister2020,Mapelli2020,Wong2021,Bouffanais2021,Mould2022,Antonelli2023,Afroz2024}, e.g. isolated binaries \cite{Postnov2014,Mapelli2020}, dense clusters, \cite{Benacquista2013,Mapelli2020,Spera2022}, or objects embedded in AGN disks \cite{Ishibashi2020,Grobner2020,Gayathri2023,Veronesi2025}, where GW data analysis provides key observables for diagnosing evolutionary histories such as the spin-orbit misalignments \cite{Rodriguez2016b,Gerosa2018,Steinle2021,Steinle2022,Gompertz2022,Gerosa2013,Talbot2017,Farr2017,Stevenson2017,Gerosa2018,Farr2018,Wysocki2018,Zevin2020b,Miller2020,Bavera2020,Callister2021,Varma2022,2022PhRvD.106j3019T,2022PhRvD.105l3024F,Johnson-McDaniel2023,Godfrey2023,Miller2024} 
and eccentricity 
\cite{Romero-Shaw2021,Saini2024}.  

The emergent new field known as multi-messenger astrophysics combines information from EM and GW observations and can provide independent measurements of astrophysically related aspects of the system's evolution \cite{Nicholl2021,Gompertz2023,2020MNRAS.494.1587C,Barbieri2019,Coughlin2019,Dietrich2020,Breschi2021,Raaijmakers2021,Ruiz2021,Bavera2022,DOrazio2022,Colombo2022,Colombo2024,Colombo2025,Bisero2025,2025A&A...697A..36L,2025arXiv250312320R,2023JCAP...07..068B,2025arXiv250312263A}. 
To better constrain these kinds of models will require future GW detector sensitivities to obtain tight constraints on the binary parameters of individual sources and on statistically large population catalogs. 
These source parameters can then be combined with binary evolution models to form these systems, and models of the mass ejection and jet launching from the merger to relate them to the observed EM transients. 
This approach enables a wide range of important science cases for future facilities. 
Some important examples include: 
(a) connecting the KN observables with the binary parameters to enable constraints on formation channels, even when lacking EM counterparts, or from wide-field time-domain surveys without a complete GW dataset; 
(b) combining the pre-merger properties of BNS mergers from the observed EM and GW transients constrains the physics of mass ejection and hence the neutron star equation of state; 
and (c) integrating all of the processes together enables forward modeling that can inform future surveys and determine the sensitivity required to probe different binary populations, and, importantly, the various uncertainties of progenitor astrophysical evolution.  

The total parameter space subtended by the modeling of those various related processes will be enormous, and the relationships between the components of such a model will have complicated inter-dependence. 
Consequently, this implies that comprehensive multi-messenger models of BNS populations will be heavily burdened by modeling systematics as the improved future datasets open our view to features in the parameter space that may still remain sparsely populated. 

Beyond this \emph{burden} of dimensionality, there will be other challenges. 
For instance, some parameters will be common between the models of the source and signals, such as the BNS masses and orbital inclination, introducing modeling systematics from differing assumptions between the models used to analyze the same source. 
Different prescriptions for the physics in any given part of the system's evolution (e.g. how to treat radiative transfer in the KN stage \cite{Bauswein2013}) can also lead to difficult-to-quantify systematics. 
These problems transcend to the population level via biases in the correlations between certain parameters across many sources. 

While a high fidelity, i.e. cosmologically and astrophysically comprehensive and self-consistent model of the formation, evolution, fate, and observables of compact binaries would be optimal for resolving the systematics and degeneracies of compact binary evolution, this is beyond our current tools. 
Nonetheless, the process of building such a model might leverage existing models of aspects of the binary evolution process which are studied in great detail. 
Fortunately there exist in the community highly specialized codes for independent analysis of these related aspects. 
There are several recent studies with comprehensive modeling for multi-messenger BNS science, but each study differs in scope and emphasis. For example, Loffredo et~al. and Patricelli et~al. and Colombo et~al. construct BNS merger populations with population synthesis and forward-model gravitational waves, kilonova light curves, and sGRB afterglows to forecast yields with future detectors \cite{2025A&A...697A..36L,2022MNRAS.513.4159P,Colombo2025}. Ronchini et~al. focus on the high-energy (prompt and X-ray/$\gamma$-ray afterglow) side of BNS counterpart searches and quantify joint detection prospects between ET and future high-energy satellites \cite{Ronchini2022}. Loffredo et~al. and Colombo et~al. also explore the dependence on progenitor formation uncertainties, the nuclear equation of state, and observation configurations \cite{2025A&A...697A..36L,Colombo2025}. 

Similar to these previous studies, we combine state-of-the-art semi-analytic models to demonstrate some difficulties, and exciting possibilities, that arise from forward modeling via an end-to-end pipeline of source astrophysical evolution and observables.
We use a forward-model pipeline that (i) ties a detailed progenitor-population model for binary neutron stars with varying progenitor properties to (ii) a unified set of multi-messenger products (GW observability and sky-localizability, kilonova light curves, short-GRB afterglow fluxes, and kilonova-remnant evolution) and then (iii) propagates progenitor uncertainties to joint observables and detection rates for future detector networks. This combination of BNS progenitor population modeling and explicit treatment of the full suite of EM channels (including the kilonova remnant evolution not treated in prior studies) and an emphasis on how progenitor assumptions map into multi-messenger modeling systematics is, to our knowledge, distinct from previous works which focused on observational systematics. 
Our work therefore complements and extends existing literature by unifying progenitor physics by identifying modeling systematics that will be important in analyzing joint GW and EM datasets from future facilities. 

We focus on networks of third-generation GW detectors and on mergers of BNSs in the isolated-binary formation channel\footnote{This is one of the main formation channels expected for stellar-mass compact binaries, where a binary star forms from a protostellar disk and evolves in isolation, i.e., such as the galactic field, through stages of stellar evolution as the stars interact.}. 
We utilize: 
(i) the rapid binary population synthesis model \textsc{COMPAS} \cite{COMPAS2022} for progenitor evolution and formation of the BNS population; 
(ii) \texttt{gwfast} \cite{gwfast2022} for estimating the signal-to-noise ratio and parameter uncertainties of the BNS merger with GW detector networks; 
(iii) the EM transients fitter \textsc{MOSFiT} \cite{MOSFiT2018} for computing the properties of the kilonova explosion from the BNS merger \cite{Nicholl2021} and of the resultant kilonova remnant, 
and (iv) an analytic model for the flux density of the sGRB afterglow. 
While designed independently, together these four models can estimate a bigger multi-messenger picture of BNS observability. 

We find that future GW detector networks will have the potential to discover thousands of BNS mergers whose signal-to-noise ratio and parameter uncertainties scale with the sensitivity of the network components. A subpopulation of these BNSs with small sky location errors (i.e. 90th percentile of the sky area uncertainty is < 1 sq. deg.) correlated with bright KNe (i.e. i-band apparent magnitude < 25) can be increased in size by an order of magnitude, i.e. from 1\% to 10\% of the total population, by including a 20km Einstein Telescope \cite{ET2025} to a network of two Cosmic Explorer \cite{CE2021} detectors. 
We recover the usual degeneracies in the GW observables, such as the luminosity distance and orbital inclination, and find that breaking these degeneracies will depend crucially on the specific configuration of third generation GW detector networks. 
Lastly, we demonstrate the dependence of the joint detection rates on the common envelope efficiency parameter, showing how its non-monotonicity and affects on the merger rate density translate into a systematic uncertainty of the detection rate. This provides new insights into how the interplay of assumptions regarding progenitor evolution and cosmological modeling affect the predicted rates of detectable BNSs, motivating further work toward understanding multi-messenger systematics in studies utilizing future datasets.

This paper is organized as follows. We present the adopted models of binary evolution and of the GW and EM signals in Sec.~\ref{sec:Methods}. In Sec.~\ref{sec:Results} we present our results, and in Sec.~\ref{sec:ConcDisc} we summarize our conclusions.

\newcolumntype{C}[1]{>{\centering\arraybackslash}p{#1}}

\begin{table*}
\caption{\label{Tab:Description}
Descriptions of the models used in this study to connect isolated 
binary neutron-star population astrophysics with the GW and EM observability of the resultant binary mergers, kilonovae, and short-duration gamma ray bursts.  
}
\vspace{0.1cm}
\def\arraystretch{1.2}  
\centering
\begin{tabular}{C{0.5in}|C{5.75in}|C{0.5in}}
  \hline
  \hline
  Code & Description & Ref. \\
  \hline
  \href{https://compas.science/}{\texttt{COMPAS}} &  \makecell[t]{Compact Object Mergers: Population Astrophysics \& Statistics is a rapid binary population \\ synthesis code that evolves a binary system from zero-age main sequence to two compact remnants}
  & \cite{COMPAS2022} \\
  \hline 
  \href{https://github.com/CosmoStatGW/gwfast}{\texttt{gwfast}}  &  \makecell{A Python package for fast Fisher Information Matrix applications \\ in GW cosmology based on automatic differentiation} &  \cite{gwfast2022} \\
  \hline
  \href{https://github.com/guillochon/MOSFiT}{\texttt{MOSFiT}} &  \makecell{Modular Open-Source Fitter for Transients is a Python module for fitting, sharing, \\ and estimating the parameters of transients via user-contributed transient models}  & \cite{MOSFiT2018,Nicholl2021} \\
  \hline
  afterglow &  \makecell{
  A python implementation of an analytical model for short-duration gamma-ray burst light curves} & 
  \cite{Sari98} \\ 
  \hline 
\end{tabular}
\end{table*}

\section{Multi-messenger population model}\label{sec:Methods}

We combine three publicly available codes \textsc{COMPAS}, \texttt{gwfast}, and \textsc{MOSFiT} with an analytical model for GRB afterglows \citep{Sari98}, for modeling progenitor evolution and multi-messenger signals associated with BNS mergers. These are described in Table~\ref{Tab:Description}. Here, we focus on estimates of the GW measurement uncertainties of BNS mergers in the context of current and future terrestrial detectors and on the EM observability of possible counterparts. 

The main parameters used in this paper are: 
$m_{1, \rm ZAMS}$ ($m_{2, \rm ZAMS}$) is the mass of the initially more (less) massive ZAMS star which we call the primary (secondary) star, and  
$m_{1, \rm BNS}$ ($m_{2, \rm BNS}$) is the mass of the resultant neutron star which may or may not be the more massive neutron star depending on the possibility for mass-ratio reversal; the BNS mass ratio $q \leq 1$, symmetric mass ratio $\eta = m_{1, \rm BNS} m_{2, \rm BNS} / (m_{1, \rm BNS} + m_{2, \rm BNS})^2$, total mass $M_{\rm BNS} = (m_{1, \rm BNS} + m_{2, \rm BNS})$, and chirp mass $\mathcal{M} = (m_{1, \rm BNS} m_{2, \rm BNS})^{3/5} / (m_{1, \rm BNS} + m_{2, \rm BNS})^{1/5}$; the BNS orbital semi-major axis $a_{\rm BNS}$ and inclination $\iota$; the timescale over which the BNS is driven to merge under GW emission $t_{\rm merge}$; and the luminosity distance between source and observer $d_{\rm L}$ or equivalently the cosmological redshift $z$. The remaining parameters are discussed further in Subsec.~\ref{subsec:Systematics}. 

A main aspect of this study is the complex \emph{hierarchy of parameters} that arises from combining models with  differing capabilities and assumptions. 
The BNS masses are predicted by the binary population model \textsc{COMPAS} and treated as inputs for the observable models \texttt{gwfast} and \textsc{MOSFiT} which require the distance and orbital inclination as inputs (we leave this choice to Sec.~\ref{sec:Results}), implying that differences in assumptions between \emph{intrinsic} and \emph{extrinsic} parameters can arise as an underlying systematic across the differing constructions of such models.
This is further compounded by, e.g. in our model, when the afterglow flux depends on the distance to the source but assumes the observer is on-axis with the jet, resulting in inconsistent treatment of the inclination between our models of the KN and afterglow. 
Such difficulties inherently arise when attempting to combine models for the broad evolution of the system, which, as we discuss at length in Sec.~\ref{subsec:Systematics}, implies the presence of immense systematic modelling biases and parameter space degeneracies. 

Throughout this work, we assume the Planck 2018 \cite{Planck2018} cosmology to convert cosmological redshift $z$ to luminosity distance $d_{\rm L}$.

\subsection{Binary Neutron-star Formation}\label{subsec:BNSformation} 

For modeling the evolution and formation of BNSs in the isolated binary channel, we employ the rapid binary population synthesis model \textsc{COMPAS} \cite{Stevenson2017b,VignaGomez2018,VignaGomez2020,COMPAS2022}. 
In models such as \textsc{COMPAS}, the initial binaries are parameterized in terms of the stellar mass and metallicity from which the evolution of stellar binaries is prescribed in terms of timescales that govern the stages of evolution, i.e. the movement of the stars on the Herzsprung-Russell diagram. 
These models are designed to produce rapid populations by folding uncertainties from a vast array of processes, where essentially the stars in a binary evolve in isolation, e.g. governed by mass and metallicity dependent formulae fitted to evolutionary tracks from detailed stellar evolution models, and the effects of binary interactions on their macroscopic parameters are tracked throughout the binary's lifetime. 
There are many population synthesis models today, and the differing aspects of their implementations of the same physics is significant enough to lead to different predictions. 
In the context of the greater sensitivity of future detectors, those differences will be small in comparison to the differences between the stellar evolution models upon which the population models are based, presenting as a complex, layered systematic that we will discuss further in Sec.~\ref{subsec:Systematics}. 

Consider a fiducial population of initial binaries in circular orbits (see \cite{VignaGomez2020}), where we assume each star has high metallicity $Z = Z_{\odot} = 0.02$ and is non-rotating. 
As \textsc{COMPAS} implements many astrophysical processes, it also comes with many free parameters. 
Most of these free parameters are fixed across the population, and we use the Python interface that comes with \textsc{COMPAS} to generate evolutionary tracks for the population whose parameters are specified in a configuration file. 
The zero-age main sequence (ZAMS) population is produced via the following initial binary parameters: 
we assume the mass of the initially more massive star, $m_{1, \rm ZAMS}$, in the binary  is drawn from the initial mass function $dN/dm_{1, \rm ZAMS} \propto m_{1, \rm ZAMS}^{-2.3}$ \cite{Kroupa2001} sampled between $5 \leq m_{1, \rm ZAMS} \leq 50 \Msol$. The mass of the secondary star ($m_{2, \rm ZAMS}$) is obtained by drawing from a flat distribution in mass ratio ($q_{\rm ZAMS} = m_{2, \rm ZAMS}/m_{1, \rm ZAMS}$), i.e. $dN/dq \propto 1$ with $0.1 < q_{\rm ZAMS} \leq 1$ \cite{Sana2012}, and the initial separation is drawn from the log uniform distribution $dN/da \propto a^{-1}$ with $0.01 < a_{\rm ZAMS}/AU < 1000$ \cite{Opik1924}. All other flags and parameters of relevance in the configuration file are left to their default \textsc{COMPAS} settings and values, respectively. For detailed explanations of the various physical processes and models implemented in \textsc{COMPAS}, we refer the interested reader to e.g. \cite{VignaGomez2020,Neijssel2019}. Here we will only briefly introduce a few main processes. We evolve $\num{1.5e8}$ ZAMS binaries until double compact object formation and obtain 10,550 BNSs that merge within the age of the Universe, where the BNS merge timescale $t_{\rm merge}$ due to GW emission is computed in \textsc{COMPAS} from the semi-major axis and eccentricity dependent orbit-averaged solutions of \cite{Peters1964}. 

\begin{figure*}
\centering
\includegraphics[width=0.9\textwidth]{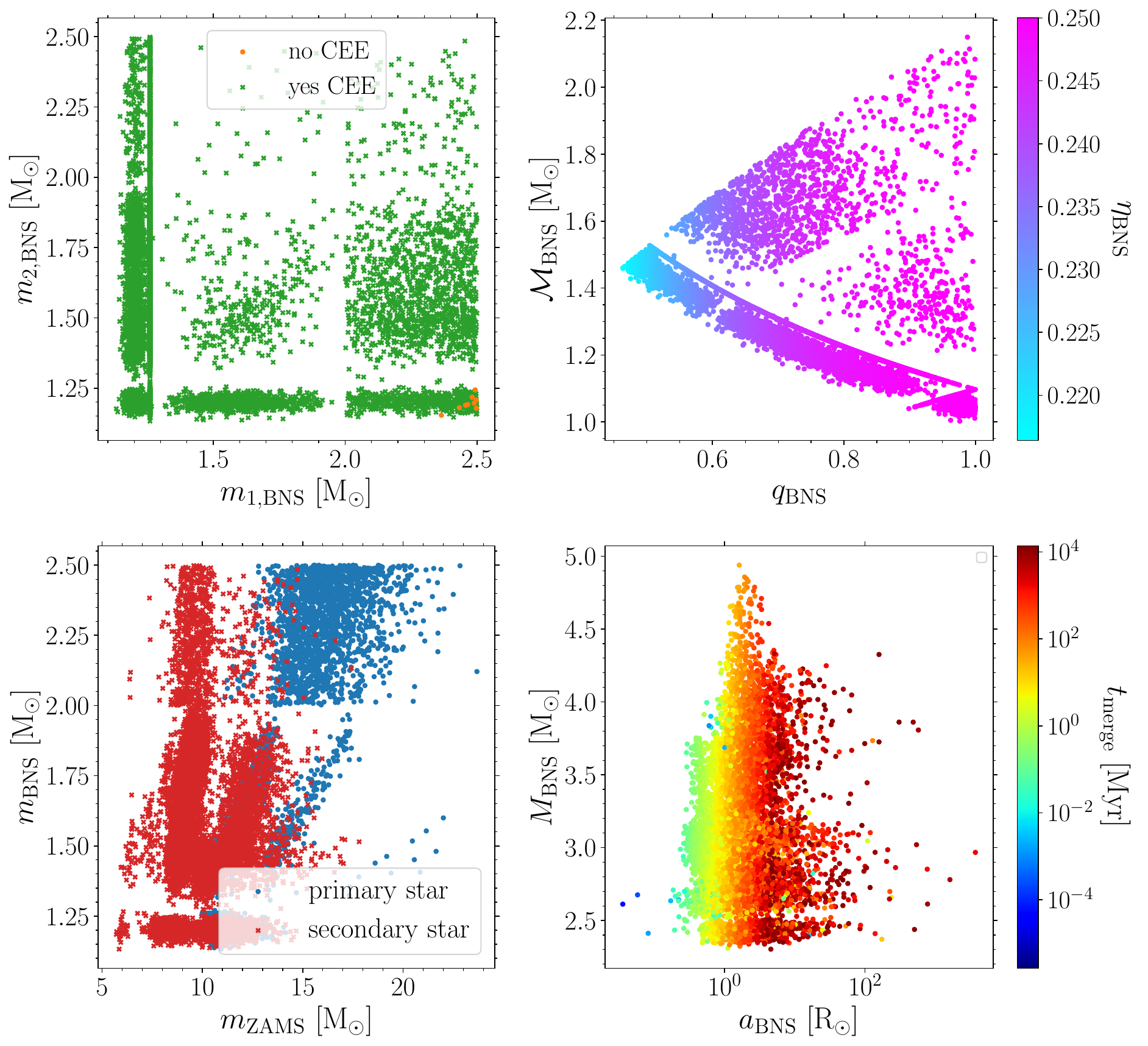}
\caption{Properties of our fiducial BNS population from the \textsc{COMPAS} rapid binary population synthesis model. \textit{Top left panel}: the masses of the neutron stars in each BNS with green x's (orange o's) for binaries that do (do not) undergo an episode of Common Envelope Evolution during their progenitor evolution history. \textit{Top right panel}: the chirp mass $\mathcal{M}_{\rm BNS}$, mass ratio $q_{\rm BNS}$, and symmetric mass ratio $\eta_{\rm BNS}$ of each BNS. \textit{Bottom left panel}: the correlation between the progenitor initial mass $m_{\rm ZAMS}$ with the mass of the neutron star in the BNS that it formed $m_{\rm BNS}$ for the primary (blue o's) and secondary (red x's) stars. 
\textit{Bottom right panel}: the distributions of BNS total mass $M_{\rm BNS}$, orbital semi-major axis $a_{\rm BNS}$, and merge timescale due to GW emission $t_{\rm merge}$. 
} \label{F:evolutions}
\end{figure*}

\begin{figure*}
\centering
\includegraphics[width=\textwidth]{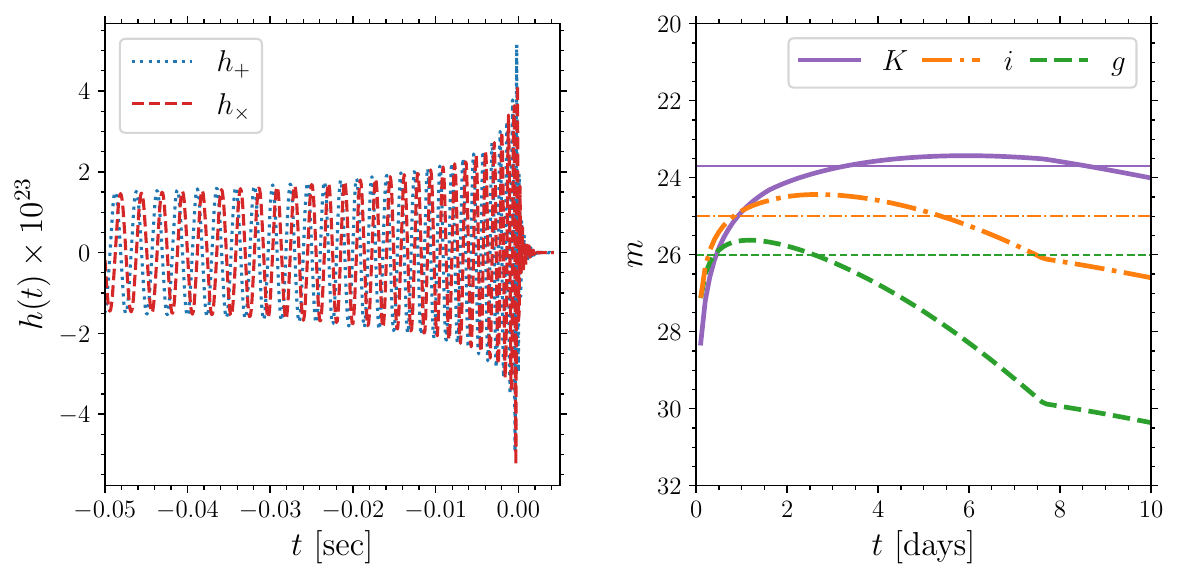}
\caption{For a single BNS of the fiducial population with parameters $m_{1, \rm BNS} = 1.3 \Msol$, $m_{2, \rm BNS} = 1.7 \Msol$, and $t_{\rm merge} = 0.4$ Myr and assuming redshift $z = 0.1$ and inclination $\cos(\iota) = \pi/4$, the left and right panels show the resulting GW polarization states $h_+(t)$ and $h_x(t)$ and the light curves from the kilonova explosion in the $i$, $g$, and $K$ bands, respectively. Horizontal lines in the right panel correspond to the $i$ and $g$ 180 second exposure with Rubin \cite{Andreoni2024a} and the $F213$ filter (similar to the $K$ band) 55 second exposure with Roman \cite{Andreoni2024b}. 
} \label{F:signals}
\end{figure*}

Prior to the GW-dominated regime, mass transfer processes dominate the evolution of the stellar binary's orbital angular momentum \cite{Postnov2014}. In principle this is a highly complicated matter for hydrodynamics and numerical simulations to more accurately model. For population synthesis models, this uncertainty is traditionally cast in terms of time-dependent log-exponents that relate the dynamical dependence of the donor's and accretor's radii and Roche lobe radii to the exchange of stellar gas \cite{Soberman1997}. Phenomenologically, this results in two main mass transfer possibilities for our stellar binaries: stable mass transfer (SMT) and common envelope evolution (CEE). In the former (the latter) the accretion rate of the accreting star is dynamically stable (unstable) w.r.t. the mass transfer rate given by the donor star. 

Most important for our study, a fraction of the donor's envelope is accreted during SMT, whereas the binary separation is drastically decreased due to expulsion of the donor's envelope in CEE (which may prematurely merge the binary). These two types of \emph{mass transfer events}, along with stellar winds mass-loss (which can be substantial for high-metallicity and high-mass stars), constitute the main ways in which the stars lose their envelopes. Loss of the envelope exposes the helium-rich core that ultimately gravitationally collapses to form a compact object, such as neutron stars. 
Together, these possibilities sketch a simplified, yet complicated landscape of possible evolutionary pathways and their imprint on the correlations between the progenitor and observed binary properties. 

The panels in Figure \ref{F:evolutions} show the main properties of our fiducial BNS population and their progenitors. In the top-left panel, the non-trivial, yet generic with respect to population synthesis models, shapes of the distributions of BNS masses reflect the discriminative combinations of the various astrophysical processes encoded by \textsc{COMPAS}, e.g.'s, premature mergers, unbinding natal kicks, mass transfer efficiencies, initial conditions, stellar wind mass loss uncertainties, etc. These can result in nontrivial predictions for the properties of compact binaries, as we shall explore in Sec.~\ref{sec:Results}.  
All of these BNSs experience multiple phases of mass transfer, most are brief episodes of SMT, and nearly every BNS underwent an episode of CEE, consistent with the results of \cite{VignaGomez2020}. 
The vast majority, ie 90\%, of the BNSs had CEE initiated by the secondary star and the remaining had double-core CEE.  
Most (66\%) underwent mass-ratio reversals due to SMT resulting in the initially more (less) massive ZAMS star forming the less (more) massive neutron star. 
The bottom-left panel of Fig.~\ref{F:evolutions} indicates which binaries experienced mass-ratio reversal, ie BNSs for which the secondary $m_{\rm BNS}$ (red x's) is larger than the primary $m_{\rm BNS}$ (blue o's) since the secondary star is defined as being initially less massive. 
The top-right panel shows the distributions of other BNS mass parameters, most notably the lower bound on the mass ratio $q_{\rm BNS} \gtrsim 0.4$ due to the bounds on the BNS masses. 
This panel also illustrates the great degeneracy between the chirp mass $\mathcal{M}_{\rm BNS}$ and mass ratio $q_{\rm BNS}$ which will be important for the resultant KN apparent magnitudes in Sec.~\ref{sec:Results}.    
The BNS semi-major axis $a_{\rm BNS}$, total mass $M_{\rm BNS}$, and corresponding GW merge timescale $t_{\rm merge}$ are shown in the bottom-right panel. Generally, smaller $a_{\rm BNS}$ ($M_{\rm BNS}$) yield smaller (larger) $t_{\rm merge}$, but the nonzero BNS eccentricities (introduced from the natal kicks in neutron star formation) cause $t_{\rm merge}$ to be non-monotonic in either $a_{\rm BNS}$ or $M_{\rm BNS}$ as higher eccentricity binaries merge more quickly \cite{Peters1964}.

The fiducial \textsc{COMPAS} population shown in Fig.~1 exhibits two main features: a pronounced concentration of component masses near $1.2$--$1.3\Msol$ and a high-mass tail reaching $\simeq2.4$--$2.5\Msol$ and total system masses up to $\sim4$--$5\Msol$. 
In comparison, the observed Galactic BNS systems display a narrower distribution, with most component masses clustered around $1.25$--$1.35\Msol$ \cite{2019ApJ...876...18F,2016ARA&A..54..401O}. The \textsc{COMPAS} low-mass peak aligns with this Galactic population, while the predicted high-mass tail extends well beyond the range currently sampled by radio pulsar binaries. GW detections occupy an intermediate regime. GW170817 ($M_{\rm tot}\simeq2.74\Msol$) lies within the low-mass cluster, whereas GW190425 ($M_{\rm tot}\simeq3.4\Msol$) matches the upper end of the \textsc{COMPAS} distribution \cite{GW170817discovery,GW190425discovery}. Population analyses of the LVK catalogs suggest an intrinsic neutron-star mass distribution spanning roughly $1.2$--$2.0\Msol$ \cite{GWTC1,GWTC2,GWTC3,GWTC4}, broader than the Galactic sample yet narrower than the full theoretical spread. The \textsc{COMPAS} population thus encompasses both observed regimes, naturally reproducing the low-mass cluster while predicting heavier systems consistent with GW190425.
Observed and theoretical samples are affected by different selection functions. The Galactic sample is shaped by radio luminosity limits, beaming geometry, and pulsar lifetimes, whereas GW detections favor high-chirp-mass systems observable to larger distances \cite{2017ApJ...846..170T}. Accounting for these biases is essential for quantitative comparison, but qualitatively, the agreement between the \textsc{COMPAS} low-mass peak and observed systems, together with its broader high-mass extension, indicates that the fiducial model captures the main features of the BNS population while predicting additional heavy systems. 

While forward evolutions of population synthesis models can map the parameter space and reveal novel insights into the progenitors of observed systems, it is challenging to obtain populations that comprehensively fill the parameter space. 
Statistical tools can help obtain sizeable populations in arbitrary regions of the parameter space in reasonable computation time and are an active area of research.

To self-consistently model the redshift and chirp-mass distribution of merging BNSs, we compute a metallicity-dependent merger efficiency following the formalism of \cite{Neijssel2019}. First, we simulate $15$ COMPAS populations with metallicities $Z\in[2\times10^{-4},\,2\times10^{-2}]$, and we compute the effective star-forming mass represented by the simulation, 
\begin{equation}
    M_{\rm sim} \;=\; 3.1\times10^8\,\left(\frac{N_{\rm bin}}{10^6}\right) \Msol\,,
\end{equation}
using the scaling relation of \cite{Neijssel2019}. From the processed COMPAS outputs we extract, for every merging BNS, the component masses, formation time, and delay time. We restrict to systems with $1.0 < m_{1,2}/\Msol < 3.2$ and compute their chirp masses 
$\mathcal{M} = (m_1 m_2)^{3/5} (m_1+m_2)^{-1/5}$.
Using logarithmically spaced delay-time bins from $t_{\rm delay}\in[10^{-5},10^{2}]\,{\rm Gyr}$ and a uniform grid in chirp mass, we obtain the merger yield
\begin{equation}
    \eta(Z, t_{\rm delay}, \mathcal{M}) \;=\;
    \frac{{\rm d}^3 N_{\rm merg}}{{\rm d}M_\star\,{\rm d}t_{\rm delay}\,{\rm d}\mathcal{M}} 
\end{equation}
with unit of ${\rm mergers}/\Msol\,{\rm Gyr}\Msol$.

We then perform a discrete cosmic integration to obtain the full merger rate density as a function of merger redshift and chirp mass. For every merger redshift $z_{\rm m}$ on a grid of $z\in[0,2]$, we compute the cosmic age $t_{\rm m}$ and determine the corresponding formation redshift for each delay time via $t_{\rm form}=t_{\rm m}-t_{\rm delay}$ and inverting the age-redshift relation of the Planck 2018 cosmology.  
The star-formation rate density $\psi(z)$ is modeled with the analytic fit of \cite{Neijssel2019}, and the metallicity distribution at formation is assumed to follow their log-normal prescription,
\begin{equation}
    p(Z|z_{\rm form}) = \frac{1}{Z\,\sigma_{\ln}\sqrt{2\pi}}
    \exp\!\left[-\frac{\left(\ln Z - \mu(z_{\rm form})\right)^2}{2\sigma_{\ln}^2}\right]\,,
\end{equation}
with $\sigma_{\ln}=0.39$ and $\mu(z)$ set by the redshift-dependent mean metallicity $\langle Z\rangle= Z_0\,10^{\alpha z}$ with $Z_0 = 0.035$ and $\alpha = -0.23$. Combining these ingredients, the contribution of $Z$ to the merger rate density in each chirp-mass bin is
\begin{equation}
    R(z_{\rm m},\mathcal{M}) 
    = \sum_Z \int \eta(Z,t_{\rm delay},\mathcal{M})\,
      \psi(z_{\rm form})\,
      p(Z|z_{\rm form})\,
      {\rm d}t_{\rm delay}\,,
\end{equation}
yielding ${\rm d}^2N/({\rm d}V_{\rm c}\,{\rm d}t_{\rm obs}\,{\rm d}\mathcal{M})$ in units of
mergers per ${\rm Gpc}^3\,{\rm yr}\Msol$.  The final redshift and mass-dependent merger rate $R(z,\mathcal{M})$ is marginalized over~$\mathcal{M}$ to obtain the volumetric merger rate $R(z)$.

Given the mass- and redshift-dependent merger rate density $R(z,\mathcal{M})$ constructed above, we generate synthetic BNS merger populations by direct Monte Carlo sampling from the implied joint distribution. 
For simplicity, we sample $10^4$ binaries for all populations. 
To construct the normalized sampling distribution, we form
\begin{equation}
    p(z,\mathcal{M})
    \propto
    R(z,\mathcal{M})\,
    \frac{{\rm d}V_{\rm c}}{{\rm d}z}\,
    \frac{1}{1+z}\,,
\end{equation}
We draw $n_{\rm samp}$ pairs $(z,\mathcal{M})$ by inverse-transform sampling of the cumulative distribution. For each synthetic merger, we next assign formation metallicity, delay time, and component masses in a manner self-consistent with the metallicity-dependent yield tables. Specifically, the metallicity $Z$ is drawn from a discrete distribution with probabilities proportional to the total yield $\sum_{i}\eta(Z,t_{{\rm delay},i},\mathcal{M})\,\Delta t_{{\rm delay},i}$ of each metallicity bin. For that $Z$, we identify the chirp-mass bin closest to the sampled $\mathcal{M}$ and draw a delay time from the corresponding one-dimensional distribution $\eta(Z,t_{\rm delay},\mathcal{M})\,\Delta t_{\rm delay}$. Finally, the component masses $(m_1,m_2)$ are drawn from the COMPAS BNS data at the same metallicity by selecting those whose chirp masses fall within the width of the sampled $\mathcal{M}$ bin. 
The resulting population of $10^4$ mergers thus preserves all correlations between redshift, chirp mass, delay time, metallicity, and component masses that are present in the underlying synthesis models and in the cosmological weighting encoded by $R(z,\mathcal{M})$.

To convert the cosmological merger population into predicted detection rates for joint GW and EM observing campaigns, we compute detection probabilities for every event on the $(z,\mathcal{M})$ grid and combine them into a single per-event detection probability. We then fold these probabilities through the fully integrated merger rate predicted by $R(z,\mathcal{M})$. 

We first evaluate the total observer-frame BNS merger rate by integrating the two-dimensional merger rate density over chirp mass and redshift,
\begin{equation}
    \mathcal{R}_{\rm tot}
    = \int {\rm d}z \int {\rm d}\mathcal{M}\;
      R(z,\mathcal{M})\,
      \frac{{\rm d}V_{\rm c}}{{\rm d}z}\,
      \frac{1}{1+z}\,,
\end{equation}
where ${\rm d}V_{\rm c}/{\rm d}z$ is the full-sky comoving-volume element. $R(z,\mathcal{M})$ is numerically integrated using trapezoidal quadrature over the mass and redshift grids. The result $\mathcal{R}_{\rm tot}$ represents the total number of BNS mergers per year across the full sky within the simulated cosmic volume. 

For each event in the sampled population we evaluate a smooth GW detection probability based on its signal-to-noise ratio,
\begin{equation}
    P_{\rm GW}(\rho)
    = \frac{1}{2}\left[1+\operatorname{erf}
        \left(\frac{\rho - \rho_{\rm thr}}{\sqrt{2}\,\sigma_{\rho}}\right)\right]\,,
\end{equation}
where $\rho_{\rm thr}=12$ is the nominal detection threshold and $\sigma_{\rho}=1$ controls the transition width. This prescription approximates a realistic triggering probability while avoiding a discontinuous step function. 

We next determine whether the kilonova associated with each merger would be observable in the Rubin $i$ and $g$ bands and the Roman F213 (K-band) filter. For each band we use a simple deterministic model for EM visibility,
\begin{equation}
    P_{\rm EM} =
    \Theta(m_{\rm lim}-m_{\rm peak})\,f_{\rm obs}\,,
\end{equation}
where $m_{\rm peak}$ is the source peak magnitude, $m_{\rm lim}$ is the limiting magnitude of the instrument (taken as $i=25$, $g=26$, $K=24$), $\Theta$ is the Heaviside function, and $f_{\rm obs}$ is an observing-duty-cycle factor (0.5 for Rubin). We require simultaneous detectability in all three bands,
\begin{equation}
    P_{\rm EM} = P_{i}\,P_{g}\,P_{K}\,.
\end{equation}
Additionally, we impose a sky-localization requirement by setting $P_{\rm EM}=0$ for events whose GW localization area exceeds $10~{\rm deg}^{2}$, reflecting the practical need for precise positions to enable rapid follow-up. 

For each event we compute a combined detection probability
\begin{equation}
    P_{\rm det} = P_{\rm GW}\,P_{\rm EM}\,.
\end{equation}
We then average this quantity over all events in the simulation, $\langle P_{\rm det}\rangle = N^{-1}\sum_{i}P_{{\rm det},i}$, to obtain the fraction of mergers detectable in a joint GW+EM campaign. Finally, the expected number of detected BNS mergers per year is
\begin{equation}
    \mathcal{R}_{\rm det}
    = \mathcal{R}_{\rm tot}\,\langle P_{\rm det}\rangle.
\end{equation}
This procedure ensures that detectability is evaluated in a manner consistent with the underlying cosmological population, GW network sensitivity, and kilonova brightness models. 

In the next two subsections, we explain how we use the output of \textsc{COMPAS} as the input for \texttt{gwfast} and \textsc{MOSFiT}.

\subsection{Third Generation Gravitational-wave Networks}
\label{subsec:GWsignal}

A single BNS can be described by the neutron star masses, distance, luminosity distance (or redshift), tidal deformabilities, eccentricity, spin magnitudes, spin-orbit misalignments, azimuthal and polar angles subtending the two spin vectors and the orbital and total angular momenta, respectively, the source's latitude and longitude on the sky, and the GW waveform polarization and the phase at coalescence, i.e. the following set of parameters, respectively: $m_1$, $m_2$, $d_{\rm L}$, $\iota$, $\Lambda_1$, $\Lambda_2$, $e$, $\chi_1$, $\chi_2$, $\theta_1$, $\theta_2$, $\theta_L$, $\Phi_L$, $\theta_{\rm s}$, $\phi_{\rm s}$, $\psi$, $\phi_{\rm c}$. 
The masses $m_1$ and $m_2$, distance $d_{\rm L}$ (or redshift $z$), sky location $\theta_{\rm s},\ \phi_{\rm s}$, and orbital inclination $\iota$ determine the leading-order dependence of the binary waveform strain amplitude in the wide-separation limit, see e.g.'s  \cite{Cutler1994,Kidder1995}, and are thus the first we can constrain with GW data from BNS mergers. In this study, we focus on these main parameters. 
Other parameters can be important at higher order post-Newtonian terms and include effects arising from the strong-gravity regime but the merger phase is ultimately computed with numerical relativity and the ringdown is understood in terms of decaying normal modes. These inspiral, merger, and ringdown portions of the BNS coalescence process are stitched together in different ways to obtain full waveform families that can produce different predictions for the same data \cite{2025PhRvX..15c1036D}. 

From a GW observer's perspective, typically, Bayesian inference with a waveform template family in a matched-filter analysis and, given prior distributions on the source parameters, is used to compute the signal-to-noise ratio (SNR) of a source and posterior distributions of the parameters (from which measurement uncertainties can be computed) for a given detector's sensitivity \cite{Abbott2020}. 
GW source data analysis is generally a great open question, and the usual 'full-glory' Bayesian approach described above is computationally limited to small sets of systems. 
Fortunately, a well-known approximation utilizing the Fisher information matrix can be used to estimate the SNR and parameter uncertainties \cite{Cutler1994} (assuming the latter are Gaussian distributed in the high-SNR limit) for populations of GW sources. 

In this work we use \texttt{gwfast} \cite{gwfast2022}, which combines computationally cutting-edge implementation of the Fisher matrix approach with modularity for various GW waveform models and possible detector networks. 
For the inspiral-merger-ringdown merger of BNSs, we use the \texttt{IMRPhenomD\_NRTidalv2} \cite{Dietrich2019} waveform model, valid for neutron masses between 1$\Msol$ to 3$\Msol$, dimensionless aligned spin magnitudes up to 0.6, and dimensionless tidal deformabilities up to 5000. 
We assume Gaussian priors for the GW polarization angle, orbital inclination, and sky location latitude and longitude, and hold the tidal deformabilities, aligned spin components, and coalescence time and phase fixed for each binary. 
We arrive at the GW SNR $\rho$, the $90^{\rm th}$ percentile of the sky area $\Omega_{90}$ \cite{Barack2004,Wen2010}, and the relative, ie fractional, uncertainties of the main BNS parameters: the chirp mass $\Delta\mathcal{M}/\mathcal{M}$, symmetric mass-ratio $\Delta\eta/\eta$, luminosity distance $\Delta d_{\rm L}/ d_{\rm L}$, and inclination $\Delta\iota/\iota$. 

In particular, for a given GW detector network, we take the BNS masses $m_{1 \rm BNS}$ and $m_{2 \rm BNS}$ from the BNS population of Subsec.~\ref{subsec:BNSformation} and pass these to the appropriate \texttt{gwfast} functions to compute the GW observables, ie SNR, sky area, and main parameter uncertainties.  
We then carry out this calculation for each binary in the population to forecast the science potential of future GW networks to a realistic BNS population. 
The true parameters of the signal are composed of the \textsc{COMPAS} output ($\mathcal{M}$, $\eta$, and $t_{\rm merge}$), we explore different cases of distance and inclination, and the remaining true parameters of each signal are held constant across the population: coalescence phase $\Phi_{\rm c} = 0$ and polarization angle $\psi = 0$, sky location latitude and longitude $\theta = \pi/2 + 0.4$ and $\phi = 3.4$, tidal deformabilities $\Lambda_1 = 368.12$ and $\Lambda_2 = 586.55$, and spin magnitudes $\chi_1 = 0$ and $\chi_2 = 0$. 
Taking one such binary from our fiducial population (ie, see Fig.~\ref{F:evolutions}), we show for illustrative purposes its GW strain amplitude $h(t)$ in the left panel of Figure~\ref{F:signals} composed of the two GW polarization states $h_{\rm +}$ and $h_{\rm x}$ computed with the \texttt{IMRPhenomD\_NRTidalv2} waveform \cite{Dietrich2019} via \textsc{PyCBC} \cite{pycbc}. The two polarization states compose the entire GW signal $h(t)$ whose amplitude $\sim 10^{-23}$ reflects the larger distance of $z = 0.1$ compared to $z \approx 0.01$ of GW170817. 

The future of GW astronomy is extremely bright with many GW detectors on the horizon. We consider the following detectors as implemented in \texttt{gwfast}: 
\begin{itemize}
    \item the O3a sensitivities of the LIGO, Virgo, and KAGRA (LVK) detectors \cite{GWTC1,GWTC2,GWTC3}. 
    \item Cosmic Explorer (CE) with arm lengths of either 40km or 20km and locations either in New Mexico or Nevada, USA \cite{CEpsd,CE2021}. 
    \item Einstein Telescope (ET) with arm length of 20km and locations either in Sardinia, Italy or The Netherlands \cite{ETpsd,ET2025}. 
\end{itemize}
From these detectors, we construct 5 GW detector networks with \texttt{gwfast}, i.e. see \cite{Iacovelli2022,Maggiore2024}, for use in our multi-messenger study: 
\begin{itemize}
    \item LVK, 
    \item CE $40{\rm km}$, 
    \item CE $40{\rm km}$ + CE $20{\rm km}$, 
    \item CE $40{\rm km}$ + ET $20{\rm km}$, 
    \item CE $40{\rm km}$ + CE $20{\rm km}$ + ET $20{\rm km}$
\end{itemize}
These choices are intended to be illustrative and are motivated by recommendations from the Next-Generation Gravitational-Wave Detector Concepts Report\footnote{\href{https://nsf-gov-resources.nsf.gov/files/mpsac-nggw-subcommittee-repor-2024-03-23-r.pdf?VersionId=uhdtblJUPYsavII5EUuSNQJcFdRr92q\_}{NSF Mathematical and Physical Sciences Advisory Committee, March 2024}}. Taking again the masses of the single BNS from Fig.~\ref{F:signals}, we can estimate the detectable horizon\footnote{See the  \href{https://gwfast.readthedocs.io/en/latest/notebooks/gwfast_tutorial.html}{\texttt{gwfast} implementation}.} of each network. Assuming an SNR threshold of 12 and using the optimal sky location of $\theta\approx 0.9$ and $\phi = 0$, the largest redshifts at which each network listed above can detect the BNS are 0.03 for LVK, 2.5 for CE 40km, 3 for CE 40km + CE 20km, 3.2 for CE 40km + ET 20km, and 3.5 for CE 40km + CE 20km + ET 20km; though, these depend on the BNS parameters. 

\begin{figure*}
\centering
\includegraphics[width=\textwidth]{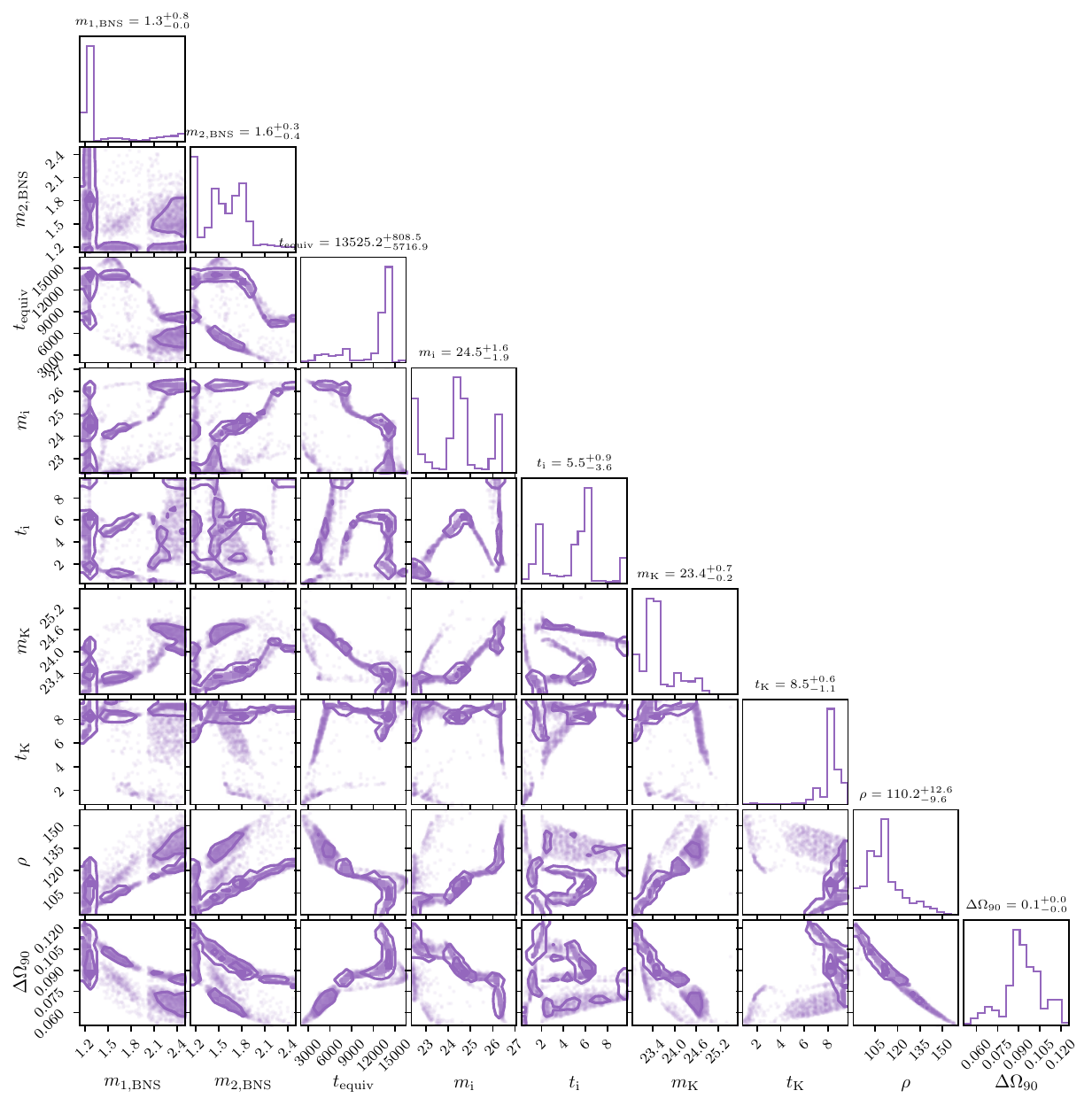}
\caption{The output of the multi-messenger pipeline for the fiducial BNS population: the BNS masses $m_{1, \rm BNS}$, $m_{2, \rm BNS}$ from \textsc{COMPAS}, the peak magnitudes for the KN optical $i$ and infrared $K$ bands (with the number days $t_{\rm i}$ and $t_{\rm K}$ it is within 1 magnitude of the peak) from \textsc{MOSFiT}, the KNR timescale $t_{\rm equiv}$ in days, and the SNR $\rho$ and sky area $90^{\rm th}$ percentile $\Omega_{90}$ from \texttt{gwfast} for GW detector network CE 40km + CE 20km + ET 20 km. In this case, we assume a single redshift ($z = 0.1$) and inclination angle ($\iota=\pi/4$) for each BNS, implying that the variance in the GW and EM observables is due to only the BNS masses. As a single distance is assumed here, the 1 keV X-ray afterglow at 0.1 days after the BNS merger is $\Phi_{1 \rm keV} \approx 5.02\,\mu{\rm Jy}$ for all the BNSs. The sGRB afterglow and KNR assume the same ISM density $\rho_{\rm ISM, fid} = \num{1.5e-25} {\rm g}/{\rm cm}^3$. 
} \label{F:cornerFiducial}
\end{figure*}

\subsection{Kilonovae: Explosions, Afterglows, \& Remnants}\label{subsec:EMsignal}

A kilonova (KN) is the associated multi-wavelength EM signature from the outflows of a BNS merger and the ensuing decay of a variety of r-process nucleosynthesis elements. 
GW170817 triggered a boon for models of BNS mergers \cite{Radice2020} and KNe \cite{Nakar2020}. 
Connecting these EM-observable phenomena to the pre-merger BNS properties is challenging as it requires the intermediary step of numerical relativity, and many models exist to address this  \cite{Coughlin2019b,Dietrich2020,Breschi2021,Raaijmakers2021b}. We use the model of \cite{Nicholl2021}, implemented in \textsc{MOSFiT} \cite{MOSFiT2018}, to compute the KN properties with fitted formulae to numerical relativity simulations that relate the dynamical and disk/wind ejecta masses to the BNS masses and radii and the theoretical maximum neutron star mass.   
These are then related to the kilonova emission by employing a semi-analytic model of the luminosity from r-process decay \cite{Villar2017,Cowperthwaite2017,Metzger2019} using analytic solutions for radiative diffusion and photospheric temperature and radius. 
This tracks multiple ejecta components and incorporates viewing angle dependence of the observed luminosity and color where the luminosity of each component scales proportionally with its mass. The important parameters of this kilonova luminosity model are: the BNS masses, the neutron star compactnesses, the maximum stable neutron star mass, and the inclination angle. 
We utilize the fiducial parameters of this model as specified in \cite{Nicholl2021}, and compute the apparent magnitude of kilonovae in the $i$, $g$, and $K$ bands. For full details of this model, see Section 2 of \cite{Nicholl2021}. 

Next we describe the model of the flux density of the sGRB afterglow. We use the analytical solutions of \citep{Sari98} to construct on-axis light curves and spectral energy distributions. 
The sGRB afterglow emerges as synchrotron emission due to the jet’s interaction with the surrounding interstellar medium. The outgoing jet sweeps up ambient particles causing it to decelerate and hence a shock forms between the jet and the interstellar medium \cite{afterglow_theory}. We use a shell model that implements synchrotron theory \cite{synchrotron_eqns} assuming the emitting particles to be electrons. This population of electrons are initialized as a power-law in terms of their individual energies \cite{Afterglow_eqns}. As the jet interacts with the interstellar medium it decelerates and hence the flux of the emission decreases with time. The model assumes the observer to be on-axis with a top hat jet profile. A top hat profile is characterized by a sharp drop off in the Lorentz factor of the jet at a pre-defined angle where inside this angle the Lorentz factor is constant. Therefore, the same emission is seen for all on-axis observers and no emission is seen off axis. Synchrotron self-absorption (where the emitted synchrotron radiation is absorbed by the surrounding medium and re-emitted) is not accounted for in the model but is unlikely to be present at the frequencies we are concerned with, appearing primarily at lower frequencies. For a given frequency, this provides the sGRB afterglow flux density $\Phi$, in units of $\mu{\rm Jy}$. 

Similar to our approach in Subsec.~\ref{subsec:GWsignal}, we use the BNS masses from the fiducial BNS population of Subsec.~\ref{subsec:BNSformation} and assume distances and inclinations as inputs for \textsc{MOSFiT} to obtain light curves of the associated KN explosion. 
To demonstrate this, we take the same BNS whose GW merger signal is shown in the left panel of Fig. \ref{F:signals} and we compute the KN light curves as shown in the right panel of Fig. \ref{F:signals}. 
For this redshift $z = 0.1$, the 1 keV (i.e., \num{2.4e17} Hz X-ray) afterglow at 0.1 days after the BNS merger is $\Phi_{1 \rm keV} \approx 5\,\mu{\rm Jy}$. 

Over longer times, i.e. hundreds of days to decades, the ejecta expand into a kilonova remnant (KNR) which we define as representing the diffuse emission from the kilonova as it interacts with its surrounding medium. We trace the KNR evolution by drawing an analogy to supernova remnants \cite{SafiHarb2019}. In particular, as the KNR expands into the surrounding medium over 100s to 1000s of years, thermal and non-thermal X-rays are expected to be emitted from both shock-heated gas and the decay of heavy r-process elements.
To characterize the KNRs, one can compute the time for the KNR shock to sweep up an amount of mass equivalent to the KN ejecta mass \cite{SafiHarb2019}, 
\begin{equation}\label{E:Tequiv}
    t_{\rm equiv} = \left( \frac{3m_{\rm ejecta}}{4\pi\rho_{\rm ISM}v_{\rm ejecta}^3} \right)^{1/3}  \,,
\end{equation}
where $m_{\rm ejecta}$ and $v_{\rm ejecta}$ are the KN ejecta mass and velocity, respectively, and we assume the KNR sweeps up a spherical shell of interstellar medium with density $\rho_{\rm ISM}$.  
To compute $t_{\rm equiv}$, we easily obtain $m_{\rm ejecta}$ and $v_{\rm ejecta}$ from the output of \textsc{MOSFiT} which depends on the BNS parameters as described above. 
Throughout this work, we assume a fiducial ISM mass density $\rho_{\rm ISM, fid} = (\num{1.67e-24} {\rm g}) (0.1/{\rm cm}^3) \approx \num{1.67e-25} {\rm g}/{\rm cm}^3$ in the model for the sGRB afterglow and in Eq.~\ref{E:Tequiv}, unless otherwise specified.

Thus we have three sources for EM observables: the early-time KN emission, the GRB afterglow, and the later KNR phase. 
In summary, the KNR timescale $t_{\rm equiv}$ indirectly depends on the BNS masses via the direct dependence of the KN ejecta mass and velocity on the BNS masses; the GRB afterglow is independent of the BNS masses but depends on the distance and shares with the KNR the ISM density as a free parameter; and the GW signal (from Subsec.~\ref{subsec:GWsignal}) and KN light curves depend directly on the BNS masses, distance, and orbital inclination. Each of the models includes further parameters that are not shared with the other models, constituting a global systematic uncertainty of the framework.  
We combine these with the GW model of Subsec.~\ref{subsec:GWsignal} to compute multi-messenger observables in  Sec.~\ref{sec:Results}.

\section{Results}
\label{sec:Results}

\begin{figure}
\centering
\includegraphics[width=0.48\textwidth]{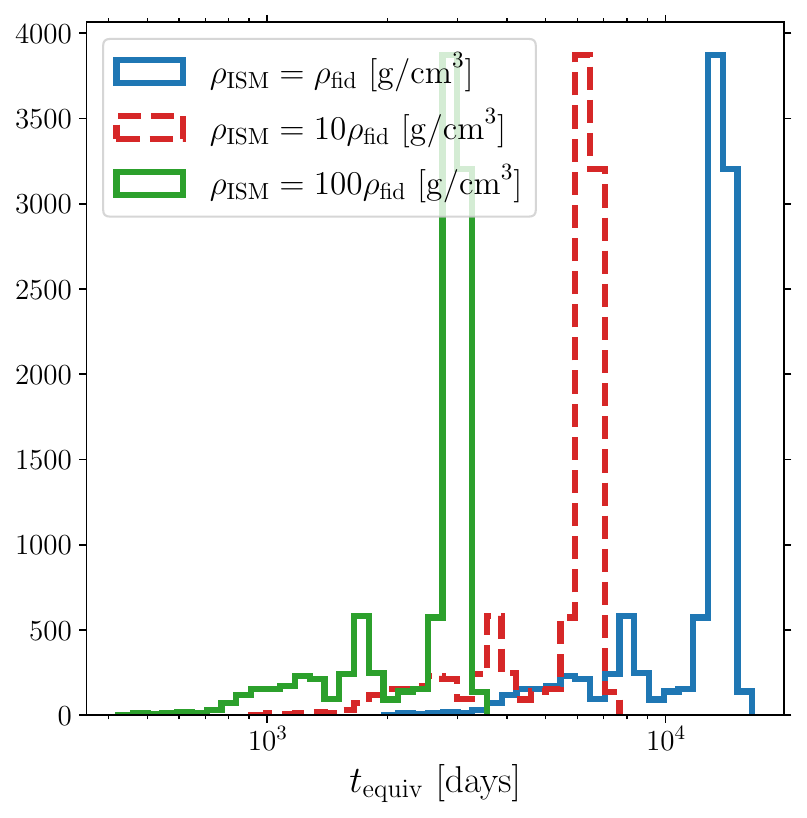}
\caption{
Histograms of the timescale $t_{\rm equiv}$ over which an amount of ISM mass equivalent to the total $m_{\rm ejecta}$ of the KN is swept-up by the KNR for the fiducial population of BNS mergers, i.e. see Fig.'s~\ref{F:evolutions}-\ref{F:cornerFiducial}. The colors blue, red, and green correspond to the three densities $\rho_{\rm ISM} = \rho_{\rm ISM, fid}$, $10\rho_{\rm ISM, fid}$, $100\rho_{\rm ISM, fid}$, and the $m_{\rm ejecta}$ of each BNS is computed with the \textsc{MOSFiT} BNS kilonova model. 
} \label{F:histTequiv}
\end{figure}

\subsection{Multi-messenger BNS Population Observables}
\label{subsec:MMresults}

The \textsc{COMPAS} population synthesis model provides intrinsic parameters for the BNSs that merge within the age of the Universe. We must assign each BNS extrinsic, ie observer-dependent, parameters to compute their multi-messenger observables. With the fiducial population from Subsec.~\ref{subsec:BNSformation}, we consider two cases in this section: (1) a constant value of redshift and of inclination for all the BNSs, and (2) uniformly assigned redshifts (weighted by volume) and inclinations for the BNSs. 
In case (2), \emph{subthreshold} binaries \cite{Magee2019} have GW signal-to-noise ratio near the boundary of detection as their distances place them near the edge of the GW detector's horizon. 
While individually less significant than target-of-opportunity sources, their sizable subpopulations form an interesting science case for future facilities such as Rubin \cite{Andreoni2024a} and Roman \cite{Andreoni2024b}. 
In the next section, we consider case (3) of redshifts and masses sampled from computed merger rate densities (assuming a constant value of inclination) following the method described in Section~\ref{subsec:BNSformation}. Case (3) is the most realistic population of these three as it incorporates metallicity and redshift dependence of the BNS cosmic formation history.

In the first case, we assume redshift $z = 0.1$ and orbital inclination $\cos(\iota) = \pi/4$ for all the BNSs in the fiducial population. We follow the procedure detailed in Sec.~\ref{sec:Methods}, and Figure~\ref{F:cornerFiducial} displays the resulting distributions of BNS masses (i.e. the same as in Fig.~\ref{F:evolutions}) and the main EM and GW observables: 
the BNS masses $m_{1, \rm BNS}$, $m_{2, \rm BNS}$, the peak magnitudes for the KN optical $i$ and infrared $K$ bands and corresponding number days $t_{\rm i}$ and $t_{\rm K}$ it is within 1 magnitude of the peak, respectively, the KNR timescale $t_{\rm equiv}$, and the SNR $\rho$ and sky area $90^{\rm th}$ percentile $\Omega_{90}$ for a GW detector network CE 40km + CE 20km + ET 20 km. 
The highly nontrivial shapes of the distributions of $m_{1, \rm BNS}$ and $m_{2, \rm BNS}$ determine specific cross-correlations between the multi-messenger observables as all else is held constant. 
For example, see the significant subpopulations with $m_{1, \rm BNS} \sim 1.2\Msol$ ($m_{2, \rm BNS} \sim 1.2\Msol$) and corresponding range of $1.2\Msol \lesssim m_{2, \rm BNS} \leq 2.5\Msol$ ($m_{1, \rm BNS} \leq 2.5\Msol$) correlated with smaller network SNR $\rho \lesssim 120$ as they have small chirp masses $\mathcal{M}_{\rm BNS} \lesssim 1.5\Msol$ (i.e., see the top-right panel of Fig.~\ref{F:evolutions}) and hence quieter GW chirps. 
However, the peak magnitudes $m_{\rm i}$ and $m_{\rm K}$ can be smaller, i.e. the KNe are brighter for fixed redshift $z$, with smaller BNS masses $m_{1, \rm BNS}$ and $m_{2, \rm BNS}$ because their smaller chirp mass $\mathcal{M}_{\rm BNS}$ yields larger $m_{\rm ejecta}$ (i.e., see Fig. 2 of \cite{Nicholl2021}), although these display degeneracy which vanishes in the high-mass limit. The simplest correlation is, as one might expect, the monotonic relation between higher  $\rho$ and smaller uncertainty in the locations of the sources on the sky $\Omega_{90}$ (see the panel in the bottom row and eighth column), as higher $\rho$ generally provides smaller GW parameter uncertainties. With the single distance $z = 0.1$ assumed here, the 1 keV X-ray afterglow at 0.1 days after the BNS merger is $\Phi_{1 \rm keV} \approx 5\,\mu{\rm Jy}$ ($\num{5e-29}$ erg/cm$^2$/s/Hz). 

\begin{figure*}
\centering
\includegraphics[width=\textwidth]{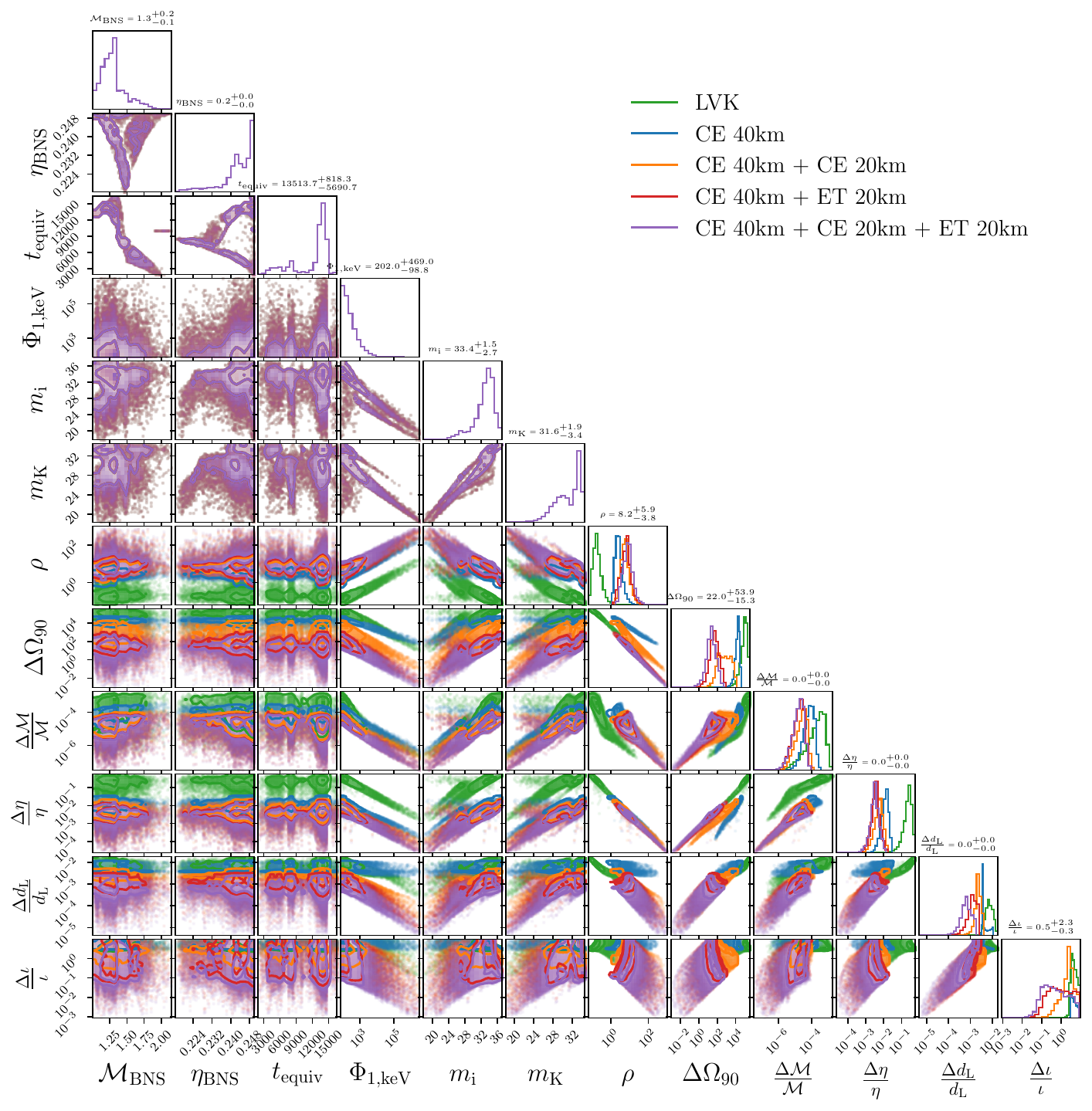}
\caption{The output of the multi-messenger pipeline for the fiducial BNS population: the BNS chirp mass $\mathcal{M}_{\rm BNS}$ and symmetric mass ratio $\eta_{\rm BNS}$, the peak magnitudes $m_{\rm i}$ and $m_{\rm K}$ for the KN optical i and infrared K bands from \textsc{MOSFiT}, the flux density in $\mu{\rm Jy}$ of the 1 keV GRB afterglow $\Phi_{1 \rm keV}$ at 0.1 days after the BNS mergers, the KNR timescale $t_{\rm equiv}$ in days, and the GW SNR $\rho$, sky area $90^{\rm th}$ percentile $\Omega_{90}$, and fractional uncertainties on the chirp mass $\Delta\mathcal{M}/\mathcal{M}$, symmetric mass-ratio $\Delta\eta/\eta$, luminosity distance $\Delta d_{\rm L}/ d_{\rm L}$, and inclination $\Delta\iota/\iota$ from \texttt{gwfast} for detector networks corresponding to colors: LVK (green), CE 40km (blue), CE 40km + CE 20km (orange), CE 40km + ET 20km (red), CE 40km + CE 20km + ET 20 km (purple). 
We uniformly assign redshift $0.01 \leq z \leq 2$ and inclination $0.1 \leq \cos(\iota) \leq 0.9$  to each BNS, implying that the variance in the GW and EM observables is due to a complicated interplay of parameters. 
The sGRB afterglow and KNR assume the same ISM density $\rho_{\rm ISM, fid} = \num{1.5e-25} {\rm g}/{\rm cm}^3$.
Also, the statistical correlations between data points displayed here are weighed by $z^2$ to mimic a more realistic population of sources. } \label{F:cornerSubthresh}
\end{figure*}

The joint dependence on $m_{1, \rm BNS}$ and $m_{2, \rm BNS}$ can result in non-trivial degeneracies in the multi-messenger observables. 
For example, BNSs with lower $\rho$, and consequently higher $\Omega_{90}$, tend to exhibit brighter peak KN magnitudes $m_{\rm i}$ and $m_{\rm K}$ (see the panels in the second row and fourth and sixth columns), but can be highly degenerate in the peak times depending on the band (fifth and seventh columns). 
This suggests an interesting observational challenge for EM followup of BNS mergers. 
Meanwhile, the KNR timescale $t_{\rm equiv}$ in Eq.~(\ref{E:Tequiv}) is indirectly (directly) correlated with $\rho$ ($\Omega_{90}$), but is highly degenerate with the EM observables $m_{\rm i}$ and $m_{\rm K}$ due to the complicated interplay of the dependence on the KN ejecta mass and thus $\mathcal{M}_{\rm BNS}$ and $q_{\rm BNS}$. The variation in the BNS mass, and hence $m_{\rm ejecta}$, distributions translate into the shape of the $t_{\rm equiv}$ distribution in our model, shown in Figures~\ref{F:cornerFiducial} and \ref{F:cornerSubthresh}.  
Figure~\ref{F:histTequiv} shows $t_{\rm equiv}$ for three ISM densities where longer $t_{\rm equiv}$ results from higher density ISM as the KNR turns on more quickly (i.e., blue histogram) and would thus evolve more slowly. 

In our second case, we consider another simplified population of BNSs with volume-weighted observables in Fig. \ref{F:cornerSubthresh}.  
The BNSs are uniformly distributed between $z = 0$ and $z = 2$, with weights given by the relative volumes at redshift $z$, i.e., the effective number of binaries out to redshift $z$ is $N(<z) \propto V(<z) \propto z^3$, implying the number density per unit redshift is $dN/dz \propto z^2$. 
Since weights in a histogram serve as a numerical proxy for the relative number of counts (i.e. sources), this results in higher redshift BNSs contributing more to the distributions in Fig. \ref{F:cornerSubthresh} compared to the naive (ie un-weighted) observables. We also checked $dN/dz \propto (dV_{\rm c}/dz)(1+z)^{-1}$, computing the differential co-moving volume element per unit redshift with Astropy \cite{Astropy}, and found similar results as in Fig. \ref{F:cornerSubthresh}. 
In addition to the quantities in Fig.~\ref{F:cornerFiducial}, Fig. \ref{F:cornerSubthresh} shows the 0.1-day flux density of the X-ray GRB afterglow $\Phi_{1 \rm keV}$ and the GW fractional uncertainties on the chirp mass $\Delta\mathcal{M}/\mathcal{M}$, symmetric mass-ratio $\Delta\eta/\eta$, luminosity distance $\Delta d_{\rm L}/ d_{\rm L}$, and inclination $\Delta\iota/\iota$ for several detector networks: LVK (green), CE 40km (blue), CE 40km + CE 20km (orange), CE 40km + ET 20km (red), and an ``optimal network'' CE 40km + CE 20km + ET 20 km (purple). 

For the LVK network in green the median SNR is $\rho < 1$ and the sky area and parameter uncertainties can be enormous, while those for networks composed of future facilities scale monotonically in $\rho$ and the parameter uncertainties. 
The median SNR of the optimal network is $\rho \approx 7$, providing an upper limit for our ability to detect \emph{subthreshold} binaries. 
A Bayesian analysis would reveal specific dependencies beyond the capabilities of the Fisher approximation. 
Nevertheless, the method is informative: e.g. the approximated sky area $\Omega_{90}$ for a single CE 40km detector is about as large as for the LVK network despite having an order of magnitude larger $\rho$ demonstrating the importance of multiple detectors for sky localization via triangulation in multi-messenger pipelines.  

The combination of the dependence on $m_{1, \rm BNS}$ and $m_{2, \rm BNS}$ with the variation from the assumed distributions of $z$ (which gives a non-uniform distribution in $d_{\rm L}$) and $\cos(\iota)$ produces interesting correlations in the multi-messenger observables. The dependence on $z$ dominates the behavior of the observables over the dependence on the BNS masses (which can be confirmed by comparing the panels of Fig.'s \ref{F:cornerFiducial} and \ref{F:cornerSubthresh}). 
The EM and GW observables vary trivially with $z$ such that systems with larger $z$ have fainter KNe and afterglows and less detectable (lower SNR) GW mergers, i.e. the EM luminosity vanishes as $1/z^2$ and GW loudness as 1/$z$. 
Consequently, as shown in the first four rows and the fourth, fifth, and sixth columns of Fig.~\ref{F:cornerSubthresh}, the KN apparent magnitudes and GRB-afterglow flux density reflect an analogous dependence on the GW parameter uncertainties $\Delta\mathcal{M}/\mathcal{M}$, $\Delta\eta/\eta$, $\Delta d_{\rm L}/ d_{\rm L}$, and $\Delta\iota/\iota$ where the brightest KNe and afterglows correspond to the closest systems with the highest GW SNR $\rho$ and smallest GW fractional uncertainties. We find the fraction of BNSs with $\rho \leq 12$ (for the optimal GW network) and KNe apparent magnitudes below the thresholds of Roman ($K$ band) and Rubin ($i$ and $g$ bands) used in Fig.~\ref{F:signals} to be $\approx 400/10000 = 0.04$ which serves as a theoretical upper bound.

Strong degeneracies are seen in Fig.~\ref{F:cornerSubthresh} between the mass parameters $m_{1, \rm BNS}$ and $m_{2, \rm BNS}$ and fractional uncertainties $\Delta\mathcal{M}/\mathcal{M}$, $\Delta\eta/\eta$, as expected from their definitions. 
The well-known degeneracy between luminosity distance and inclination angle burdens the correlations between the fractional uncertainties $\Delta d_{\rm L}/ d_{\rm L}$ and $\Delta\iota/\iota$. Interestingly, the optimal network (CE 40km + CE 20km + ET 20 km (purple)) and its similar but less optimal alternatives (CE 40km + CE 20km (orange) and CE 40km + ET 20km (red)) can generally achieve similar SNR $\rho$, but notably these three networks can differ in their ability to measure the binary parameters. For example, between the two-detector networks, the one with an ET detector (red) provides better sky location errors than the one with two CE detectors (orange). This is because in the frequency range of 100 to 1000 Hz a 20km ET has better sensitivity than a 20km CE but oppositely so in the frequency range of 10 to 100 Hz (e.g. see Fig. 3 of \cite{Iacovelli2022}), resulting in similar $\rho$ for the two networks across the entire BNS coalescence but better sky localizability for the network with an ET detector (red) as it provides better sensitivity in the frequency range where the noise-weighted energy flux peaks \cite{Wen2010}. Also, these three detector networks can provide similar mass uncertainties, but differ in their ability to break degeneracy between the luminosity distance and inclination, which requires small fractional errors, where e.g. the network with two CE detectors (orange) provides similar $\Delta\iota/\iota$ as the network with a single CE detector while the peaks in the distributions of $\Delta d_{\rm L}/ d_{\rm L}$ across all networks decrease monotonically with increasing network sensitivity. 

As the KNR timescale $t_{\rm equiv}$ is independent of the extrinsic parameters $z$ and $\cos(\iota)$, it is highly degenerate with the GW fractional uncertainties due to their shared BNS mass dependence, as seen in the third column of Fig.~\ref{F:cornerSubthresh}. It is possible that some of these degeneracies can be broken by, in principle, combining EM and GW datasets, e.g. knowing the BNS masses from the GW signal and the ejecta mass from the KN light curve in the K band, or knowing the KNR magnitude and KN magnitudes to tightly constrain the KN evolution to improve multi-messenger constraints on the BNS ejecta mass. 

Together, these showcase the complicated combined parameter space of multi-messenger BNS science. The first two cases of $z$-dependence were chosen to show the correlations that arise with approximately cosmological populations. In the next section, we examine our third case of a realistic $z$-dependence that includes the full metallicity-dependent star formation rate that determines the merger rate of BNSs which are key uncertainties in population modeling of cosmological compact binaries \cite{Neijssel2019,Broekgaarden2022,Mandel2022,vanSon2022,Fishbach2023,SchiebelbeinZwack2024,deSa2024,Fishbach2025,Sgalletta2025,2025ApJ...985L..33R}.

\subsection{Multi-messenger Modeling Systematics}
\label{subsec:Systematics}

Even within a single model, the effects of the underlying astrophysical degeneracies and systematics on data analysis of these sources are not well understood, but this has been sufficient due to the low sensitivities of the GW and EM detectors. This implies that the improvements of future detectors will open the proverbial floodgates to such challenges, especially for multi-messenger studies, as demonstrated in the previous section, that involve combining models into a unified framework and can result in interesting arrangements of parameter dependencies. Such frameworks would transmit the fundamental modeling systematics that exist in the individual models into \emph{multi-messenger modeling systematics}.  
Classic sources of fundamental systematics are the enormity of uncertainties in a single star's evolution, dubbed 'stellar multiplicity' \cite{Breivik2019}, and the well-studied systematics in GW waveforms \cite{Littenberg2013,2025PhRvX..15c1036D,Kapil2024}. For another example, the classification of a GW source as a BNS merger, e.g. based on a detection statistic, can affect the predicted distributions of binary parameters from the GW measurement and hence followup searches for EM counterparts, which are already hampered by degeneracy in the GW sky map and the possibility for offsets from host galaxies \cite{Gaspari2024}. 

\begin{figure}
\centering
\includegraphics[width=0.5\textwidth]{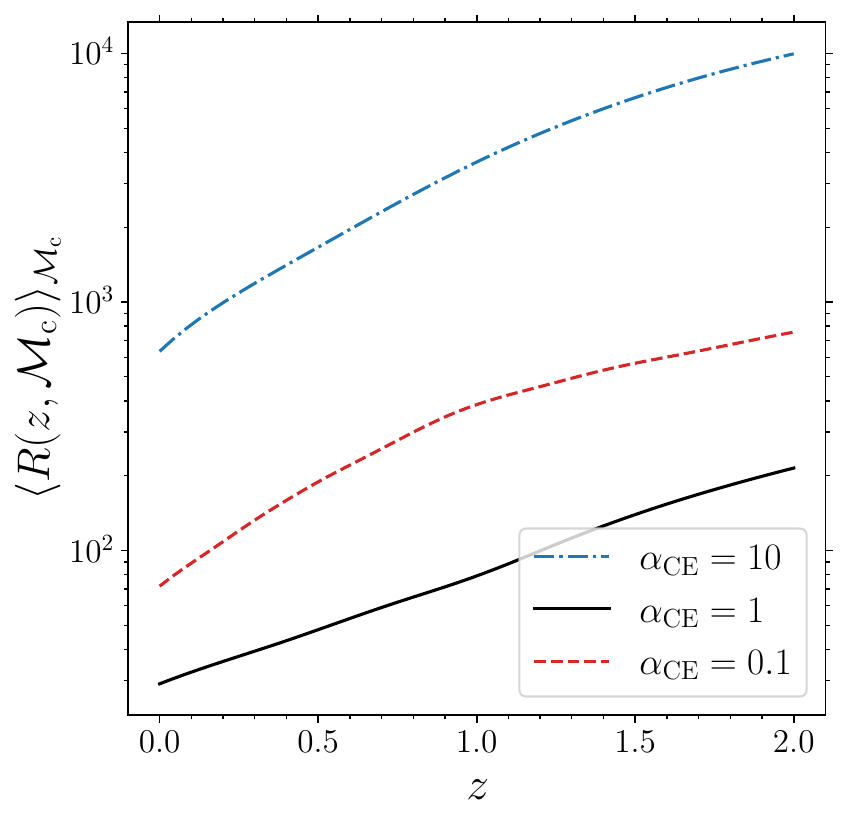}
\caption{
The merger rate density $R(z,\mathcal{M})$ marginalized over chirp mass $\mathcal{M}$ for three cosmologically realistic BNS populations, where each population assumes a different value of the common envelope accretion efficiency ($\alpha_{\rm CE} = 0.1$ is the red dashed line, $\alpha_{\rm CE} = 1$ is the black solid line, $\alpha_{\rm CE} = 10$ is the blue dash-dot line). These intrinsic merger rates are computed as detailed in Section~\ref{subsec:BNSformation}.
} \label{F:MergerRateDensities}
\end{figure}

The multi-messenger modeling systematics transmitted from combinations of specialized models for full source evolution and analysis produce structures in the parameter space whose complexity precludes precise diagnosis of their impact on analysis of future datasets with current theoretical tools. 
We show an example of how an important uncertainty of binary stellar evolution can be transmitted as a systematic in the modeling of multi-messenger observables, using our fiducial-sampled population as a baseline for comparison. 
Keeping all else equal to the fiducial case, we vary the efficiency of CEE and compare to the fiducial value of $\alpha_{\rm CE} = 1$ with two cases of $\alpha_{\rm CE} = 0.1$ and $\alpha_{\rm CE} = 10$. 
A comparison of the computed merger rate densities, marginalized over $\mathcal{M}$, is shown in Figure~\ref{F:MergerRateDensities} where the sampled population that assumed $\alpha_{\rm CE} = 0.1$ ($\alpha_{\rm CE} = 10$) is the red dashed (blue dot-dashed) line and the fiducial-sampled population with $\alpha_{\rm CE} = 1.0$ is the black line. The non-monotonicity of the merger rate density amplitude with respect to $\alpha_{\rm CE}$ is consistent with a recent study that utilized COMPAS \cite{2024ApJ...976...24B}. 
This arises from a complicated interplay of astrophysical effects: increasing $\alpha_{\rm CE}$ raises the probability that a binary ejects its envelope and thus survives the common-envelope phase, but it also typically produces larger post-CE separations and hence longer gravitational-radiation inspiral times. Consequently, changes in $\alpha_{\rm CE}$ can either increase or decrease the number of systems that merge within a Hubble time depending on the detailed balance between survival fraction, post-CE separation distribution, delay-time weighting, and metallicity-dependent formation efficiency \cite{2012ApJ...759...52D,2018MNRAS.480.2011G,2021MNRAS.508.5028B}. 
While $R(z,\mathcal{M})$ differs in amplitude for these three $\alpha_{\rm CE}$ cases, in principle it can also differ in shape by varying other parameters that control the progenitor astrophysics or by using different models for the star formation rate redshift and metallicity dependence \cite{Neijssel2019}. 

\begin{figure*}
\centering
\includegraphics[width=\textwidth]{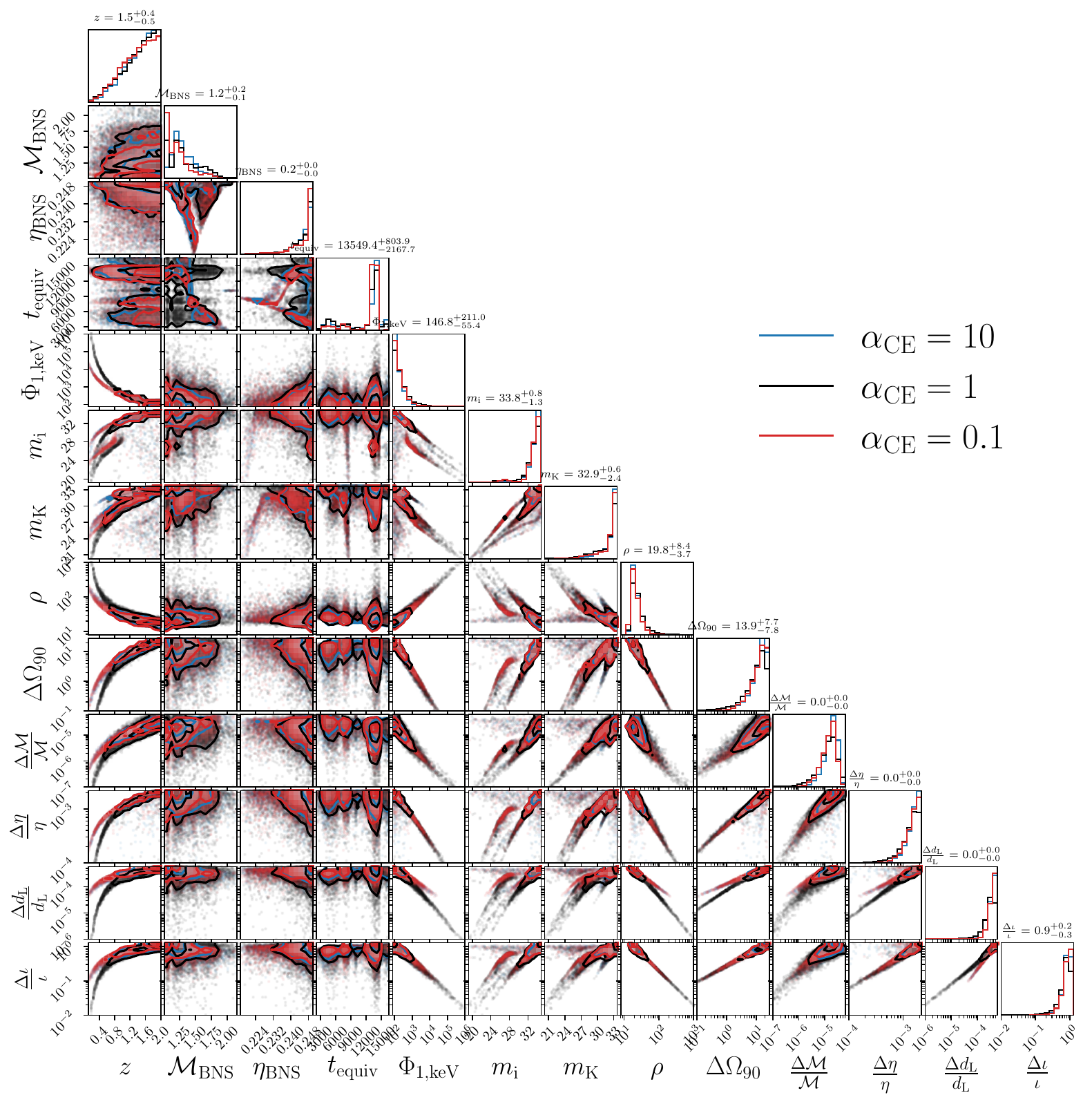}
\caption{
The output of the multi-messenger pipeline for the BNSs sampled from the thre merger rate densities of Fig.~\ref{F:MergerRateDensities}. In addition to all of the same GW and EM observables plotted in Fig.~\ref{F:cornerSubthresh}, we also show the distributions of sampled redshifts and masses used to compute the EM and GW observables. A single value of orbital inclination, $\iota = \pi/4$ is assumed for all BNSs. Only one GW detector network from Fig.~\ref{F:cornerSubthresh}, CE 40km + CE 20km + ET 20 km, is used for each of the three cosmological populations, and the fiducial ISM density $\rho_{\rm ISM, fid} = \num{1.5e-25} {\rm g}/{\rm cm}^3$ is assumed for the sGRB afterglow and KNR.} \label{F:cornerSampledObservables}
\end{figure*}
    
We then sample BNS populations from these merger rate densities, and apply our multi-messenger modeling pipeline to compute the corresponding EM and GW observables of the three $\alpha_{\rm CE}$ cases. 
In Figure \ref{F:cornerSampledObservables} we show the sampled redshift and mass distributions along with their corresponding EM and GW observables, where we show equal numbers of BNSs in each case, despite the difference in amplitude of the merger rate density, to clearly compare intrinsic EM and GW distributions across $\alpha_{\rm CE}$ independent of cosmic rates. 
The distributions of redshifts between the three $\alpha_{\rm CE}$ cases shown in Fig.~\ref{F:cornerSampledObservables} are similar due to the similar shape of the merger rate densities, and their differences in amplitude are integrated away. However, the distributions of masses reflect the differences in the BNS masses from the COMPAS predictions for the three $\alpha_{\rm CE}$ cases, which result in subtle differences in the EM and GW observables. The fiducial case ($\alpha_{\rm CE} = 1.0$) produces more higher mass BNSs resulting in more KN that are brighter and GW mergers with higher SNR (and hence smaller uncertainty on the luminosity distance) compared to the low and high $\alpha_{\rm CE}$ cases. Similarly, higher chirp mass BNSs yield higher ejecta velocity and lower ejecta mass, producing shorter KNR timescale $t_{\rm equiv}$. 
Compared with the BNS population in Fig.~\ref{F:cornerSubthresh} which assumed a range of orbital inclinations, these distributions of observables have lower variance due to a fixed inclination of $\pi/4$. 

Finally, we compute detection rates as detailed in Section~\ref{subsec:BNSformation}. The resulting detection-rate estimates for the three common-envelope efficiencies are summarized in Table~\ref{tab:rates_detection}. Across all models, the joint GW and EM detection fractions remain at the percent level, mostly dominated by the requirements on EM brightness. Nevertheless, the large differences in the underlying volumetric merger rates lead to order-of-magnitude variations in the expected annual detection counts. In particular, the $\alpha_{\rm CE}=10$ model yields the highest intrinsic merger rate and, correspondingly, the largest multi-messenger detection rate, while $\alpha_{\rm CE}=1$ produces the lowest rate. This illustrates the sensitivity of joint detection expectations to the population-synthesis assumptions that regulate binary evolution pathways.

Our analysis builds on recent work by \cite{2025A&A...697A..36L} and \cite{2022MNRAS.513.4159P}. Like those studies, we explore how a main uncertainty in binary evolution, embodied by the common-envelope efficiency $\alpha_{\rm CE}$, affects multi-messenger detection expectations. For example, \cite{2022MNRAS.513.4159P} performed a set of population-synthesis runs for multiple $\alpha_{\rm CE}$ choices and folded the resulting catalogs through GW network sensitivities, low-latency sky-localization estimates, and GRB/KN models to predict joint detection counts, and report SNRs, localization areas and detection rates but do not present GW parameter-estimation distributions (e.g. per-event posteriors or relative uncertainties). For another example, \cite{2025A&A...697A..36L} present an end-to-end forecast that includes Fisher-matrix GW parameter-uncertainty estimates together with Rubin. 
They compare populations for two CE efficiencies ($\alpha_{\rm CE}=0.5$ and $\alpha_{\rm CE}=1.0$) and quantify how the detection rate changes between these choices. However, Loffredo \emph{et al.} do not explore the wide $\alpha_{\rm CE}> 1$ regime, and their population construction differs from ours: they generate BNS catalogs with the SEVN machinery and a continuous metallicity evolution prescription tied to cosmic star formation, rather than assembling discrete, metallicity-resolved yield tables on a $Z$ grid and performing an explicit cosmic integration as we do, and they restrict their EM forecasts to Rubin $(g,i)$ follow-up, whereas we include \textit{Roman} K-band detectability. Generally, these studies find that variations in CE efficiency produce order-of-magnitude changes in merger rates and corresponding detection counts. 

\begin{table}[t]
\centering
\caption{Summary of volumetric merger rates and multi-messenger detection expectations for three common-envelope efficiencies $\alpha_{\rm CE}$.  $R_{\rm tot}$ is the total volumetric merger rate; $f_{\rm det}$ is the fraction of simulated events that satisfy the joint GW and EM detection criteria; $R_{\rm det}$ is the expected number of detections per year.}
\label{tab:rates_detection}
\begin{tabular}{>{\centering\arraybackslash}p{0.3in}|>{\centering\arraybackslash}p{0.7in}>{\centering\arraybackslash}p{0.4in}>{\centering\arraybackslash}p{0.7in}}
$\alpha_{\rm CE}$ & $R_{\rm tot}$ [yr$^{-1}$] & $f_{\rm det}$ & $R_{\rm det}$ [yr$^{-1}$] \\
\hline
0.1  & $1.29905\times10^{5}$  & 0.03  & $3.90\times10^{3}$ \\
1    & $3.18474\times10^{4}$  & 0.02  & $6.37\times10^{2}$ \\
10   & $1.46247\times10^{6}$  & 0.04  & $5.85\times10^{4}$ \\
\end{tabular}
\end{table}

In our summary results (Table~\ref{tab:rates_detection}), moving from $\alpha_{\rm CE}=1$ to $\alpha_{\rm CE}=0.1$ corresponds to an increase in detected events by a factor of $\approx 6$, while moving to $\alpha_{\rm CE}=10$ yields a relative increase of $\approx 92$. By contrast, Loffredo et al.\ compare $\alpha_{\rm CE}=0.5$ and $\alpha_{\rm CE}=1.0$ and report a noticeable uplift in their predicted optical yields when moving from $\alpha_{\rm CE}=0.5$ to $\alpha_{\rm CE}=1.0$; however, they do not explore $\alpha_{\rm CE}$ values greater than 1, and their reported increase is not easily comparable to our analysis due to differences in population construction and metallicity treatment. 
Also, Loffredo \emph{et al.} utilize a more realistic model of the sGRB afterglow as ours does not include inclination effects, further complicating direct comparisons. Along with a better model for the KNR, these are important directions for future work in our model. 
While both our study and Loffredo \emph{et al.} agree qualitatively that CE physics strongly modulates multi-messenger yields, the quantitative sensitivity to $\alpha_{\rm CE}$ depends on the population synthesis implementation, the range of $\alpha_{\rm CE}$ explored, and the adopted metallicity and star formation rate modeling. Therefore, these ingredients constitute important multi-messenger modeling systematics in studies of BNSs that will need to be better understood in preparation for future facilities. 

We emphasize that progenitor uncertainties can affect the shape and amplitude of $R(z,M)$ and the distributions of EM and GW observables. Both of these present as systematics in the computed detection rates. In our realistic (hierarchically sampled) BNS populations, these distinctions are blurred, representing a complicated systematic from untangling the dominant mechanisms that determine $R(z,M)$ coupled with those that determine the observables. 
A full accounting of multi-messenger modeling systematics arising from progenitor population assumptions is beyond the scope of the present study. 
Our analysis above shows that astrophysical processes that control the shape and amplitude of the BNS mass and formation time distributions can affect the merger rate density and hence the EM and GW observables. 
Some further important examples, which we leave for a future study, include: 
(1) uncertain mass transfer parameters such as the critical mass ratio (related to the mass-radius exponents of stellar donors) used to determine the stability and outcome of mass transfer episodes \cite{2024MNRAS.530.3706D}
and the accreted fraction $f_{\rm acc}$ which controls the efficiency of accretion during SMT \cite{Broekgaarden2022};  
(2) different prescriptions for modeling of stellar winds as they control mass loss and can be strongly metallicity-dependent for high mass ZAMS stars, implying a coupling with models for the redshift evolution of cosmic star formation history that may produce a degeneracy with other processes (e.g. stellar evolution timescales and formation time) in the redshift-metallicity space; 
(3) and parameters that are important for models of the EM and GW observables such as the maximum mass, i.e. the Tolman-Oppenheimer-Volkoff mass $m_{\rm TOV}$, and equation of state of neutron stars. For example, a higher $m_{\rm TOV}$ or softer equation of state may produce higher merger rates of BNSs (as more stars can avoid forming into black holes \cite{2018MNRAS.480.2011G,2018MNRAS.474.2937C}, or can survive their natal kicks due to smaller gravitational mass \cite{2020MNRAS.499.3214M} and possibly shorter delay times, respectively), and may affect the shapes of the neutron star mass distributions and merger rates and hence the distributions of EM and GW observables and detection rates in nontrivial ways. 
These systematics reflect the complex parameter hierarchy that can complicate multi-messenger modeling and thus data analysis of future detectors in the joint progenitor-EM-GW parameter space.

Our results highlight that uncertainties in the common-envelope efficiency $\alpha_{\rm CE}$ propagate non-trivially into every stage of multi-messenger forecasting, affecting the intrinsic merger rate, the redshift and mass distributions, the GW signal strengths, the EM peak magnitudes, and ultimately the joint detection rate by more than two orders of magnitude across the range $\alpha_{\rm CE}=0.1$ to $10$.  This sensitivity illustrates the core point of this study: as next-generation facilities will detect $\sim10^{3}$ to $10^{5}$ BNS mergers per year, systematic uncertainties in population synthesis and binary evolution will dominate the interpretation of those datasets unless explicitly incorporated into modeling methods and observational strategies.  Our findings imply that multi-messenger detection programs must be designed to remain robust against large variations in the merger rate density and small variations in the EM luminosity function and GW SNR distribution induced by different CE assumptions.  For instance, optical-only strategies optimized for a high-$\alpha_{\rm CE}$ universe (with intrinsically brighter, more abundant KNe) risk underperforming if Nature favors lower CE efficiencies, whereas incorporating NIR facilities (e.g., Roman) mitigates this sensitivity by stabilizing detection completeness across a wider range of progenitor models.  Likewise, the subtle $\alpha_{\rm CE}$-dependent variation in localization areas and distance uncertainties suggests that follow-up resource allocation, tiling depth, and cadence should be dynamically optimized using hierarchical population models rather than fixed assumptions.  More broadly, our analysis indicates that  detection strategies for future GW–EM networks will benefit from treating population synthesis as a source of astrophysical systematics, to ensure both robust event recovery and unbiased inference of BNS physics in the multi-messenger era. 

Other main examples of multi-messenger modeling systematics indicated in our work include: 
\begin{itemize}
    \item As the BNS chirp mass $\mathcal{M}_{\rm BNS}$ and mass ratio $q$ for a specific realization (ie a single hyperplane in the $\gtrsim 30$ dimensional progenitor binary parameter space) from \textsc{COMPAS} are passed to \textsc{MOSFiT} and \texttt{gwfast} to compute the KN and GW observables, which introduce further free parameters such as the distance and inclination, the observables carry imprints of the specific realization of the binary parameter space. This motivates the creation of a catalogue of theoretical multi-messenger sources to aid exploration of the binary parameter space and anticipate systematics. 
    
    \item Numerical relativity simulations serve as a linchpin for connecting the pre- and post-merger systems, and underlie models of the KN light curves and GW waveform. 
    This implies that the assumptions of the numerical relativity simulations, e.g. scope of coverage in the parameter space, numerical stability, etc., are translated into multi-messenger systematics with complicated common origin but that manifest differently, e.g. the GW waveforms are sensitive to the spin but scale trivially with the total mass, whereas the KN simulations dependence on the total mass is complicated and the spin dependence is unknown. 
    
    \item The models we use have different assumptions for the orbital inclination: \textsc{COMPAS} evolves the neutron star spin directions which may evolve under relativistic spin precession to modulate the inclination, \texttt{gwfast} self-consistently treats the inclination for such precession effects, \textsc{MOSFiT} incorporates the affect of inclination as viewing different components of the KN geometry and hence emission, and we currently assume in our model that only on-axis sGRBs are observed (which we discuss further in Section~\ref{sec:ConcDisc}). 
    
    \item Predictions for KNRs rely on detailed simulations \cite{Liu2024} and semi-analytic models are in a nascient stage of development, a topic we leave to work in preparation \cite{Steinle2025}. Coupled with the GRB afterglow via the ISM density $\rho_{\rm ISM}$, our work implies a way to jointly constrain $\rho_{\rm ISM}$ via GW and EM data from BNS mergers and a model such as ours for the KNR and GRB afterglow; conversely, not accounting for this in a multi-messenger model would cause systematic uncertainty in the KNR and afterglow parameter determinations.  
\end{itemize}

Differences between binary population synthesis approaches are not well understood and result in different astrophysical source inference. For example, numerous methods exist for modeling mass transfer processes such as CEE and SMT. 
These are crucial for determining the compact object mass and spin distributions, and the number of observable BNSs, which can quickly vanish depending on the parameters governing the mass transfer suggesting steep features in the population parameter space. Other processes are important for determining the number of expected BNS mergers, such as stellar winds which are vital for determining compact object mass and spin distributions \cite{Renzo2017}. 
Similar neutron star masses can occur for high and low metallicity stars \cite{vanSon2025}, a degeneracy that may be broken by a multi-messenger approach with independent measurements of the neutron star masses through GWs and KNe observations and of the metallicity in the proximity of the source from EM observations. 
Also, natal kicks can unbind binaries, drastically decreasing their observable number and affecting orbital parameters. In the course of a binary's evolution, both stars may experience natal kicks when they form compact objects in gravitational collapse. 
Mass transfer likely circularizes the binary after the first kick, but the binary is generically with nonzero eccentricity after the second kick which precedes BNS formation. 
These are well-studied uncertainties, but their impact on multi-messenger population studies is an open problem \cite{Richards2023}. 
While isolated binaries are expected to mostly circularize due to GW emission during the long inspiral phase before entering a terrestrial GW detection band, residual eccentricity from the natal kick of the secondary star will introduce a systematic that couples \cite{Fumagalli2024} with uncertainties of the spin evolution \cite{Steinle2021}. We considered the canonical isolated formation channel for simplicity, but a generic analysis would examine mixed contributions from multiple channels.   

Individual-binary systematics from GW analysis transcend to population-level systematics in the usual compact-binary  population analysis, i.e., hierarchical Bayesian inference with selection effects \cite{Vitale2022}. As the number of observable mergers is theoretically sensitive to the region of the progenitor parameter space from which the mergers originated, computational limitations prevent replacement of the power laws that govern the population parameters with robust astrophysical population models, injecting artificial biases in the inferred population hyper-posteriors. 
Consequently, there is great reciprocity in the development of  hierarchical population analysis and source astrophysical models that consistently account for the formation of BNSs through cosmic expansion history, metallicity-dependent star formation rate evolution, and mixed populations from various channels. A main uncertainty are the formation and delay times used to parameterize the time of formation from ZAMS to BNS merger which depend on the uncertainties of binary evolution. The population synthesis framework \textsc{POSYDON} \cite{Fragos2023,Andrews2024} attempts to bridge some of these hurdles by combining detailed stellar evolution physics with binary population astrophysics (via machine learning) and post-processed cosmological evolution, and serves as an interesting case of next generation population synthesis frameworks. 

Determining the equation of state of dense matter is a prime open question of modern physics and is probed in nuclear experiments and astrophysical observations of neutron stars \cite{Lattimer2010,Lattimer2021}, with multi-messenger astronomy leading an important contribution \cite{Radice2018,Bauswein2019,Margalit2019a,Margalit2019b,Marczenko2020,Guven2020,Dietrich2020,Nicholl2021,Lund2024,Breschi2024,Guo2024,Ecker2025,Perna2025}. The equation of state presents a particularly challenging systematic \cite{Legred2024,BabiucHamilton2025}, since it determines the masses and radii of the neutron stars but is probed via measurements of the masses and radii, implying a feedback loop between probing and modeling the equation of state. These challenges transcend to analyses that depend on GW equation of state constraints, such as tests of nuclear physics \cite{Burgio2021} and of dark matter \cite{Ellis2018}. 
Also, $m_{\rm TOV}$ serves as the upper limit before the neutron star collapses into a black hole. In binary population synthesis modeling, this is commonly chosen to be $m_{\rm TOV} = 2.5\Msol$ motivated by observations \cite{Alsing2018,Ai2020,Rocha2021}, but is uncertain and can range from $\sim 2$ to $\sim$ 2.5 \cite{Kalogera1996,Lattimer2004,Chamel2013}. 
$m_{\rm TOV}$ is important for the BNS GW waveform and KN light curve as these rely on numerical relativity simulations of BNS mergers, and can cause systematic uncertainty in multi-messenger analyses \cite{Margalit2017}. When applied to data analysis of BNS mergers, this uncertainty translates into bias in parameter estimation of the tidal deformability parameters, which depend on the masses, radii, and spins of the neutron stars but the latter two have similar challenges as the binary eccentricity and the former two are intimately coupled with the equation of state, itself a great open problem that requires accurate GW waveforms \cite{Williams2024}.  

Astrophysical frameworks such as ours will be critical for understanding the array of systematic errors and uncertainties that can influence data analysis of multi-messenger sources.

\section{Conclusions}
\label{sec:ConcDisc}

The emergence of multi-messenger astrophysics has rejuvenated studies of stellar populations across the cosmos. While generating much optimism for the prospects of discovery with future facilities, it also reveals the necessity for advancing beyond our current theoretical tools in preparation of the future. Although just the tip of the iceberg, our study is the first to show specific examples of the complicated parameter space that is constructed by the assumptions and parameterizations of these theoretical models. 
Combining state-of-the-art models for binary population astrophysics and multi-messenger observables, our results illuminate key areas where modeling biases and systematic uncertainties are likely to arise when utilizing future datasets of multiple messengers, so called \emph{multi-messenger systematics}. 

For our fiducial BNS population, obtained with a binary population synthesis model, we examined two cases. In the first case where we assume the same distance and inclination for each BNS, we find inverse correlation between brighter KN and louder GWs, e.g. as smaller $\mathcal{M}_{\rm BNS}$ gives lower GW SNR and larger sky area uncertainty but brighter KN due to larger ejecta mass \cite{Nicholl2021}. 

In the second case, we assign a distribution of distances and inclinations for the BNS population, whose observables are volume-weighted for distributions of parameters of a realistic source population. The SNRs of these BNSs for various GW detector networks can straddle the boundary of detectability, revealing subpopulations of \emph{subthreshold} binaries which are important for multi-messenger campaigns \cite{Du2024}. We demonstrate that GW170817-like binaries from the isolated channel are rarely observable with LVK (i.e. the tail of the green distribution in the seventh column of Fig.~\ref{F:cornerSubthresh}) but large fractions are more observable with future GW networks. 
The correlations between the EM and GW observables are primarily determined by their dependence on the distance, where brighter KNe and afterglows arise from systems with smaller distances and are thus correlated with the highest GW signal-to-noise ratio and smallest fractional uncertainties.  
Likewise, the existence of a GW network with at least two next generation detectors is crucial for sky localizability with $\Delta\Omega_{90} \lesssim 100$ ${\rm sq.}^2$, consistent with detailed estimations of EM counterparts \cite{Nicholl2025,Andreoni2024a,Andreoni2024b,Bisero2025}, and breaking parameter degeneracies such as the well-known general relativistic degeneracies between the mass parameters and between the distance and inclination angle. 
However, this is model dependent and might be sensitive to the shape of the distributions of redshift and inclination assigned to the BNS population. 
We find that inclusion of a 20km ET detector in future networks composed of CE detectors will enable improved sky localization, even when there's only two detectors in the network, due to differences in detector sensitivity. 
As described in Sec.~\ref{subsec:Systematics}, aspects of the models can impact the parameter correlations, implying a great potential for systematic biases in progenitor analysis of data from future facilities. 
The observables can be degenerate in the BNS binary parameters, for example the KNe light curve magnitudes and the time intervals that the KNe light curves remain within 1 magnitude of their peaks which nontrivially fill the parameter space. 
We contend that such degeneracies can be broken by, in principle, combining EM and GW datasets. 

In our third case, we employ state-of-the-art merger rate calculations by varying the metallicity of the binary stars and convolving their redshift formation efficiency with their delay and formation times. This allows us to self-consistently sample a new population from the merger rate and \textsc{COMPAS} output, and to compute the resulting EM and GW observables and the joint detection rate. 
Importantly, we apply this procedure to quantitatively demonstrate how binary evolution uncertainties can transmit to the detection rate through the merger rate and EM and GW observables. The example we explore is the efficiency of CEE which is known to control the number of observable BNS mergers. By exploring this efficiency at three values (our fiducial $\alpha_{\rm CE} =1$, and $\alpha_{\rm CE} = 0.1$ and $\alpha_{\rm CE} = 10$), we show that the merger rate density amplitude can change significantly and that the the EM and GW observables can also differ: for example, the non-monotonic dependence of the merger rate density on $\alpha_{\rm CE}$ is translated into non-monotonicity of the detection rate, while the dependence of the observables on the mass distributions on varying $\alpha_{\rm CE}$ is sub-dominant. 
Across our populations, we find that most KNe produce KNRs with characteristic timescale $\sim 10^4$ days with a long tail of shorter-lived KNRs that correspond to BNSs with higher chirp masses, lower ejecta masses, and higher ejecta velocities - demonstrating the KNR population's dependence on the BNS merger and KN explosion properties. 
This also indicates that degeneracies can arise in the parameter space due to the same EM and GW observables arising from different microphysics assumptions. We compare our results to other similar studies, and we provide a roadmap for future analyses of multi-messenger modeling systematics with a detailed discussion of astrophysical sources of uncertainty and implications for observational strategies. 

While we adopt an on–axis top–hat jet approximation for simplicity, observational studies indicate that sGRB jet half-opening angles are typically small (median values of a few degrees; see \cite{2015ApJ...815..102F} and references therein), implying that many detected bursts are viewed within or near the jet core.  High-resolution radio observations of GW170817 further demonstrate that at least some BNS mergers launch structured relativistic jets \cite{2019Sci...363..968G}.  Analytic and numerical studies show that, for the steep angular luminosity profiles commonly inferred for short GRBs, both prompt and afterglow peak fluxes decrease sharply with viewing angle \cite{2020MNRAS.493.3521B}, and population-level modeling of structured jets suggests similarly rapid angular fall-offs \cite{2023A&A...680A..45S}.  Under these steep-profile assumptions, strongly off-axis afterglows contribute only a modest fraction to flux-limited cosmological samples, so our on-axis approximation is appropriate for capturing the bulk of the detectable population at $z \gtrsim 0.1$.  Nonetheless, we note that the quantitative off-axis contribution remains model-dependent, and incorporating structured-jet and inclination effects in future work will refine flux predictions in a survey-specific manner.  

Frameworks such as ours will be important for
\begin{itemize}
    \item quantifying the systematics between models of progenitor and source evolution and signals, e.g. the KN light curves and GW waveform rely on numerical relativity in related but different ways with differing assumptions and conventions which would lead to systematic biases, e.g. \cite{2025PhRvD.111b3049G}, in multi-messenger studies with future facilities; 
    
    \item finding selection effects that span the entire process of multi-messenger source evolution and emission, which will be used to inform detection campaigns and population data analysis as is done for current detectors and theoretical pipelines; 
    
    \item mapping the full parameter space of binary stellar evolution from formation to merger and parameter estimation of GW and EM signals to be encoded by machine learning algorithms for feasible multi-messenger data interpretation of statistically large populations of BNSs \cite{Dax2025}; and
    
    \item jointly constraining the local ISM density of BNS mergers with multi-messenger modeling of GRB afterglows \cite{Duque2020} and KNRs (and possibly KN afterglows). Prospects for observing KNRs are optimistic with current and future detectors \cite{KumarAcharya2025}, which can be used to search for GWs in archival GW detector data \cite{Liu2022}. 
\end{itemize}

Ultimately, our work reveals that, in order to prepare for the future detectors, we will need to either start making models to be more reliably integrated together or building cross-disciplinary and collaborative models together -- or both -- and the success of analyses of future observing campaigns depends on our ability to do so.

\acknowledgements
The Authors thank Michael Kesden and Michael Landry for insightful discussions, and the anonymous referee for constructive feedback. 
N.S. and S.S.H. are supported by the Natural Sciences and Engineering Research Council of Canada through the Canada Research Chairs and Discovery Grants programs.
MN is supported by the European Research Council (ERC) under the European Union’s Horizon 2020 research and innovation programme (grant agreement No.~948381).
IW is supported by the UKRI Science and Technology Facilities Council (STFC). BPG acknowledges support from STFC grant No. ST/Y002253/1 and The Leverhulme Trust grant No. RPG-2024-117.
Simulations in this paper made use of the COMPAS rapid binary population synthesis code (version 02.42.01) which is freely available at \href{http://github.com/TeamCOMPAS/COMPAS}{http://github.com/TeamCOMPAS/COMPAS}, and of \textsc{PyCBC} \cite{pycbc}, \textsc{Astropy} \cite{Astropy}, \textsc{h5py} \cite{h5py}, \textsc{corner} \cite{corner}, and \textsc{matplotlib} \cite{matplotlib}.
This research was enabled by use of the GREX cluster (https://um-grex.github.io/grex-docs/friends/alliancecan/) and the Digital Research Alliance of Canada (alliancecan.ca).

\newpage

\bibliography{multimess.bib}

\providecommand{\noopsort}[1]{}
\begin{thebibliography}{219}%
\makeatletter
\providecommand \@ifxundefined [1]{%
 \@ifx{#1\undefined}
}%
\providecommand \@ifnum [1]{%
 \ifnum #1\expandafter \@firstoftwo
 \else \expandafter \@secondoftwo
 \fi
}%
\providecommand \@ifx [1]{%
 \ifx #1\expandafter \@firstoftwo
 \else \expandafter \@secondoftwo
 \fi
}%
\providecommand \natexlab [1]{#1}%
\providecommand \enquote  [1]{``#1''}%
\providecommand \bibnamefont  [1]{#1}%
\providecommand \bibfnamefont [1]{#1}%
\providecommand \citenamefont [1]{#1}%
\providecommand \href@noop [0]{\@secondoftwo}%
\providecommand \href [0]{\begingroup \@sanitize@url \@href}%
\providecommand \@href[1]{\@@startlink{#1}\@@href}%
\providecommand \@@href[1]{\endgroup#1\@@endlink}%
\providecommand \@sanitize@url [0]{\catcode `\\12\catcode `\$12\catcode `\&12\catcode `\#12\catcode `\^12\catcode `\_12\catcode `\%12\relax}%
\providecommand \@@startlink[1]{}%
\providecommand \@@endlink[0]{}%
\providecommand \url  [0]{\begingroup\@sanitize@url \@url }%
\providecommand \@url [1]{\endgroup\@href {#1}{\urlprefix }}%
\providecommand \urlprefix  [0]{URL }%
\providecommand \Eprint [0]{\href }%
\providecommand \doibase [0]{https://doi.org/}%
\providecommand \selectlanguage [0]{\@gobble}%
\providecommand \bibinfo  [0]{\@secondoftwo}%
\providecommand \bibfield  [0]{\@secondoftwo}%
\providecommand \translation [1]{[#1]}%
\providecommand \BibitemOpen [0]{}%
\providecommand \bibitemStop [0]{}%
\providecommand \bibitemNoStop [0]{.\EOS\space}%
\providecommand \EOS [0]{\spacefactor3000\relax}%
\providecommand \BibitemShut  [1]{\csname bibitem#1\endcsname}%
\let\auto@bib@innerbib\@empty
\bibitem [{\citenamefont {{Abbott}}\ \emph {et~al.}(2017{\natexlab{a}})\citenamefont {{Abbott}} \emph {et~al.}}]{GW170817discovery}%
  \BibitemOpen
  \bibfield  {author} {\bibinfo {author} {\bibfnamefont {B.~P.}\ \bibnamefont {{Abbott}}} \emph {et~al.},\ }\href {https://doi.org/10.1103/PhysRevLett.119.161101} {\bibfield  {journal} {\bibinfo  {journal} {\prl}\ }\textbf {\bibinfo {volume} {119}},\ \bibinfo {eid} {161101} (\bibinfo {year} {2017}{\natexlab{a}})},\ \Eprint {https://arxiv.org/abs/1710.05832} {arXiv:1710.05832 [gr-qc]} \BibitemShut {NoStop}%
\bibitem [{\citenamefont {{Abbott}}\ \emph {et~al.}(2019{\natexlab{a}})\citenamefont {{Abbott}} \emph {et~al.}}]{GW170817redeux}%
  \BibitemOpen
  \bibfield  {author} {\bibinfo {author} {\bibfnamefont {B.~P.}\ \bibnamefont {{Abbott}}} \emph {et~al.},\ }\href {https://doi.org/10.1103/PhysRevX.9.011001} {\bibfield  {journal} {\bibinfo  {journal} {Physical Review X}\ }\textbf {\bibinfo {volume} {9}},\ \bibinfo {eid} {011001} (\bibinfo {year} {2019}{\natexlab{a}})},\ \Eprint {https://arxiv.org/abs/1805.11579} {arXiv:1805.11579 [gr-qc]} \BibitemShut {NoStop}%
\bibitem [{\citenamefont {{Abbott}}\ \emph {et~al.}(2017{\natexlab{b}})\citenamefont {{Abbott}} \emph {et~al.}}]{GW170817mmobs}%
  \BibitemOpen
  \bibfield  {author} {\bibinfo {author} {\bibfnamefont {B.~P.}\ \bibnamefont {{Abbott}}} \emph {et~al.},\ }\href {https://doi.org/10.3847/2041-8213/aa91c9} {\bibfield  {journal} {\bibinfo  {journal} {\apjl}\ }\textbf {\bibinfo {volume} {848}},\ \bibinfo {eid} {L12} (\bibinfo {year} {2017}{\natexlab{b}})},\ \Eprint {https://arxiv.org/abs/1710.05833} {arXiv:1710.05833 [astro-ph.HE]} \BibitemShut {NoStop}%
\bibitem [{\citenamefont {{Abbott}}\ \emph {et~al.}(2017{\natexlab{c}})\citenamefont {{Abbott}} \emph {et~al.}}]{GW170817grb}%
  \BibitemOpen
  \bibfield  {author} {\bibinfo {author} {\bibfnamefont {B.~P.}\ \bibnamefont {{Abbott}}} \emph {et~al.},\ }\href {https://doi.org/10.3847/2041-8213/aa920c} {\bibfield  {journal} {\bibinfo  {journal} {\apjl}\ }\textbf {\bibinfo {volume} {848}},\ \bibinfo {eid} {L13} (\bibinfo {year} {2017}{\natexlab{c}})},\ \Eprint {https://arxiv.org/abs/1710.05834} {arXiv:1710.05834 [astro-ph.HE]} \BibitemShut {NoStop}%
\bibitem [{\citenamefont {{Goldstein}}\ \emph {et~al.}(2017)\citenamefont {{Goldstein}}, \citenamefont {{Veres}}, \citenamefont {{Burns}} \emph {et~al.}}]{Goldstein2017}%
  \BibitemOpen
  \bibfield  {author} {\bibinfo {author} {\bibfnamefont {A.}~\bibnamefont {{Goldstein}}}, \bibinfo {author} {\bibfnamefont {P.}~\bibnamefont {{Veres}}}, \bibinfo {author} {\bibfnamefont {E.}~\bibnamefont {{Burns}}}, \emph {et~al.},\ }\href {https://doi.org/10.3847/2041-8213/aa8f41} {\bibfield  {journal} {\bibinfo  {journal} {\apjl}\ }\textbf {\bibinfo {volume} {848}},\ \bibinfo {eid} {L14} (\bibinfo {year} {2017})},\ \Eprint {https://arxiv.org/abs/1710.05446} {arXiv:1710.05446 [astro-ph.HE]} \BibitemShut {NoStop}%
\bibitem [{\citenamefont {{Savchenko}}\ \emph {et~al.}(2017)\citenamefont {{Savchenko}}, \citenamefont {{Ferrigno}}, \citenamefont {{Kuulkers}} \emph {et~al.}}]{Savchenko2017}%
  \BibitemOpen
  \bibfield  {author} {\bibinfo {author} {\bibfnamefont {V.}~\bibnamefont {{Savchenko}}}, \bibinfo {author} {\bibfnamefont {C.}~\bibnamefont {{Ferrigno}}}, \bibinfo {author} {\bibfnamefont {E.}~\bibnamefont {{Kuulkers}}}, \emph {et~al.},\ }\href {https://doi.org/10.3847/2041-8213/aa8f94} {\bibfield  {journal} {\bibinfo  {journal} {\apjl}\ }\textbf {\bibinfo {volume} {848}},\ \bibinfo {eid} {L15} (\bibinfo {year} {2017})},\ \Eprint {https://arxiv.org/abs/1710.05449} {arXiv:1710.05449 [astro-ph.HE]} \BibitemShut {NoStop}%
\bibitem [{\citenamefont {{Valenti}}\ \emph {et~al.}(2017{\natexlab{a}})\citenamefont {{Valenti}}, \citenamefont {{Sand}}, \citenamefont {{Yang}}, \citenamefont {{Cappellaro}}, \citenamefont {{Tartaglia}}, \citenamefont {{Corsi}}, \citenamefont {{Jha}}, \citenamefont {{Reichart}}, \citenamefont {{Haislip}},\ and\ \citenamefont {{Kouprianov}}}]{GW170817kn}%
  \BibitemOpen
  \bibfield  {author} {\bibinfo {author} {\bibfnamefont {S.}~\bibnamefont {{Valenti}}}, \bibinfo {author} {\bibfnamefont {D.~J.}\ \bibnamefont {{Sand}}}, \bibinfo {author} {\bibfnamefont {S.}~\bibnamefont {{Yang}}}, \bibinfo {author} {\bibfnamefont {E.}~\bibnamefont {{Cappellaro}}}, \bibinfo {author} {\bibfnamefont {L.}~\bibnamefont {{Tartaglia}}}, \bibinfo {author} {\bibfnamefont {A.}~\bibnamefont {{Corsi}}}, \bibinfo {author} {\bibfnamefont {S.~W.}\ \bibnamefont {{Jha}}}, \bibinfo {author} {\bibfnamefont {D.~E.}\ \bibnamefont {{Reichart}}}, \bibinfo {author} {\bibfnamefont {J.}~\bibnamefont {{Haislip}}},\ and\ \bibinfo {author} {\bibfnamefont {V.}~\bibnamefont {{Kouprianov}}},\ }\href {https://doi.org/10.3847/2041-8213/aa8edf} {\bibfield  {journal} {\bibinfo  {journal} {\apjl}\ }\textbf {\bibinfo {volume} {848}},\ \bibinfo {eid} {L24} (\bibinfo {year} {2017}{\natexlab{a}})},\ \Eprint {https://arxiv.org/abs/1710.05854} {arXiv:1710.05854 [astro-ph.HE]} \BibitemShut {NoStop}%
\bibitem [{\citenamefont {{Arcavi}}\ \emph {et~al.}(2017)\citenamefont {{Arcavi}}, \citenamefont {{Hosseinzadeh}}, \citenamefont {{Howell}}, \citenamefont {{McCully}}, \citenamefont {{Poznanski}}, \citenamefont {{Kasen}}, \citenamefont {{Barnes}}, \citenamefont {{Zaltzman}}, \citenamefont {{Vasylyev}}, \citenamefont {{Maoz}},\ and\ \citenamefont {{Valenti}}}]{Arcavi2017}%
  \BibitemOpen
  \bibfield  {author} {\bibinfo {author} {\bibfnamefont {I.}~\bibnamefont {{Arcavi}}}, \bibinfo {author} {\bibfnamefont {G.}~\bibnamefont {{Hosseinzadeh}}}, \bibinfo {author} {\bibfnamefont {D.~A.}\ \bibnamefont {{Howell}}}, \bibinfo {author} {\bibfnamefont {C.}~\bibnamefont {{McCully}}}, \bibinfo {author} {\bibfnamefont {D.}~\bibnamefont {{Poznanski}}}, \bibinfo {author} {\bibfnamefont {D.}~\bibnamefont {{Kasen}}}, \bibinfo {author} {\bibfnamefont {J.}~\bibnamefont {{Barnes}}}, \bibinfo {author} {\bibfnamefont {M.}~\bibnamefont {{Zaltzman}}}, \bibinfo {author} {\bibfnamefont {S.}~\bibnamefont {{Vasylyev}}}, \bibinfo {author} {\bibfnamefont {D.}~\bibnamefont {{Maoz}}},\ and\ \bibinfo {author} {\bibfnamefont {S.}~\bibnamefont {{Valenti}}},\ }\href {https://doi.org/10.1038/nature24291} {\bibfield  {journal} {\bibinfo  {journal} {\nat}\ }\textbf {\bibinfo {volume} {551}},\ \bibinfo {pages} {64} (\bibinfo {year} {2017})},\ \Eprint {https://arxiv.org/abs/1710.05843} {arXiv:1710.05843 [astro-ph.HE]} \BibitemShut
  {NoStop}%
\bibitem [{\citenamefont {{Coulter}}\ \emph {et~al.}(2017)\citenamefont {{Coulter}}, \citenamefont {{Foley}}, \citenamefont {{Kilpatrick}} \emph {et~al.}}]{Coulter2017}%
  \BibitemOpen
  \bibfield  {author} {\bibinfo {author} {\bibfnamefont {D.~A.}\ \bibnamefont {{Coulter}}}, \bibinfo {author} {\bibfnamefont {R.~J.}\ \bibnamefont {{Foley}}}, \bibinfo {author} {\bibfnamefont {C.~D.}\ \bibnamefont {{Kilpatrick}}}, \emph {et~al.},\ }\href {https://doi.org/10.1126/science.aap9811} {\bibfield  {journal} {\bibinfo  {journal} {Science}\ }\textbf {\bibinfo {volume} {358}},\ \bibinfo {pages} {1556} (\bibinfo {year} {2017})},\ \Eprint {https://arxiv.org/abs/1710.05452} {arXiv:1710.05452 [astro-ph.HE]} \BibitemShut {NoStop}%
\bibitem [{\citenamefont {{Nicholl}}\ \emph {et~al.}(2017)\citenamefont {{Nicholl}}, \citenamefont {{Berger}}, \citenamefont {{Kasen}} \emph {et~al.}}]{Nicholl2017}%
  \BibitemOpen
  \bibfield  {author} {\bibinfo {author} {\bibfnamefont {M.}~\bibnamefont {{Nicholl}}}, \bibinfo {author} {\bibfnamefont {E.}~\bibnamefont {{Berger}}}, \bibinfo {author} {\bibfnamefont {D.}~\bibnamefont {{Kasen}}}, \emph {et~al.},\ }\href {https://doi.org/10.3847/2041-8213/aa9029} {\bibfield  {journal} {\bibinfo  {journal} {\apjl}\ }\textbf {\bibinfo {volume} {848}},\ \bibinfo {eid} {L18} (\bibinfo {year} {2017})},\ \Eprint {https://arxiv.org/abs/1710.05456} {arXiv:1710.05456 [astro-ph.HE]} \BibitemShut {NoStop}%
\bibitem [{\citenamefont {{Lipunov}}\ \emph {et~al.}(2018)\citenamefont {{Lipunov}}, \citenamefont {{Kornilov}}, \citenamefont {{Gorbovskoy}}, \citenamefont {{Lipunova}}, \citenamefont {{Vlasenko}}, \citenamefont {{Panchenko}}, \citenamefont {{Tyurina}},\ and\ \citenamefont {{Grinshpun}}}]{Lipunov2018}%
  \BibitemOpen
  \bibfield  {author} {\bibinfo {author} {\bibfnamefont {V.}~\bibnamefont {{Lipunov}}}, \bibinfo {author} {\bibfnamefont {V.}~\bibnamefont {{Kornilov}}}, \bibinfo {author} {\bibfnamefont {E.}~\bibnamefont {{Gorbovskoy}}}, \bibinfo {author} {\bibfnamefont {G.}~\bibnamefont {{Lipunova}}}, \bibinfo {author} {\bibfnamefont {D.}~\bibnamefont {{Vlasenko}}}, \bibinfo {author} {\bibfnamefont {I.}~\bibnamefont {{Panchenko}}}, \bibinfo {author} {\bibfnamefont {N.}~\bibnamefont {{Tyurina}}},\ and\ \bibinfo {author} {\bibfnamefont {V.}~\bibnamefont {{Grinshpun}}},\ }\href {https://doi.org/10.1016/j.newast.2018.02.004} {\bibfield  {journal} {\bibinfo  {journal} {\na}\ }\textbf {\bibinfo {volume} {63}},\ \bibinfo {pages} {48} (\bibinfo {year} {2018})},\ \Eprint {https://arxiv.org/abs/1710.05911} {arXiv:1710.05911 [astro-ph.HE]} \BibitemShut {NoStop}%
\bibitem [{\citenamefont {{Soares-Santos}}\ \emph {et~al.}(2017)\citenamefont {{Soares-Santos}}, \citenamefont {{Holz}}, \citenamefont {{Annis}} \emph {et~al.}}]{SoaresSantos2017}%
  \BibitemOpen
  \bibfield  {author} {\bibinfo {author} {\bibfnamefont {M.}~\bibnamefont {{Soares-Santos}}}, \bibinfo {author} {\bibfnamefont {D.~E.}\ \bibnamefont {{Holz}}}, \bibinfo {author} {\bibfnamefont {J.}~\bibnamefont {{Annis}}}, \emph {et~al.},\ }\href {https://doi.org/10.3847/2041-8213/aa9059} {\bibfield  {journal} {\bibinfo  {journal} {\apjl}\ }\textbf {\bibinfo {volume} {848}},\ \bibinfo {eid} {L16} (\bibinfo {year} {2017})},\ \Eprint {https://arxiv.org/abs/1710.05459} {arXiv:1710.05459 [astro-ph.HE]} \BibitemShut {NoStop}%
\bibitem [{\citenamefont {{Cowperthwaite}}\ \emph {et~al.}(2017)\citenamefont {{Cowperthwaite}}, \citenamefont {{Berger}}, \citenamefont {{Villar}} \emph {et~al.}}]{Cowperthwaite2017}%
  \BibitemOpen
  \bibfield  {author} {\bibinfo {author} {\bibfnamefont {P.~S.}\ \bibnamefont {{Cowperthwaite}}}, \bibinfo {author} {\bibfnamefont {E.}~\bibnamefont {{Berger}}}, \bibinfo {author} {\bibfnamefont {V.~A.}\ \bibnamefont {{Villar}}}, \emph {et~al.},\ }\href {https://doi.org/10.3847/2041-8213/aa8fc7} {\bibfield  {journal} {\bibinfo  {journal} {\apjl}\ }\textbf {\bibinfo {volume} {848}},\ \bibinfo {eid} {L17} (\bibinfo {year} {2017})},\ \Eprint {https://arxiv.org/abs/1710.05840} {arXiv:1710.05840 [astro-ph.HE]} \BibitemShut {NoStop}%
\bibitem [{\citenamefont {{Smartt}}\ \emph {et~al.}(2017)\citenamefont {{Smartt}}, \citenamefont {{Chen}}, \citenamefont {{Jerkstrand}} \emph {et~al.}}]{Smartt2017}%
  \BibitemOpen
  \bibfield  {author} {\bibinfo {author} {\bibfnamefont {S.~J.}\ \bibnamefont {{Smartt}}}, \bibinfo {author} {\bibfnamefont {T.~W.}\ \bibnamefont {{Chen}}}, \bibinfo {author} {\bibfnamefont {A.}~\bibnamefont {{Jerkstrand}}}, \emph {et~al.},\ }\href {https://doi.org/10.1038/nature24303} {\bibfield  {journal} {\bibinfo  {journal} {\nat}\ }\textbf {\bibinfo {volume} {551}},\ \bibinfo {pages} {75} (\bibinfo {year} {2017})},\ \Eprint {https://arxiv.org/abs/1710.05841} {arXiv:1710.05841 [astro-ph.HE]} \BibitemShut {NoStop}%
\bibitem [{\citenamefont {{Tanvir}}\ \emph {et~al.}(2017)\citenamefont {{Tanvir}}, \citenamefont {{Levan}}, \citenamefont {{Gonz{\'a}lez-Fern{\'a}ndez}} \emph {et~al.}}]{Tanvir2017}%
  \BibitemOpen
  \bibfield  {author} {\bibinfo {author} {\bibfnamefont {N.~R.}\ \bibnamefont {{Tanvir}}}, \bibinfo {author} {\bibfnamefont {A.~J.}\ \bibnamefont {{Levan}}}, \bibinfo {author} {\bibfnamefont {C.}~\bibnamefont {{Gonz{\'a}lez-Fern{\'a}ndez}}}, \emph {et~al.},\ }\href {https://doi.org/10.3847/2041-8213/aa90b6} {\bibfield  {journal} {\bibinfo  {journal} {\apjl}\ }\textbf {\bibinfo {volume} {848}},\ \bibinfo {eid} {L27} (\bibinfo {year} {2017})},\ \Eprint {https://arxiv.org/abs/1710.05455} {arXiv:1710.05455 [astro-ph.HE]} \BibitemShut {NoStop}%
\bibitem [{\citenamefont {{Valenti}}\ \emph {et~al.}(2017{\natexlab{b}})\citenamefont {{Valenti}}, \citenamefont {{Sand}}, \citenamefont {{Yang}}, \citenamefont {{Cappellaro}}, \citenamefont {{Tartaglia}}, \citenamefont {{Corsi}}, \citenamefont {{Jha}}, \citenamefont {{Reichart}}, \citenamefont {{Haislip}},\ and\ \citenamefont {{Kouprianov}}}]{Valenti2017}%
  \BibitemOpen
  \bibfield  {author} {\bibinfo {author} {\bibfnamefont {S.}~\bibnamefont {{Valenti}}}, \bibinfo {author} {\bibfnamefont {D.~J.}\ \bibnamefont {{Sand}}}, \bibinfo {author} {\bibfnamefont {S.}~\bibnamefont {{Yang}}}, \bibinfo {author} {\bibfnamefont {E.}~\bibnamefont {{Cappellaro}}}, \bibinfo {author} {\bibfnamefont {L.}~\bibnamefont {{Tartaglia}}}, \bibinfo {author} {\bibfnamefont {A.}~\bibnamefont {{Corsi}}}, \bibinfo {author} {\bibfnamefont {S.~W.}\ \bibnamefont {{Jha}}}, \bibinfo {author} {\bibfnamefont {D.~E.}\ \bibnamefont {{Reichart}}}, \bibinfo {author} {\bibfnamefont {J.}~\bibnamefont {{Haislip}}},\ and\ \bibinfo {author} {\bibfnamefont {V.}~\bibnamefont {{Kouprianov}}},\ }\href {https://doi.org/10.3847/2041-8213/aa8edf} {\bibfield  {journal} {\bibinfo  {journal} {\apjl}\ }\textbf {\bibinfo {volume} {848}},\ \bibinfo {eid} {L24} (\bibinfo {year} {2017}{\natexlab{b}})},\ \Eprint {https://arxiv.org/abs/1710.05854} {arXiv:1710.05854 [astro-ph.HE]} \BibitemShut {NoStop}%
\bibitem [{\citenamefont {{Kasen}}\ \emph {et~al.}(2017)\citenamefont {{Kasen}}, \citenamefont {{Metzger}}, \citenamefont {{Barnes}}, \citenamefont {{Quataert}},\ and\ \citenamefont {{Ramirez-Ruiz}}}]{2017Natur.551...80K}%
  \BibitemOpen
  \bibfield  {author} {\bibinfo {author} {\bibfnamefont {D.}~\bibnamefont {{Kasen}}}, \bibinfo {author} {\bibfnamefont {B.}~\bibnamefont {{Metzger}}}, \bibinfo {author} {\bibfnamefont {J.}~\bibnamefont {{Barnes}}}, \bibinfo {author} {\bibfnamefont {E.}~\bibnamefont {{Quataert}}},\ and\ \bibinfo {author} {\bibfnamefont {E.}~\bibnamefont {{Ramirez-Ruiz}}},\ }\href {https://doi.org/10.1038/nature24453} {\bibfield  {journal} {\bibinfo  {journal} {\nat}\ }\textbf {\bibinfo {volume} {551}},\ \bibinfo {pages} {80} (\bibinfo {year} {2017})},\ \Eprint {https://arxiv.org/abs/1710.05463} {arXiv:1710.05463 [astro-ph.HE]} \BibitemShut {NoStop}%
\bibitem [{\citenamefont {{Drout}}\ \emph {et~al.}(2017)\citenamefont {{Drout}}, \citenamefont {{Piro}}, \citenamefont {{Shappee}} \emph {et~al.}}]{2017Sci...358.1570D}%
  \BibitemOpen
  \bibfield  {author} {\bibinfo {author} {\bibfnamefont {M.~R.}\ \bibnamefont {{Drout}}}, \bibinfo {author} {\bibfnamefont {A.~L.}\ \bibnamefont {{Piro}}}, \bibinfo {author} {\bibfnamefont {B.~J.}\ \bibnamefont {{Shappee}}}, \emph {et~al.},\ }\href {https://doi.org/10.1126/science.aaq0049} {\bibfield  {journal} {\bibinfo  {journal} {Science}\ }\textbf {\bibinfo {volume} {358}},\ \bibinfo {pages} {1570} (\bibinfo {year} {2017})},\ \Eprint {https://arxiv.org/abs/1710.05443} {arXiv:1710.05443 [astro-ph.HE]} \BibitemShut {NoStop}%
\bibitem [{\citenamefont {{D'Avanzo}}\ \emph {et~al.}(2018)\citenamefont {{D'Avanzo}}, \citenamefont {{Campana}}, \citenamefont {{Salafia}} \emph {et~al.}}]{DAvanzo2018}%
  \BibitemOpen
  \bibfield  {author} {\bibinfo {author} {\bibfnamefont {P.}~\bibnamefont {{D'Avanzo}}}, \bibinfo {author} {\bibfnamefont {S.}~\bibnamefont {{Campana}}}, \bibinfo {author} {\bibfnamefont {O.~S.}\ \bibnamefont {{Salafia}}}, \emph {et~al.},\ }\href {https://doi.org/10.1051/0004-6361/201832664} {\bibfield  {journal} {\bibinfo  {journal} {\aap}\ }\textbf {\bibinfo {volume} {613}},\ \bibinfo {eid} {L1} (\bibinfo {year} {2018})},\ \Eprint {https://arxiv.org/abs/1801.06164} {arXiv:1801.06164 [astro-ph.HE]} \BibitemShut {NoStop}%
\bibitem [{\citenamefont {{Lyman}}\ \emph {et~al.}(2018)\citenamefont {{Lyman}}, \citenamefont {{Lamb}}, \citenamefont {{Levan}} \emph {et~al.}}]{Lyman2018}%
  \BibitemOpen
  \bibfield  {author} {\bibinfo {author} {\bibfnamefont {J.~D.}\ \bibnamefont {{Lyman}}}, \bibinfo {author} {\bibfnamefont {G.~P.}\ \bibnamefont {{Lamb}}}, \bibinfo {author} {\bibfnamefont {A.~J.}\ \bibnamefont {{Levan}}}, \emph {et~al.},\ }\href {https://doi.org/10.1038/s41550-018-0511-3} {\bibfield  {journal} {\bibinfo  {journal} {Nature Astronomy}\ }\textbf {\bibinfo {volume} {2}},\ \bibinfo {pages} {751} (\bibinfo {year} {2018})},\ \Eprint {https://arxiv.org/abs/1801.02669} {arXiv:1801.02669 [astro-ph.HE]} \BibitemShut {NoStop}%
\bibitem [{\citenamefont {{Lamb}}\ \emph {et~al.}(2019)\citenamefont {{Lamb}}, \citenamefont {{Lyman}}, \citenamefont {{Levan}}, \citenamefont {{Tanvir}}, \citenamefont {{Kangas}}, \citenamefont {{Fruchter}}, \citenamefont {{Gompertz}}, \citenamefont {{Hjorth}}, \citenamefont {{Mandel}}, \citenamefont {{Oates}}, \citenamefont {{Steeghs}},\ and\ \citenamefont {{Wiersema}}}]{Lamb2019}%
  \BibitemOpen
  \bibfield  {author} {\bibinfo {author} {\bibfnamefont {G.~P.}\ \bibnamefont {{Lamb}}}, \bibinfo {author} {\bibfnamefont {J.~D.}\ \bibnamefont {{Lyman}}}, \bibinfo {author} {\bibfnamefont {A.~J.}\ \bibnamefont {{Levan}}}, \bibinfo {author} {\bibfnamefont {N.~R.}\ \bibnamefont {{Tanvir}}}, \bibinfo {author} {\bibfnamefont {T.}~\bibnamefont {{Kangas}}}, \bibinfo {author} {\bibfnamefont {A.~S.}\ \bibnamefont {{Fruchter}}}, \bibinfo {author} {\bibfnamefont {B.}~\bibnamefont {{Gompertz}}}, \bibinfo {author} {\bibfnamefont {J.}~\bibnamefont {{Hjorth}}}, \bibinfo {author} {\bibfnamefont {I.}~\bibnamefont {{Mandel}}}, \bibinfo {author} {\bibfnamefont {S.~R.}\ \bibnamefont {{Oates}}}, \bibinfo {author} {\bibfnamefont {D.}~\bibnamefont {{Steeghs}}},\ and\ \bibinfo {author} {\bibfnamefont {K.}~\bibnamefont {{Wiersema}}},\ }\href {https://doi.org/10.3847/2041-8213/aaf96b} {\bibfield  {journal} {\bibinfo  {journal} {\apjl}\ }\textbf {\bibinfo {volume} {870}},\ \bibinfo {eid} {L15} (\bibinfo {year} {2019})},\ \Eprint
  {https://arxiv.org/abs/1811.11491} {arXiv:1811.11491 [astro-ph.HE]} \BibitemShut {NoStop}%
\bibitem [{\citenamefont {{Balasubramanian}}\ \emph {et~al.}(2022)\citenamefont {{Balasubramanian}}, \citenamefont {{Corsi}}, \citenamefont {{Mooley}}, \citenamefont {{Hotokezaka}}, \citenamefont {{Kaplan}}, \citenamefont {{Frail}}, \citenamefont {{Hallinan}}, \citenamefont {{Lazzati}},\ and\ \citenamefont {{Murphy}}}]{Balasubramanian2022}%
  \BibitemOpen
  \bibfield  {author} {\bibinfo {author} {\bibfnamefont {A.}~\bibnamefont {{Balasubramanian}}}, \bibinfo {author} {\bibfnamefont {A.}~\bibnamefont {{Corsi}}}, \bibinfo {author} {\bibfnamefont {K.~P.}\ \bibnamefont {{Mooley}}}, \bibinfo {author} {\bibfnamefont {K.}~\bibnamefont {{Hotokezaka}}}, \bibinfo {author} {\bibfnamefont {D.~L.}\ \bibnamefont {{Kaplan}}}, \bibinfo {author} {\bibfnamefont {D.~A.}\ \bibnamefont {{Frail}}}, \bibinfo {author} {\bibfnamefont {G.}~\bibnamefont {{Hallinan}}}, \bibinfo {author} {\bibfnamefont {D.}~\bibnamefont {{Lazzati}}},\ and\ \bibinfo {author} {\bibfnamefont {E.~J.}\ \bibnamefont {{Murphy}}},\ }\href {https://doi.org/10.3847/1538-4357/ac9133} {\bibfield  {journal} {\bibinfo  {journal} {\apj}\ }\textbf {\bibinfo {volume} {938}},\ \bibinfo {eid} {12} (\bibinfo {year} {2022})},\ \Eprint {https://arxiv.org/abs/2205.14788} {arXiv:2205.14788 [astro-ph.HE]} \BibitemShut {NoStop}%
\bibitem [{\citenamefont {{Safi-Harb}}\ \emph {et~al.}(2019{\natexlab{a}})\citenamefont {{Safi-Harb}}, \citenamefont {{Doerksen}}, \citenamefont {{Rogers}},\ and\ \citenamefont {{Fryer}}}]{Safi-Harb2019}%
  \BibitemOpen
  \bibfield  {author} {\bibinfo {author} {\bibfnamefont {S.}~\bibnamefont {{Safi-Harb}}}, \bibinfo {author} {\bibfnamefont {N.}~\bibnamefont {{Doerksen}}}, \bibinfo {author} {\bibfnamefont {A.}~\bibnamefont {{Rogers}}},\ and\ \bibinfo {author} {\bibfnamefont {C.~L.}\ \bibnamefont {{Fryer}}},\ }\href {https://doi.org/10.48550/arXiv.1812.11320} {\bibfield  {journal} {\bibinfo  {journal} {\jrasc}\ }\textbf {\bibinfo {volume} {113}},\ \bibinfo {pages} {7} (\bibinfo {year} {2019}{\natexlab{a}})},\ \Eprint {https://arxiv.org/abs/1812.11320} {arXiv:1812.11320 [astro-ph.HE]} \BibitemShut {NoStop}%
\bibitem [{\citenamefont {{Troja}}\ \emph {et~al.}(2020)\citenamefont {{Troja}}, \citenamefont {{van Eerten}}, \citenamefont {{Zhang}}, \citenamefont {{Ryan}}, \citenamefont {{Piro}}, \citenamefont {{Ricci}}, \citenamefont {{O'Connor}}, \citenamefont {{Wieringa}}, \citenamefont {{Cenko}},\ and\ \citenamefont {{Sakamoto}}}]{Troja2020}%
  \BibitemOpen
  \bibfield  {author} {\bibinfo {author} {\bibfnamefont {E.}~\bibnamefont {{Troja}}}, \bibinfo {author} {\bibfnamefont {H.}~\bibnamefont {{van Eerten}}}, \bibinfo {author} {\bibfnamefont {B.}~\bibnamefont {{Zhang}}}, \bibinfo {author} {\bibfnamefont {G.}~\bibnamefont {{Ryan}}}, \bibinfo {author} {\bibfnamefont {L.}~\bibnamefont {{Piro}}}, \bibinfo {author} {\bibfnamefont {R.}~\bibnamefont {{Ricci}}}, \bibinfo {author} {\bibfnamefont {B.}~\bibnamefont {{O'Connor}}}, \bibinfo {author} {\bibfnamefont {M.~H.}\ \bibnamefont {{Wieringa}}}, \bibinfo {author} {\bibfnamefont {S.~B.}\ \bibnamefont {{Cenko}}},\ and\ \bibinfo {author} {\bibfnamefont {T.}~\bibnamefont {{Sakamoto}}},\ }\href {https://doi.org/10.1093/mnras/staa2626} {\bibfield  {journal} {\bibinfo  {journal} {\mnras}\ }\textbf {\bibinfo {volume} {498}},\ \bibinfo {pages} {5643} (\bibinfo {year} {2020})},\ \Eprint {https://arxiv.org/abs/2006.01150} {arXiv:2006.01150 [astro-ph.HE]} \BibitemShut {NoStop}%
\bibitem [{\citenamefont {{Ren}}\ and\ \citenamefont {{Dai}}(2022)}]{Ren2022}%
  \BibitemOpen
  \bibfield  {author} {\bibinfo {author} {\bibfnamefont {J.}~\bibnamefont {{Ren}}}\ and\ \bibinfo {author} {\bibfnamefont {Z.~G.}\ \bibnamefont {{Dai}}},\ }\href {https://doi.org/10.1093/mnras/stac797} {\bibfield  {journal} {\bibinfo  {journal} {\mnras}\ }\textbf {\bibinfo {volume} {512}},\ \bibinfo {pages} {5572} (\bibinfo {year} {2022})},\ \Eprint {https://arxiv.org/abs/2203.08576} {arXiv:2203.08576 [astro-ph.HE]} \BibitemShut {NoStop}%
\bibitem [{\citenamefont {{Hajela}}\ \emph {et~al.}(2022)\citenamefont {{Hajela}}, \citenamefont {{Margutti}}, \citenamefont {{Bright}} \emph {et~al.}}]{Hajela2022}%
  \BibitemOpen
  \bibfield  {author} {\bibinfo {author} {\bibfnamefont {A.}~\bibnamefont {{Hajela}}}, \bibinfo {author} {\bibfnamefont {R.}~\bibnamefont {{Margutti}}}, \bibinfo {author} {\bibfnamefont {J.~S.}\ \bibnamefont {{Bright}}}, \emph {et~al.},\ }\href {https://doi.org/10.3847/2041-8213/ac504a} {\bibfield  {journal} {\bibinfo  {journal} {\apjl}\ }\textbf {\bibinfo {volume} {927}},\ \bibinfo {eid} {L17} (\bibinfo {year} {2022})},\ \Eprint {https://arxiv.org/abs/2104.02070} {arXiv:2104.02070 [astro-ph.HE]} \BibitemShut {NoStop}%
\bibitem [{\citenamefont {{Margalit}}\ and\ \citenamefont {{Metzger}}(2017)}]{Margalit2017}%
  \BibitemOpen
  \bibfield  {author} {\bibinfo {author} {\bibfnamefont {B.}~\bibnamefont {{Margalit}}}\ and\ \bibinfo {author} {\bibfnamefont {B.~D.}\ \bibnamefont {{Metzger}}},\ }\href {https://doi.org/10.3847/2041-8213/aa991c} {\bibfield  {journal} {\bibinfo  {journal} {\apjl}\ }\textbf {\bibinfo {volume} {850}},\ \bibinfo {eid} {L19} (\bibinfo {year} {2017})},\ \Eprint {https://arxiv.org/abs/1710.05938} {arXiv:1710.05938 [astro-ph.HE]} \BibitemShut {NoStop}%
\bibitem [{\citenamefont {{Abbott}}\ \emph {et~al.}(2017{\natexlab{d}})\citenamefont {{Abbott}} \emph {et~al.}}]{GW170817postmerger}%
  \BibitemOpen
  \bibfield  {author} {\bibinfo {author} {\bibfnamefont {B.~P.}\ \bibnamefont {{Abbott}}} \emph {et~al.},\ }\href {https://doi.org/10.3847/2041-8213/aa9a35} {\bibfield  {journal} {\bibinfo  {journal} {\apjl}\ }\textbf {\bibinfo {volume} {851}},\ \bibinfo {eid} {L16} (\bibinfo {year} {2017}{\natexlab{d}})},\ \Eprint {https://arxiv.org/abs/1710.09320} {arXiv:1710.09320 [astro-ph.HE]} \BibitemShut {NoStop}%
\bibitem [{\citenamefont {{Albert}}\ \emph {et~al.}(2017)\citenamefont {{Albert}} \emph {et~al.}}]{GW170817cosmicrays}%
  \BibitemOpen
  \bibfield  {author} {\bibinfo {author} {\bibfnamefont {A.}~\bibnamefont {{Albert}}} \emph {et~al.},\ }\href {https://doi.org/10.3847/2041-8213/aa9aed} {\bibfield  {journal} {\bibinfo  {journal} {\apjl}\ }\textbf {\bibinfo {volume} {850}},\ \bibinfo {eid} {L35} (\bibinfo {year} {2017})},\ \Eprint {https://arxiv.org/abs/1710.05839} {arXiv:1710.05839 [astro-ph.HE]} \BibitemShut {NoStop}%
\bibitem [{\citenamefont {{Rodrigues}}\ \emph {et~al.}(2019)\citenamefont {{Rodrigues}}, \citenamefont {{Biehl}}, \citenamefont {{Boncioli}},\ and\ \citenamefont {{Taylor}}}]{Rodrigues2019}%
  \BibitemOpen
  \bibfield  {author} {\bibinfo {author} {\bibfnamefont {X.}~\bibnamefont {{Rodrigues}}}, \bibinfo {author} {\bibfnamefont {D.}~\bibnamefont {{Biehl}}}, \bibinfo {author} {\bibfnamefont {D.}~\bibnamefont {{Boncioli}}},\ and\ \bibinfo {author} {\bibfnamefont {A.~M.}\ \bibnamefont {{Taylor}}},\ }\href {https://doi.org/10.1016/j.astropartphys.2018.10.007} {\bibfield  {journal} {\bibinfo  {journal} {Astroparticle Physics}\ }\textbf {\bibinfo {volume} {106}},\ \bibinfo {pages} {10} (\bibinfo {year} {2019})},\ \Eprint {https://arxiv.org/abs/1806.01624} {arXiv:1806.01624 [astro-ph.HE]} \BibitemShut {NoStop}%
\bibitem [{\citenamefont {{Burns}}(2020)}]{Burns2020}%
  \BibitemOpen
  \bibfield  {author} {\bibinfo {author} {\bibfnamefont {E.}~\bibnamefont {{Burns}}},\ }\href {https://doi.org/10.1007/s41114-020-00028-7} {\bibfield  {journal} {\bibinfo  {journal} {Living Reviews in Relativity}\ }\textbf {\bibinfo {volume} {23}},\ \bibinfo {eid} {4} (\bibinfo {year} {2020})},\ \Eprint {https://arxiv.org/abs/1909.06085} {arXiv:1909.06085 [astro-ph.HE]} \BibitemShut {NoStop}%
\bibitem [{\citenamefont {{Margutti}}\ and\ \citenamefont {{Chornock}}(2021)}]{Margutti2021}%
  \BibitemOpen
  \bibfield  {author} {\bibinfo {author} {\bibfnamefont {R.}~\bibnamefont {{Margutti}}}\ and\ \bibinfo {author} {\bibfnamefont {R.}~\bibnamefont {{Chornock}}},\ }\href {https://doi.org/10.1146/annurev-astro-112420-030742} {\bibfield  {journal} {\bibinfo  {journal} {\araa}\ }\textbf {\bibinfo {volume} {59}},\ \bibinfo {pages} {155} (\bibinfo {year} {2021})},\ \Eprint {https://arxiv.org/abs/2012.04810} {arXiv:2012.04810 [astro-ph.HE]} \BibitemShut {NoStop}%
\bibitem [{\citenamefont {{Nicholl}}\ and\ \citenamefont {{Andreoni}}(2025)}]{Nicholl2025}%
  \BibitemOpen
  \bibfield  {author} {\bibinfo {author} {\bibfnamefont {M.}~\bibnamefont {{Nicholl}}}\ and\ \bibinfo {author} {\bibfnamefont {I.}~\bibnamefont {{Andreoni}}},\ }\href {https://doi.org/10.1098/rsta.2024.0126} {\bibfield  {journal} {\bibinfo  {journal} {Philosophical Transactions of the Royal Society of London Series A}\ }\textbf {\bibinfo {volume} {383}},\ \bibinfo {eid} {20240126} (\bibinfo {year} {2025})},\ \Eprint {https://arxiv.org/abs/2410.18274} {arXiv:2410.18274 [astro-ph.HE]} \BibitemShut {NoStop}%
\bibitem [{\citenamefont {{Chan}}\ \emph {et~al.}(2018)\citenamefont {{Chan}}, \citenamefont {{Messenger}}, \citenamefont {{Heng}},\ and\ \citenamefont {{Hendry}}}]{Chan2018}%
  \BibitemOpen
  \bibfield  {author} {\bibinfo {author} {\bibfnamefont {M.~L.}\ \bibnamefont {{Chan}}}, \bibinfo {author} {\bibfnamefont {C.}~\bibnamefont {{Messenger}}}, \bibinfo {author} {\bibfnamefont {I.~S.}\ \bibnamefont {{Heng}}},\ and\ \bibinfo {author} {\bibfnamefont {M.}~\bibnamefont {{Hendry}}},\ }\href {https://doi.org/10.1103/PhysRevD.97.123014} {\bibfield  {journal} {\bibinfo  {journal} {\prd}\ }\textbf {\bibinfo {volume} {97}},\ \bibinfo {eid} {123014} (\bibinfo {year} {2018})},\ \Eprint {https://arxiv.org/abs/1803.09680} {arXiv:1803.09680 [astro-ph.HE]} \BibitemShut {NoStop}%
\bibitem [{\citenamefont {{Colombo}}\ \emph {et~al.}(2022)\citenamefont {{Colombo}}, \citenamefont {{Salafia}}, \citenamefont {{Gabrielli}}, \citenamefont {{Ghirlanda}}, \citenamefont {{Giacomazzo}}, \citenamefont {{Perego}},\ and\ \citenamefont {{Colpi}}}]{Colombo2022}%
  \BibitemOpen
  \bibfield  {author} {\bibinfo {author} {\bibfnamefont {A.}~\bibnamefont {{Colombo}}}, \bibinfo {author} {\bibfnamefont {O.~S.}\ \bibnamefont {{Salafia}}}, \bibinfo {author} {\bibfnamefont {F.}~\bibnamefont {{Gabrielli}}}, \bibinfo {author} {\bibfnamefont {G.}~\bibnamefont {{Ghirlanda}}}, \bibinfo {author} {\bibfnamefont {B.}~\bibnamefont {{Giacomazzo}}}, \bibinfo {author} {\bibfnamefont {A.}~\bibnamefont {{Perego}}},\ and\ \bibinfo {author} {\bibfnamefont {M.}~\bibnamefont {{Colpi}}},\ }\href {https://doi.org/10.3847/1538-4357/ac8d00} {\bibfield  {journal} {\bibinfo  {journal} {\apj}\ }\textbf {\bibinfo {volume} {937}},\ \bibinfo {eid} {79} (\bibinfo {year} {2022})},\ \Eprint {https://arxiv.org/abs/2204.07592} {arXiv:2204.07592 [astro-ph.HE]} \BibitemShut {NoStop}%
\bibitem [{\citenamefont {{Ronchini}}\ \emph {et~al.}(2022)\citenamefont {{Ronchini}}, \citenamefont {{Branchesi}}, \citenamefont {{Oganesyan}}, \citenamefont {{Banerjee}}, \citenamefont {{Dupletsa}}, \citenamefont {{Ghirlanda}}, \citenamefont {{Harms}}, \citenamefont {{Mapelli}},\ and\ \citenamefont {{Santoliquido}}}]{Ronchini2022}%
  \BibitemOpen
  \bibfield  {author} {\bibinfo {author} {\bibfnamefont {S.}~\bibnamefont {{Ronchini}}}, \bibinfo {author} {\bibfnamefont {M.}~\bibnamefont {{Branchesi}}}, \bibinfo {author} {\bibfnamefont {G.}~\bibnamefont {{Oganesyan}}}, \bibinfo {author} {\bibfnamefont {B.}~\bibnamefont {{Banerjee}}}, \bibinfo {author} {\bibfnamefont {U.}~\bibnamefont {{Dupletsa}}}, \bibinfo {author} {\bibfnamefont {G.}~\bibnamefont {{Ghirlanda}}}, \bibinfo {author} {\bibfnamefont {J.}~\bibnamefont {{Harms}}}, \bibinfo {author} {\bibfnamefont {M.}~\bibnamefont {{Mapelli}}},\ and\ \bibinfo {author} {\bibfnamefont {F.}~\bibnamefont {{Santoliquido}}},\ }\href {https://doi.org/10.1051/0004-6361/202243705} {\bibfield  {journal} {\bibinfo  {journal} {\aap}\ }\textbf {\bibinfo {volume} {665}},\ \bibinfo {eid} {A97} (\bibinfo {year} {2022})},\ \Eprint {https://arxiv.org/abs/2204.01746} {arXiv:2204.01746 [astro-ph.HE]} \BibitemShut {NoStop}%
\bibitem [{\citenamefont {{Li}}\ \emph {et~al.}(2022)\citenamefont {{Li}}, \citenamefont {{Heng}}, \citenamefont {{Chan}}, \citenamefont {{Messenger}},\ and\ \citenamefont {{Fan}}}]{Li2022}%
  \BibitemOpen
  \bibfield  {author} {\bibinfo {author} {\bibfnamefont {Y.}~\bibnamefont {{Li}}}, \bibinfo {author} {\bibfnamefont {I.~S.}\ \bibnamefont {{Heng}}}, \bibinfo {author} {\bibfnamefont {M.~L.}\ \bibnamefont {{Chan}}}, \bibinfo {author} {\bibfnamefont {C.}~\bibnamefont {{Messenger}}},\ and\ \bibinfo {author} {\bibfnamefont {X.}~\bibnamefont {{Fan}}},\ }\href {https://doi.org/10.1103/PhysRevD.105.043010} {\bibfield  {journal} {\bibinfo  {journal} {\prd}\ }\textbf {\bibinfo {volume} {105}},\ \bibinfo {eid} {043010} (\bibinfo {year} {2022})},\ \Eprint {https://arxiv.org/abs/2109.07389} {arXiv:2109.07389 [astro-ph.IM]} \BibitemShut {NoStop}%
\bibitem [{\citenamefont {{Corsi}}\ \emph {et~al.}(2024)\citenamefont {{Corsi}}, \citenamefont {{Barsotti}}, \citenamefont {{Berti}} \emph {et~al.}}]{Corsi2024}%
  \BibitemOpen
  \bibfield  {author} {\bibinfo {author} {\bibfnamefont {A.}~\bibnamefont {{Corsi}}}, \bibinfo {author} {\bibfnamefont {L.}~\bibnamefont {{Barsotti}}}, \bibinfo {author} {\bibfnamefont {E.}~\bibnamefont {{Berti}}}, \emph {et~al.},\ }\href {https://doi.org/10.3389/fspas.2024.1386748} {\bibfield  {journal} {\bibinfo  {journal} {Frontiers in Astronomy and Space Sciences}\ }\textbf {\bibinfo {volume} {11}},\ \bibinfo {eid} {1386748} (\bibinfo {year} {2024})},\ \Eprint {https://arxiv.org/abs/2402.13445} {arXiv:2402.13445 [astro-ph.HE]} \BibitemShut {NoStop}%
\bibitem [{\citenamefont {{Li}}\ \emph {et~al.}(2024)\citenamefont {{Li}}, \citenamefont {{Heng}}, \citenamefont {{Chan}}, \citenamefont {{Fan}},\ and\ \citenamefont {{Gou}}}]{Li2024}%
  \BibitemOpen
  \bibfield  {author} {\bibinfo {author} {\bibfnamefont {Y.}~\bibnamefont {{Li}}}, \bibinfo {author} {\bibfnamefont {I.~S.}\ \bibnamefont {{Heng}}}, \bibinfo {author} {\bibfnamefont {M.~L.}\ \bibnamefont {{Chan}}}, \bibinfo {author} {\bibfnamefont {X.}~\bibnamefont {{Fan}}},\ and\ \bibinfo {author} {\bibfnamefont {L.}~\bibnamefont {{Gou}}},\ }\href {https://doi.org/10.1103/PhysRevD.110.043001} {\bibfield  {journal} {\bibinfo  {journal} {\prd}\ }\textbf {\bibinfo {volume} {110}},\ \bibinfo {eid} {043001} (\bibinfo {year} {2024})},\ \Eprint {https://arxiv.org/abs/2406.18228} {arXiv:2406.18228 [astro-ph.CO]} \BibitemShut {NoStop}%
\bibitem [{\citenamefont {{Colombo}}\ \emph {et~al.}(2025)\citenamefont {{Colombo}}, \citenamefont {{Sharan Salafia}}, \citenamefont {{Ghirlanda}}, \citenamefont {{Iacovelli}}, \citenamefont {{Mancarella}}, \citenamefont {{Broekgaarden}}, \citenamefont {{Nava}}, \citenamefont {{Giacomazzo}},\ and\ \citenamefont {{Colpi}}}]{Colombo2025}%
  \BibitemOpen
  \bibfield  {author} {\bibinfo {author} {\bibfnamefont {A.}~\bibnamefont {{Colombo}}}, \bibinfo {author} {\bibfnamefont {O.}~\bibnamefont {{Sharan Salafia}}}, \bibinfo {author} {\bibfnamefont {G.}~\bibnamefont {{Ghirlanda}}}, \bibinfo {author} {\bibfnamefont {F.}~\bibnamefont {{Iacovelli}}}, \bibinfo {author} {\bibfnamefont {M.}~\bibnamefont {{Mancarella}}}, \bibinfo {author} {\bibfnamefont {F.~S.}\ \bibnamefont {{Broekgaarden}}}, \bibinfo {author} {\bibfnamefont {L.}~\bibnamefont {{Nava}}}, \bibinfo {author} {\bibfnamefont {B.}~\bibnamefont {{Giacomazzo}}},\ and\ \bibinfo {author} {\bibfnamefont {M.}~\bibnamefont {{Colpi}}},\ }\href {https://doi.org/10.48550/arXiv.2503.00116} {\bibfield  {journal} {\bibinfo  {journal} {arXiv e-prints}\ ,\ \bibinfo {eid} {arXiv:2503.00116}} (\bibinfo {year} {2025})},\ \Eprint {https://arxiv.org/abs/2503.00116} {arXiv:2503.00116 [astro-ph.HE]} \BibitemShut {NoStop}%
\bibitem [{\citenamefont {{O'Shaughnessy}}\ \emph {et~al.}(2005)\citenamefont {{O'Shaughnessy}}, \citenamefont {{Kim}}, \citenamefont {{Fragos}}, \citenamefont {{Kalogera}},\ and\ \citenamefont {{Belczynski}}}]{OShaughnessy2005}%
  \BibitemOpen
  \bibfield  {author} {\bibinfo {author} {\bibfnamefont {R.}~\bibnamefont {{O'Shaughnessy}}}, \bibinfo {author} {\bibfnamefont {C.}~\bibnamefont {{Kim}}}, \bibinfo {author} {\bibfnamefont {T.}~\bibnamefont {{Fragos}}}, \bibinfo {author} {\bibfnamefont {V.}~\bibnamefont {{Kalogera}}},\ and\ \bibinfo {author} {\bibfnamefont {K.}~\bibnamefont {{Belczynski}}},\ }\href {https://doi.org/10.1086/468180} {\bibfield  {journal} {\bibinfo  {journal} {\apj}\ }\textbf {\bibinfo {volume} {633}},\ \bibinfo {pages} {1076} (\bibinfo {year} {2005})},\ \Eprint {https://arxiv.org/abs/astro-ph/0504479} {arXiv:astro-ph/0504479 [astro-ph]} \BibitemShut {NoStop}%
\bibitem [{\citenamefont {{de Mink}}\ and\ \citenamefont {{Mandel}}(2016)}]{deMink2016}%
  \BibitemOpen
  \bibfield  {author} {\bibinfo {author} {\bibfnamefont {S.~E.}\ \bibnamefont {{de Mink}}}\ and\ \bibinfo {author} {\bibfnamefont {I.}~\bibnamefont {{Mandel}}},\ }\href {https://doi.org/10.1093/mnras/stw1219} {\bibfield  {journal} {\bibinfo  {journal} {\mnras}\ }\textbf {\bibinfo {volume} {460}},\ \bibinfo {pages} {3545} (\bibinfo {year} {2016})},\ \Eprint {https://arxiv.org/abs/1603.02291} {arXiv:1603.02291 [astro-ph.HE]} \BibitemShut {NoStop}%
\bibitem [{\citenamefont {{Vitale}}\ \emph {et~al.}(2017)\citenamefont {{Vitale}}, \citenamefont {{Lynch}}, \citenamefont {{Sturani}},\ and\ \citenamefont {{Graff}}}]{Vitale2017}%
  \BibitemOpen
  \bibfield  {author} {\bibinfo {author} {\bibfnamefont {S.}~\bibnamefont {{Vitale}}}, \bibinfo {author} {\bibfnamefont {R.}~\bibnamefont {{Lynch}}}, \bibinfo {author} {\bibfnamefont {R.}~\bibnamefont {{Sturani}}},\ and\ \bibinfo {author} {\bibfnamefont {P.}~\bibnamefont {{Graff}}},\ }\href {https://doi.org/10.1088/1361-6382/aa552e} {\bibfield  {journal} {\bibinfo  {journal} {Classical and Quantum Gravity}\ }\textbf {\bibinfo {volume} {34}},\ \bibinfo {eid} {03LT01} (\bibinfo {year} {2017})},\ \Eprint {https://arxiv.org/abs/1503.04307} {arXiv:1503.04307 [gr-qc]} \BibitemShut {NoStop}%
\bibitem [{\citenamefont {{Zevin}}\ \emph {et~al.}(2017)\citenamefont {{Zevin}}, \citenamefont {{Pankow}}, \citenamefont {{Rodriguez}}, \citenamefont {{Sampson}}, \citenamefont {{Chase}}, \citenamefont {{Kalogera}},\ and\ \citenamefont {{Rasio}}}]{Zevin2017}%
  \BibitemOpen
  \bibfield  {author} {\bibinfo {author} {\bibfnamefont {M.}~\bibnamefont {{Zevin}}}, \bibinfo {author} {\bibfnamefont {C.}~\bibnamefont {{Pankow}}}, \bibinfo {author} {\bibfnamefont {C.~L.}\ \bibnamefont {{Rodriguez}}}, \bibinfo {author} {\bibfnamefont {L.}~\bibnamefont {{Sampson}}}, \bibinfo {author} {\bibfnamefont {E.}~\bibnamefont {{Chase}}}, \bibinfo {author} {\bibfnamefont {V.}~\bibnamefont {{Kalogera}}},\ and\ \bibinfo {author} {\bibfnamefont {F.~A.}\ \bibnamefont {{Rasio}}},\ }\href {https://doi.org/10.3847/1538-4357/aa8408} {\bibfield  {journal} {\bibinfo  {journal} {\apj}\ }\textbf {\bibinfo {volume} {846}},\ \bibinfo {eid} {82} (\bibinfo {year} {2017})},\ \Eprint {https://arxiv.org/abs/1704.07379} {arXiv:1704.07379 [astro-ph.HE]} \BibitemShut {NoStop}%
\bibitem [{\citenamefont {{Taylor}}\ and\ \citenamefont {{Gerosa}}(2018)}]{Taylor2018}%
  \BibitemOpen
  \bibfield  {author} {\bibinfo {author} {\bibfnamefont {S.~R.}\ \bibnamefont {{Taylor}}}\ and\ \bibinfo {author} {\bibfnamefont {D.}~\bibnamefont {{Gerosa}}},\ }\href {https://doi.org/10.1103/PhysRevD.98.083017} {\bibfield  {journal} {\bibinfo  {journal} {\prd}\ }\textbf {\bibinfo {volume} {98}},\ \bibinfo {eid} {083017} (\bibinfo {year} {2018})},\ \Eprint {https://arxiv.org/abs/1806.08365} {arXiv:1806.08365 [astro-ph.HE]} \BibitemShut {NoStop}%
\bibitem [{\citenamefont {{Belczynski}}\ \emph {et~al.}(2018)\citenamefont {{Belczynski}}, \citenamefont {{Askar}}, \citenamefont {{Arca-Sedda}} \emph {et~al.}}]{Belczynski2018}%
  \BibitemOpen
  \bibfield  {author} {\bibinfo {author} {\bibfnamefont {K.}~\bibnamefont {{Belczynski}}}, \bibinfo {author} {\bibfnamefont {A.}~\bibnamefont {{Askar}}}, \bibinfo {author} {\bibfnamefont {M.}~\bibnamefont {{Arca-Sedda}}}, \emph {et~al.},\ }\href {https://doi.org/10.1051/0004-6361/201732428} {\bibfield  {journal} {\bibinfo  {journal} {\aap}\ }\textbf {\bibinfo {volume} {615}},\ \bibinfo {eid} {A91} (\bibinfo {year} {2018})},\ \Eprint {https://arxiv.org/abs/1712.00632} {arXiv:1712.00632 [astro-ph.HE]} \BibitemShut {NoStop}%
\bibitem [{\citenamefont {{Bouffanais}}\ \emph {et~al.}(2019)\citenamefont {{Bouffanais}}, \citenamefont {{Mapelli}}, \citenamefont {{Gerosa}}, \citenamefont {{Di Carlo}}, \citenamefont {{Giacobbo}}, \citenamefont {{Berti}},\ and\ \citenamefont {{Baibhav}}}]{Bouffanais2019}%
  \BibitemOpen
  \bibfield  {author} {\bibinfo {author} {\bibfnamefont {Y.}~\bibnamefont {{Bouffanais}}}, \bibinfo {author} {\bibfnamefont {M.}~\bibnamefont {{Mapelli}}}, \bibinfo {author} {\bibfnamefont {D.}~\bibnamefont {{Gerosa}}}, \bibinfo {author} {\bibfnamefont {U.~N.}\ \bibnamefont {{Di Carlo}}}, \bibinfo {author} {\bibfnamefont {N.}~\bibnamefont {{Giacobbo}}}, \bibinfo {author} {\bibfnamefont {E.}~\bibnamefont {{Berti}}},\ and\ \bibinfo {author} {\bibfnamefont {V.}~\bibnamefont {{Baibhav}}},\ }\href {https://doi.org/10.3847/1538-4357/ab4a79} {\bibfield  {journal} {\bibinfo  {journal} {\apj}\ }\textbf {\bibinfo {volume} {886}},\ \bibinfo {eid} {25} (\bibinfo {year} {2019})},\ \Eprint {https://arxiv.org/abs/1905.11054} {arXiv:1905.11054 [astro-ph.HE]} \BibitemShut {NoStop}%
\bibitem [{\citenamefont {{Zevin}}\ \emph {et~al.}(2021)\citenamefont {{Zevin}}, \citenamefont {{Bavera}}, \citenamefont {{Berry}}, \citenamefont {{Kalogera}}, \citenamefont {{Fragos}}, \citenamefont {{Marchant}}, \citenamefont {{Rodriguez}}, \citenamefont {{Antonini}}, \citenamefont {{Holz}},\ and\ \citenamefont {{Pankow}}}]{Zevin2020b}%
  \BibitemOpen
  \bibfield  {author} {\bibinfo {author} {\bibfnamefont {M.}~\bibnamefont {{Zevin}}}, \bibinfo {author} {\bibfnamefont {S.~S.}\ \bibnamefont {{Bavera}}}, \bibinfo {author} {\bibfnamefont {C.~P.~L.}\ \bibnamefont {{Berry}}}, \bibinfo {author} {\bibfnamefont {V.}~\bibnamefont {{Kalogera}}}, \bibinfo {author} {\bibfnamefont {T.}~\bibnamefont {{Fragos}}}, \bibinfo {author} {\bibfnamefont {P.}~\bibnamefont {{Marchant}}}, \bibinfo {author} {\bibfnamefont {C.~L.}\ \bibnamefont {{Rodriguez}}}, \bibinfo {author} {\bibfnamefont {F.}~\bibnamefont {{Antonini}}}, \bibinfo {author} {\bibfnamefont {D.~E.}\ \bibnamefont {{Holz}}},\ and\ \bibinfo {author} {\bibfnamefont {C.}~\bibnamefont {{Pankow}}},\ }\href {https://doi.org/10.3847/1538-4357/abe40e} {\bibfield  {journal} {\bibinfo  {journal} {\apj}\ }\textbf {\bibinfo {volume} {910}},\ \bibinfo {eid} {152} (\bibinfo {year} {2021})},\ \Eprint {https://arxiv.org/abs/2011.10057} {arXiv:2011.10057 [astro-ph.HE]} \BibitemShut {NoStop}%
\bibitem [{\citenamefont {{Callister}}\ \emph {et~al.}(2020)\citenamefont {{Callister}}, \citenamefont {{Fishbach}}, \citenamefont {{Holz}},\ and\ \citenamefont {{Farr}}}]{Callister2020}%
  \BibitemOpen
  \bibfield  {author} {\bibinfo {author} {\bibfnamefont {T.}~\bibnamefont {{Callister}}}, \bibinfo {author} {\bibfnamefont {M.}~\bibnamefont {{Fishbach}}}, \bibinfo {author} {\bibfnamefont {D.~E.}\ \bibnamefont {{Holz}}},\ and\ \bibinfo {author} {\bibfnamefont {W.~M.}\ \bibnamefont {{Farr}}},\ }\href {https://doi.org/10.3847/2041-8213/ab9743} {\bibfield  {journal} {\bibinfo  {journal} {\apjl}\ }\textbf {\bibinfo {volume} {896}},\ \bibinfo {eid} {L32} (\bibinfo {year} {2020})},\ \Eprint {https://arxiv.org/abs/2003.12152} {arXiv:2003.12152 [astro-ph.HE]} \BibitemShut {NoStop}%
\bibitem [{\citenamefont {{Mapelli}}(2020)}]{Mapelli2020}%
  \BibitemOpen
  \bibfield  {author} {\bibinfo {author} {\bibfnamefont {M.}~\bibnamefont {{Mapelli}}},\ }\href {https://doi.org/10.3389/fspas.2020.00038} {\bibfield  {journal} {\bibinfo  {journal} {Frontiers in Astronomy and Space Sciences}\ }\textbf {\bibinfo {volume} {7}},\ \bibinfo {eid} {38} (\bibinfo {year} {2020})},\ \Eprint {https://arxiv.org/abs/2105.12455} {arXiv:2105.12455 [astro-ph.HE]} \BibitemShut {NoStop}%
\bibitem [{\citenamefont {{Wong}}\ \emph {et~al.}(2021)\citenamefont {{Wong}}, \citenamefont {{Breivik}}, \citenamefont {{Kremer}},\ and\ \citenamefont {{Callister}}}]{Wong2021}%
  \BibitemOpen
  \bibfield  {author} {\bibinfo {author} {\bibfnamefont {K.~W.~K.}\ \bibnamefont {{Wong}}}, \bibinfo {author} {\bibfnamefont {K.}~\bibnamefont {{Breivik}}}, \bibinfo {author} {\bibfnamefont {K.}~\bibnamefont {{Kremer}}},\ and\ \bibinfo {author} {\bibfnamefont {T.}~\bibnamefont {{Callister}}},\ }\href {https://doi.org/10.1103/PhysRevD.103.083021} {\bibfield  {journal} {\bibinfo  {journal} {\prd}\ }\textbf {\bibinfo {volume} {103}},\ \bibinfo {eid} {083021} (\bibinfo {year} {2021})},\ \Eprint {https://arxiv.org/abs/2011.03564} {arXiv:2011.03564 [astro-ph.HE]} \BibitemShut {NoStop}%
\bibitem [{\citenamefont {{Bouffanais}}\ \emph {et~al.}(2021)\citenamefont {{Bouffanais}}, \citenamefont {{Mapelli}}, \citenamefont {{Santoliquido}}, \citenamefont {{Giacobbo}}, \citenamefont {{Di Carlo}}, \citenamefont {{Rastello}}, \citenamefont {{Artale}},\ and\ \citenamefont {{Iorio}}}]{Bouffanais2021}%
  \BibitemOpen
  \bibfield  {author} {\bibinfo {author} {\bibfnamefont {Y.}~\bibnamefont {{Bouffanais}}}, \bibinfo {author} {\bibfnamefont {M.}~\bibnamefont {{Mapelli}}}, \bibinfo {author} {\bibfnamefont {F.}~\bibnamefont {{Santoliquido}}}, \bibinfo {author} {\bibfnamefont {N.}~\bibnamefont {{Giacobbo}}}, \bibinfo {author} {\bibfnamefont {U.~N.}\ \bibnamefont {{Di Carlo}}}, \bibinfo {author} {\bibfnamefont {S.}~\bibnamefont {{Rastello}}}, \bibinfo {author} {\bibfnamefont {M.~C.}\ \bibnamefont {{Artale}}},\ and\ \bibinfo {author} {\bibfnamefont {G.}~\bibnamefont {{Iorio}}},\ }\href {https://doi.org/10.1093/mnras/stab2438} {\bibfield  {journal} {\bibinfo  {journal} {\mnras}\ }\textbf {\bibinfo {volume} {507}},\ \bibinfo {pages} {5224} (\bibinfo {year} {2021})},\ \Eprint {https://arxiv.org/abs/2102.12495} {arXiv:2102.12495 [astro-ph.HE]} \BibitemShut {NoStop}%
\bibitem [{\citenamefont {{Mould}}\ \emph {et~al.}(2022)\citenamefont {{Mould}}, \citenamefont {{Gerosa}},\ and\ \citenamefont {{Taylor}}}]{Mould2022}%
  \BibitemOpen
  \bibfield  {author} {\bibinfo {author} {\bibfnamefont {M.}~\bibnamefont {{Mould}}}, \bibinfo {author} {\bibfnamefont {D.}~\bibnamefont {{Gerosa}}},\ and\ \bibinfo {author} {\bibfnamefont {S.~R.}\ \bibnamefont {{Taylor}}},\ }\href@noop {} {\bibfield  {journal} {\bibinfo  {journal} {arXiv e-prints}\ ,\ \bibinfo {eid} {arXiv:2203.03651}} (\bibinfo {year} {2022})},\ \Eprint {https://arxiv.org/abs/2203.03651} {arXiv:2203.03651 [astro-ph.HE]} \BibitemShut {NoStop}%
\bibitem [{\citenamefont {{Antonelli}}\ \emph {et~al.}(2023)\citenamefont {{Antonelli}}, \citenamefont {{Kritos}}, \citenamefont {{Ng}}, \citenamefont {{Cotesta}},\ and\ \citenamefont {{Berti}}}]{Antonelli2023}%
  \BibitemOpen
  \bibfield  {author} {\bibinfo {author} {\bibfnamefont {A.}~\bibnamefont {{Antonelli}}}, \bibinfo {author} {\bibfnamefont {K.}~\bibnamefont {{Kritos}}}, \bibinfo {author} {\bibfnamefont {K.~K.~Y.}\ \bibnamefont {{Ng}}}, \bibinfo {author} {\bibfnamefont {R.}~\bibnamefont {{Cotesta}}},\ and\ \bibinfo {author} {\bibfnamefont {E.}~\bibnamefont {{Berti}}},\ }\href {https://doi.org/10.1103/PhysRevD.108.084044} {\bibfield  {journal} {\bibinfo  {journal} {\prd}\ }\textbf {\bibinfo {volume} {108}},\ \bibinfo {eid} {084044} (\bibinfo {year} {2023})},\ \Eprint {https://arxiv.org/abs/2306.11088} {arXiv:2306.11088 [gr-qc]} \BibitemShut {NoStop}%
\bibitem [{\citenamefont {{Afroz}}\ and\ \citenamefont {{Mukherjee}}(2024)}]{Afroz2024}%
  \BibitemOpen
  \bibfield  {author} {\bibinfo {author} {\bibfnamefont {S.}~\bibnamefont {{Afroz}}}\ and\ \bibinfo {author} {\bibfnamefont {S.}~\bibnamefont {{Mukherjee}}},\ }\href {https://doi.org/10.48550/arXiv.2411.07304} {\bibfield  {journal} {\bibinfo  {journal} {arXiv e-prints}\ ,\ \bibinfo {eid} {arXiv:2411.07304}} (\bibinfo {year} {2024})},\ \Eprint {https://arxiv.org/abs/2411.07304} {arXiv:2411.07304 [astro-ph.HE]} \BibitemShut {NoStop}%
\bibitem [{\citenamefont {{Postnov}}\ and\ \citenamefont {{Yungelson}}(2014)}]{Postnov2014}%
  \BibitemOpen
  \bibfield  {author} {\bibinfo {author} {\bibfnamefont {K.~A.}\ \bibnamefont {{Postnov}}}\ and\ \bibinfo {author} {\bibfnamefont {L.~R.}\ \bibnamefont {{Yungelson}}},\ }\href {https://doi.org/10.12942/lrr-2014-3} {\bibfield  {journal} {\bibinfo  {journal} {Living Reviews in Relativity}\ }\textbf {\bibinfo {volume} {17}},\ \bibinfo {eid} {3} (\bibinfo {year} {2014})},\ \Eprint {https://arxiv.org/abs/1403.4754} {arXiv:1403.4754 [astro-ph.HE]} \BibitemShut {NoStop}%
\bibitem [{\citenamefont {{Benacquista}}\ and\ \citenamefont {{Downing}}(2013)}]{Benacquista2013}%
  \BibitemOpen
  \bibfield  {author} {\bibinfo {author} {\bibfnamefont {M.~J.}\ \bibnamefont {{Benacquista}}}\ and\ \bibinfo {author} {\bibfnamefont {J.~M.~B.}\ \bibnamefont {{Downing}}},\ }\href {https://doi.org/10.12942/lrr-2013-4} {\bibfield  {journal} {\bibinfo  {journal} {Living Reviews in Relativity}\ }\textbf {\bibinfo {volume} {16}},\ \bibinfo {eid} {4} (\bibinfo {year} {2013})},\ \Eprint {https://arxiv.org/abs/1110.4423} {arXiv:1110.4423 [astro-ph.SR]} \BibitemShut {NoStop}%
\bibitem [{\citenamefont {{Spera}}\ \emph {et~al.}(2022)\citenamefont {{Spera}}, \citenamefont {{Trani}},\ and\ \citenamefont {{Mencagli}}}]{Spera2022}%
  \BibitemOpen
  \bibfield  {author} {\bibinfo {author} {\bibfnamefont {M.}~\bibnamefont {{Spera}}}, \bibinfo {author} {\bibfnamefont {A.~A.}\ \bibnamefont {{Trani}}},\ and\ \bibinfo {author} {\bibfnamefont {M.}~\bibnamefont {{Mencagli}}},\ }\href {https://doi.org/10.3390/galaxies10040076} {\bibfield  {journal} {\bibinfo  {journal} {Galaxies}\ }\textbf {\bibinfo {volume} {10}},\ \bibinfo {eid} {76} (\bibinfo {year} {2022})},\ \Eprint {https://arxiv.org/abs/2206.15392} {arXiv:2206.15392 [astro-ph.HE]} \BibitemShut {NoStop}%
\bibitem [{\citenamefont {{Ishibashi}}\ and\ \citenamefont {{Gr{\"o}bner}}(2020)}]{Ishibashi2020}%
  \BibitemOpen
  \bibfield  {author} {\bibinfo {author} {\bibfnamefont {W.}~\bibnamefont {{Ishibashi}}}\ and\ \bibinfo {author} {\bibfnamefont {M.}~\bibnamefont {{Gr{\"o}bner}}},\ }\href {https://doi.org/10.1051/0004-6361/202037799} {\bibfield  {journal} {\bibinfo  {journal} {\aap}\ }\textbf {\bibinfo {volume} {639}},\ \bibinfo {eid} {A108} (\bibinfo {year} {2020})},\ \Eprint {https://arxiv.org/abs/2006.07407} {arXiv:2006.07407 [astro-ph.GA]} \BibitemShut {NoStop}%
\bibitem [{\citenamefont {{Gr{\"o}bner}}\ \emph {et~al.}(2020)\citenamefont {{Gr{\"o}bner}}, \citenamefont {{Ishibashi}}, \citenamefont {{Tiwari}}, \citenamefont {{Haney}},\ and\ \citenamefont {{Jetzer}}}]{Grobner2020}%
  \BibitemOpen
  \bibfield  {author} {\bibinfo {author} {\bibfnamefont {M.}~\bibnamefont {{Gr{\"o}bner}}}, \bibinfo {author} {\bibfnamefont {W.}~\bibnamefont {{Ishibashi}}}, \bibinfo {author} {\bibfnamefont {S.}~\bibnamefont {{Tiwari}}}, \bibinfo {author} {\bibfnamefont {M.}~\bibnamefont {{Haney}}},\ and\ \bibinfo {author} {\bibfnamefont {P.}~\bibnamefont {{Jetzer}}},\ }\href {https://doi.org/10.1051/0004-6361/202037681} {\bibfield  {journal} {\bibinfo  {journal} {\aap}\ }\textbf {\bibinfo {volume} {638}},\ \bibinfo {eid} {A119} (\bibinfo {year} {2020})},\ \Eprint {https://arxiv.org/abs/2005.03571} {arXiv:2005.03571 [astro-ph.GA]} \BibitemShut {NoStop}%
\bibitem [{\citenamefont {{Gayathri}}\ \emph {et~al.}(2023)\citenamefont {{Gayathri}}, \citenamefont {{Wysocki}}, \citenamefont {{Yang}}, \citenamefont {{Delfavero}}, \citenamefont {{O'Shaughnessy}}, \citenamefont {{Haiman}}, \citenamefont {{Tagawa}},\ and\ \citenamefont {{Bartos}}}]{Gayathri2023}%
  \BibitemOpen
  \bibfield  {author} {\bibinfo {author} {\bibfnamefont {V.}~\bibnamefont {{Gayathri}}}, \bibinfo {author} {\bibfnamefont {D.}~\bibnamefont {{Wysocki}}}, \bibinfo {author} {\bibfnamefont {Y.}~\bibnamefont {{Yang}}}, \bibinfo {author} {\bibfnamefont {V.}~\bibnamefont {{Delfavero}}}, \bibinfo {author} {\bibfnamefont {R.}~\bibnamefont {{O'Shaughnessy}}}, \bibinfo {author} {\bibfnamefont {Z.}~\bibnamefont {{Haiman}}}, \bibinfo {author} {\bibfnamefont {H.}~\bibnamefont {{Tagawa}}},\ and\ \bibinfo {author} {\bibfnamefont {I.}~\bibnamefont {{Bartos}}},\ }\href {https://doi.org/10.3847/2041-8213/acbfb8} {\bibfield  {journal} {\bibinfo  {journal} {\apjl}\ }\textbf {\bibinfo {volume} {945}},\ \bibinfo {eid} {L29} (\bibinfo {year} {2023})},\ \Eprint {https://arxiv.org/abs/2301.04187} {arXiv:2301.04187 [gr-qc]} \BibitemShut {NoStop}%
\bibitem [{\citenamefont {{Veronesi}}\ \emph {et~al.}(2025)\citenamefont {{Veronesi}}, \citenamefont {{van Velzen}}, \citenamefont {{Rossi}},\ and\ \citenamefont {{Storey-Fisher}}}]{Veronesi2025}%
  \BibitemOpen
  \bibfield  {author} {\bibinfo {author} {\bibfnamefont {N.}~\bibnamefont {{Veronesi}}}, \bibinfo {author} {\bibfnamefont {S.}~\bibnamefont {{van Velzen}}}, \bibinfo {author} {\bibfnamefont {E.~M.}\ \bibnamefont {{Rossi}}},\ and\ \bibinfo {author} {\bibfnamefont {K.}~\bibnamefont {{Storey-Fisher}}},\ }\href {https://doi.org/10.1093/mnras/stae2575} {\bibfield  {journal} {\bibinfo  {journal} {\mnras}\ }\textbf {\bibinfo {volume} {536}},\ \bibinfo {pages} {375} (\bibinfo {year} {2025})},\ \Eprint {https://arxiv.org/abs/2407.21568} {arXiv:2407.21568 [astro-ph.HE]} \BibitemShut {NoStop}%
\bibitem [{\citenamefont {{Rodriguez}}\ \emph {et~al.}(2016)\citenamefont {{Rodriguez}}, \citenamefont {{Zevin}}, \citenamefont {{Pankow}}, \citenamefont {{Kalogera}},\ and\ \citenamefont {{Rasio}}}]{Rodriguez2016b}%
  \BibitemOpen
  \bibfield  {author} {\bibinfo {author} {\bibfnamefont {C.~L.}\ \bibnamefont {{Rodriguez}}}, \bibinfo {author} {\bibfnamefont {M.}~\bibnamefont {{Zevin}}}, \bibinfo {author} {\bibfnamefont {C.}~\bibnamefont {{Pankow}}}, \bibinfo {author} {\bibfnamefont {V.}~\bibnamefont {{Kalogera}}},\ and\ \bibinfo {author} {\bibfnamefont {F.~A.}\ \bibnamefont {{Rasio}}},\ }\href {https://doi.org/10.3847/2041-8205/832/1/L2} {\bibfield  {journal} {\bibinfo  {journal} {\apjl}\ }\textbf {\bibinfo {volume} {832}},\ \bibinfo {eid} {L2} (\bibinfo {year} {2016})},\ \Eprint {https://arxiv.org/abs/1609.05916} {arXiv:1609.05916 [astro-ph.HE]} \BibitemShut {NoStop}%
\bibitem [{\citenamefont {{Gerosa}}\ \emph {et~al.}(2018)\citenamefont {{Gerosa}}, \citenamefont {{Berti}}, \citenamefont {{O'Shaughnessy}}, \citenamefont {{Belczynski}}, \citenamefont {{Kesden}}, \citenamefont {{Wysocki}},\ and\ \citenamefont {{Gladysz}}}]{Gerosa2018}%
  \BibitemOpen
  \bibfield  {author} {\bibinfo {author} {\bibfnamefont {D.}~\bibnamefont {{Gerosa}}}, \bibinfo {author} {\bibfnamefont {E.}~\bibnamefont {{Berti}}}, \bibinfo {author} {\bibfnamefont {R.}~\bibnamefont {{O'Shaughnessy}}}, \bibinfo {author} {\bibfnamefont {K.}~\bibnamefont {{Belczynski}}}, \bibinfo {author} {\bibfnamefont {M.}~\bibnamefont {{Kesden}}}, \bibinfo {author} {\bibfnamefont {D.}~\bibnamefont {{Wysocki}}},\ and\ \bibinfo {author} {\bibfnamefont {W.}~\bibnamefont {{Gladysz}}},\ }\href {https://doi.org/10.1103/PhysRevD.98.084036} {\bibfield  {journal} {\bibinfo  {journal} {\prd}\ }\textbf {\bibinfo {volume} {98}},\ \bibinfo {eid} {084036} (\bibinfo {year} {2018})},\ \Eprint {https://arxiv.org/abs/1808.02491} {arXiv:1808.02491 [astro-ph.HE]} \BibitemShut {NoStop}%
\bibitem [{\citenamefont {{Steinle}}\ and\ \citenamefont {{Kesden}}(2021)}]{Steinle2021}%
  \BibitemOpen
  \bibfield  {author} {\bibinfo {author} {\bibfnamefont {N.}~\bibnamefont {{Steinle}}}\ and\ \bibinfo {author} {\bibfnamefont {M.}~\bibnamefont {{Kesden}}},\ }\href {https://doi.org/10.1103/PhysRevD.103.063032} {\bibfield  {journal} {\bibinfo  {journal} {\prd}\ }\textbf {\bibinfo {volume} {103}},\ \bibinfo {eid} {063032} (\bibinfo {year} {2021})},\ \Eprint {https://arxiv.org/abs/2010.00078} {arXiv:2010.00078 [astro-ph.HE]} \BibitemShut {NoStop}%
\bibitem [{\citenamefont {{Steinle}}\ and\ \citenamefont {{Kesden}}(2022)}]{Steinle2022}%
  \BibitemOpen
  \bibfield  {author} {\bibinfo {author} {\bibfnamefont {N.}~\bibnamefont {{Steinle}}}\ and\ \bibinfo {author} {\bibfnamefont {M.}~\bibnamefont {{Kesden}}},\ }\href {https://doi.org/10.1103/PhysRevD.106.063028} {\bibfield  {journal} {\bibinfo  {journal} {\prd}\ }\textbf {\bibinfo {volume} {106}},\ \bibinfo {eid} {063028} (\bibinfo {year} {2022})},\ \Eprint {https://arxiv.org/abs/2206.00391} {arXiv:2206.00391 [astro-ph.HE]} \BibitemShut {NoStop}%
\bibitem [{\citenamefont {{Gompertz}}\ \emph {et~al.}(2022)\citenamefont {{Gompertz}}, \citenamefont {{Nicholl}}, \citenamefont {{Schmidt}}, \citenamefont {{Pratten}},\ and\ \citenamefont {{Vecchio}}}]{Gompertz2022}%
  \BibitemOpen
  \bibfield  {author} {\bibinfo {author} {\bibfnamefont {B.~P.}\ \bibnamefont {{Gompertz}}}, \bibinfo {author} {\bibfnamefont {M.}~\bibnamefont {{Nicholl}}}, \bibinfo {author} {\bibfnamefont {P.}~\bibnamefont {{Schmidt}}}, \bibinfo {author} {\bibfnamefont {G.}~\bibnamefont {{Pratten}}},\ and\ \bibinfo {author} {\bibfnamefont {A.}~\bibnamefont {{Vecchio}}},\ }\href {https://doi.org/10.1093/mnras/stac029} {\bibfield  {journal} {\bibinfo  {journal} {\mnras}\ }\textbf {\bibinfo {volume} {511}},\ \bibinfo {pages} {1454} (\bibinfo {year} {2022})},\ \Eprint {https://arxiv.org/abs/2108.10184} {arXiv:2108.10184 [astro-ph.HE]} \BibitemShut {NoStop}%
\bibitem [{\citenamefont {{Gerosa}}\ \emph {et~al.}(2013)\citenamefont {{Gerosa}}, \citenamefont {{Kesden}}, \citenamefont {{Berti}}, \citenamefont {{O'Shaughnessy}},\ and\ \citenamefont {{Sperhake}}}]{Gerosa2013}%
  \BibitemOpen
  \bibfield  {author} {\bibinfo {author} {\bibfnamefont {D.}~\bibnamefont {{Gerosa}}}, \bibinfo {author} {\bibfnamefont {M.}~\bibnamefont {{Kesden}}}, \bibinfo {author} {\bibfnamefont {E.}~\bibnamefont {{Berti}}}, \bibinfo {author} {\bibfnamefont {R.}~\bibnamefont {{O'Shaughnessy}}},\ and\ \bibinfo {author} {\bibfnamefont {U.}~\bibnamefont {{Sperhake}}},\ }\href {https://doi.org/10.1103/PhysRevD.87.104028} {\bibfield  {journal} {\bibinfo  {journal} {\prd}\ }\textbf {\bibinfo {volume} {87}},\ \bibinfo {eid} {104028} (\bibinfo {year} {2013})},\ \Eprint {https://arxiv.org/abs/1302.4442} {arXiv:1302.4442 [gr-qc]} \BibitemShut {NoStop}%
\bibitem [{\citenamefont {{Talbot}}\ and\ \citenamefont {{Thrane}}(2017)}]{Talbot2017}%
  \BibitemOpen
  \bibfield  {author} {\bibinfo {author} {\bibfnamefont {C.}~\bibnamefont {{Talbot}}}\ and\ \bibinfo {author} {\bibfnamefont {E.}~\bibnamefont {{Thrane}}},\ }\href {https://doi.org/10.1103/PhysRevD.96.023012} {\bibfield  {journal} {\bibinfo  {journal} {\prd}\ }\textbf {\bibinfo {volume} {96}},\ \bibinfo {eid} {023012} (\bibinfo {year} {2017})},\ \Eprint {https://arxiv.org/abs/1704.08370} {arXiv:1704.08370 [astro-ph.HE]} \BibitemShut {NoStop}%
\bibitem [{\citenamefont {{Farr}}\ \emph {et~al.}(2017)\citenamefont {{Farr}}, \citenamefont {{Stevenson}}, \citenamefont {{Miller}}, \citenamefont {{Mandel}}, \citenamefont {{Farr}},\ and\ \citenamefont {{Vecchio}}}]{Farr2017}%
  \BibitemOpen
  \bibfield  {author} {\bibinfo {author} {\bibfnamefont {W.~M.}\ \bibnamefont {{Farr}}}, \bibinfo {author} {\bibfnamefont {S.}~\bibnamefont {{Stevenson}}}, \bibinfo {author} {\bibfnamefont {M.~C.}\ \bibnamefont {{Miller}}}, \bibinfo {author} {\bibfnamefont {I.}~\bibnamefont {{Mandel}}}, \bibinfo {author} {\bibfnamefont {B.}~\bibnamefont {{Farr}}},\ and\ \bibinfo {author} {\bibfnamefont {A.}~\bibnamefont {{Vecchio}}},\ }\href {https://doi.org/10.1038/nature23453} {\bibfield  {journal} {\bibinfo  {journal} {\nat}\ }\textbf {\bibinfo {volume} {548}},\ \bibinfo {pages} {426} (\bibinfo {year} {2017})},\ \Eprint {https://arxiv.org/abs/1706.01385} {arXiv:1706.01385 [astro-ph.HE]} \BibitemShut {NoStop}%
\bibitem [{\citenamefont {{Stevenson}}\ \emph {et~al.}(2017{\natexlab{a}})\citenamefont {{Stevenson}}, \citenamefont {{Berry}},\ and\ \citenamefont {{Mandel}}}]{Stevenson2017}%
  \BibitemOpen
  \bibfield  {author} {\bibinfo {author} {\bibfnamefont {S.}~\bibnamefont {{Stevenson}}}, \bibinfo {author} {\bibfnamefont {C.~P.~L.}\ \bibnamefont {{Berry}}},\ and\ \bibinfo {author} {\bibfnamefont {I.}~\bibnamefont {{Mandel}}},\ }\href {https://doi.org/10.1093/mnras/stx1764} {\bibfield  {journal} {\bibinfo  {journal} {\mnras}\ }\textbf {\bibinfo {volume} {471}},\ \bibinfo {pages} {2801} (\bibinfo {year} {2017}{\natexlab{a}})},\ \Eprint {https://arxiv.org/abs/1703.06873} {arXiv:1703.06873 [astro-ph.HE]} \BibitemShut {NoStop}%
\bibitem [{\citenamefont {{Farr}}\ \emph {et~al.}(2018)\citenamefont {{Farr}}, \citenamefont {{Holz}},\ and\ \citenamefont {{Farr}}}]{Farr2018}%
  \BibitemOpen
  \bibfield  {author} {\bibinfo {author} {\bibfnamefont {B.}~\bibnamefont {{Farr}}}, \bibinfo {author} {\bibfnamefont {D.~E.}\ \bibnamefont {{Holz}}},\ and\ \bibinfo {author} {\bibfnamefont {W.~M.}\ \bibnamefont {{Farr}}},\ }\href {https://doi.org/10.3847/2041-8213/aaaa64} {\bibfield  {journal} {\bibinfo  {journal} {\apjl}\ }\textbf {\bibinfo {volume} {854}},\ \bibinfo {eid} {L9} (\bibinfo {year} {2018})},\ \Eprint {https://arxiv.org/abs/1709.07896} {arXiv:1709.07896 [astro-ph.HE]} \BibitemShut {NoStop}%
\bibitem [{\citenamefont {{Wysocki}}\ \emph {et~al.}(2018)\citenamefont {{Wysocki}}, \citenamefont {{Gerosa}}, \citenamefont {{O'Shaughnessy}}, \citenamefont {{Belczynski}}, \citenamefont {{Gladysz}}, \citenamefont {{Berti}}, \citenamefont {{Kesden}},\ and\ \citenamefont {{Holz}}}]{Wysocki2018}%
  \BibitemOpen
  \bibfield  {author} {\bibinfo {author} {\bibfnamefont {D.}~\bibnamefont {{Wysocki}}}, \bibinfo {author} {\bibfnamefont {D.}~\bibnamefont {{Gerosa}}}, \bibinfo {author} {\bibfnamefont {R.}~\bibnamefont {{O'Shaughnessy}}}, \bibinfo {author} {\bibfnamefont {K.}~\bibnamefont {{Belczynski}}}, \bibinfo {author} {\bibfnamefont {W.}~\bibnamefont {{Gladysz}}}, \bibinfo {author} {\bibfnamefont {E.}~\bibnamefont {{Berti}}}, \bibinfo {author} {\bibfnamefont {M.}~\bibnamefont {{Kesden}}},\ and\ \bibinfo {author} {\bibfnamefont {D.~E.}\ \bibnamefont {{Holz}}},\ }\href {https://doi.org/10.1103/PhysRevD.97.043014} {\bibfield  {journal} {\bibinfo  {journal} {\prd}\ }\textbf {\bibinfo {volume} {97}},\ \bibinfo {eid} {043014} (\bibinfo {year} {2018})},\ \Eprint {https://arxiv.org/abs/1709.01943} {arXiv:1709.01943 [astro-ph.HE]} \BibitemShut {NoStop}%
\bibitem [{\citenamefont {{Miller}}\ \emph {et~al.}(2020)\citenamefont {{Miller}}, \citenamefont {{Callister}},\ and\ \citenamefont {{Farr}}}]{Miller2020}%
  \BibitemOpen
  \bibfield  {author} {\bibinfo {author} {\bibfnamefont {S.}~\bibnamefont {{Miller}}}, \bibinfo {author} {\bibfnamefont {T.~A.}\ \bibnamefont {{Callister}}},\ and\ \bibinfo {author} {\bibfnamefont {W.~M.}\ \bibnamefont {{Farr}}},\ }\href {https://doi.org/10.3847/1538-4357/ab80c0} {\bibfield  {journal} {\bibinfo  {journal} {\apj}\ }\textbf {\bibinfo {volume} {895}},\ \bibinfo {eid} {128} (\bibinfo {year} {2020})},\ \Eprint {https://arxiv.org/abs/2001.06051} {arXiv:2001.06051 [astro-ph.HE]} \BibitemShut {NoStop}%
\bibitem [{\citenamefont {{Bavera}}\ \emph {et~al.}(2020)\citenamefont {{Bavera}}, \citenamefont {{Fragos}}, \citenamefont {{Qin}}, \citenamefont {{Zapartas}}, \citenamefont {{Neijssel}}, \citenamefont {{Mandel}}, \citenamefont {{Batta}}, \citenamefont {{Gaebel}}, \citenamefont {{Kimball}},\ and\ \citenamefont {{Stevenson}}}]{Bavera2020}%
  \BibitemOpen
  \bibfield  {author} {\bibinfo {author} {\bibfnamefont {S.~S.}\ \bibnamefont {{Bavera}}}, \bibinfo {author} {\bibfnamefont {T.}~\bibnamefont {{Fragos}}}, \bibinfo {author} {\bibfnamefont {Y.}~\bibnamefont {{Qin}}}, \bibinfo {author} {\bibfnamefont {E.}~\bibnamefont {{Zapartas}}}, \bibinfo {author} {\bibfnamefont {C.~J.}\ \bibnamefont {{Neijssel}}}, \bibinfo {author} {\bibfnamefont {I.}~\bibnamefont {{Mandel}}}, \bibinfo {author} {\bibfnamefont {A.}~\bibnamefont {{Batta}}}, \bibinfo {author} {\bibfnamefont {S.~M.}\ \bibnamefont {{Gaebel}}}, \bibinfo {author} {\bibfnamefont {C.}~\bibnamefont {{Kimball}}},\ and\ \bibinfo {author} {\bibfnamefont {S.}~\bibnamefont {{Stevenson}}},\ }\href {https://doi.org/10.1051/0004-6361/201936204} {\bibfield  {journal} {\bibinfo  {journal} {\aap}\ }\textbf {\bibinfo {volume} {635}},\ \bibinfo {eid} {A97} (\bibinfo {year} {2020})},\ \Eprint {https://arxiv.org/abs/1906.12257} {arXiv:1906.12257 [astro-ph.HE]} \BibitemShut {NoStop}%
\bibitem [{\citenamefont {{Callister}}\ \emph {et~al.}(2021)\citenamefont {{Callister}}, \citenamefont {{Farr}},\ and\ \citenamefont {{Renzo}}}]{Callister2021}%
  \BibitemOpen
  \bibfield  {author} {\bibinfo {author} {\bibfnamefont {T.~A.}\ \bibnamefont {{Callister}}}, \bibinfo {author} {\bibfnamefont {W.~M.}\ \bibnamefont {{Farr}}},\ and\ \bibinfo {author} {\bibfnamefont {M.}~\bibnamefont {{Renzo}}},\ }\href {https://doi.org/10.3847/1538-4357/ac1347} {\bibfield  {journal} {\bibinfo  {journal} {\apj}\ }\textbf {\bibinfo {volume} {920}},\ \bibinfo {eid} {157} (\bibinfo {year} {2021})},\ \Eprint {https://arxiv.org/abs/2011.09570} {arXiv:2011.09570 [astro-ph.HE]} \BibitemShut {NoStop}%
\bibitem [{\citenamefont {{Varma}}\ \emph {et~al.}(2022)\citenamefont {{Varma}}, \citenamefont {{Biscoveanu}}, \citenamefont {{Isi}}, \citenamefont {{Farr}},\ and\ \citenamefont {{Vitale}}}]{Varma2022}%
  \BibitemOpen
  \bibfield  {author} {\bibinfo {author} {\bibfnamefont {V.}~\bibnamefont {{Varma}}}, \bibinfo {author} {\bibfnamefont {S.}~\bibnamefont {{Biscoveanu}}}, \bibinfo {author} {\bibfnamefont {M.}~\bibnamefont {{Isi}}}, \bibinfo {author} {\bibfnamefont {W.~M.}\ \bibnamefont {{Farr}}},\ and\ \bibinfo {author} {\bibfnamefont {S.}~\bibnamefont {{Vitale}}},\ }\href {https://doi.org/10.1103/PhysRevLett.128.031101} {\bibfield  {journal} {\bibinfo  {journal} {\prl}\ }\textbf {\bibinfo {volume} {128}},\ \bibinfo {eid} {031101} (\bibinfo {year} {2022})},\ \Eprint {https://arxiv.org/abs/2107.09693} {arXiv:2107.09693 [astro-ph.HE]} \BibitemShut {NoStop}%
\bibitem [{\citenamefont {{Tong}}\ \emph {et~al.}(2022)\citenamefont {{Tong}}, \citenamefont {{Galaudage}},\ and\ \citenamefont {{Thrane}}}]{2022PhRvD.106j3019T}%
  \BibitemOpen
  \bibfield  {author} {\bibinfo {author} {\bibfnamefont {H.}~\bibnamefont {{Tong}}}, \bibinfo {author} {\bibfnamefont {S.}~\bibnamefont {{Galaudage}}},\ and\ \bibinfo {author} {\bibfnamefont {E.}~\bibnamefont {{Thrane}}},\ }\href {https://doi.org/10.1103/PhysRevD.106.103019} {\bibfield  {journal} {\bibinfo  {journal} {\prd}\ }\textbf {\bibinfo {volume} {106}},\ \bibinfo {eid} {103019} (\bibinfo {year} {2022})},\ \Eprint {https://arxiv.org/abs/2209.02206} {arXiv:2209.02206 [astro-ph.HE]} \BibitemShut {NoStop}%
\bibitem [{\citenamefont {{Franciolini}}\ and\ \citenamefont {{Pani}}(2022)}]{2022PhRvD.105l3024F}%
  \BibitemOpen
  \bibfield  {author} {\bibinfo {author} {\bibfnamefont {G.}~\bibnamefont {{Franciolini}}}\ and\ \bibinfo {author} {\bibfnamefont {P.}~\bibnamefont {{Pani}}},\ }\href {https://doi.org/10.1103/PhysRevD.105.123024} {\bibfield  {journal} {\bibinfo  {journal} {\prd}\ }\textbf {\bibinfo {volume} {105}},\ \bibinfo {eid} {123024} (\bibinfo {year} {2022})},\ \Eprint {https://arxiv.org/abs/2201.13098} {arXiv:2201.13098 [astro-ph.HE]} \BibitemShut {NoStop}%
\bibitem [{\citenamefont {{Johnson-McDaniel}}\ \emph {et~al.}(2023)\citenamefont {{Johnson-McDaniel}}, \citenamefont {{Phukon}}, \citenamefont {{Krishnendu}},\ and\ \citenamefont {{Gupta}}}]{Johnson-McDaniel2023}%
  \BibitemOpen
  \bibfield  {author} {\bibinfo {author} {\bibfnamefont {N.~K.}\ \bibnamefont {{Johnson-McDaniel}}}, \bibinfo {author} {\bibfnamefont {K.~S.}\ \bibnamefont {{Phukon}}}, \bibinfo {author} {\bibfnamefont {N.~V.}\ \bibnamefont {{Krishnendu}}},\ and\ \bibinfo {author} {\bibfnamefont {A.}~\bibnamefont {{Gupta}}},\ }\href {https://doi.org/10.1103/PhysRevD.108.103003} {\bibfield  {journal} {\bibinfo  {journal} {\prd}\ }\textbf {\bibinfo {volume} {108}},\ \bibinfo {eid} {103003} (\bibinfo {year} {2023})},\ \Eprint {https://arxiv.org/abs/2301.10125} {arXiv:2301.10125 [astro-ph.HE]} \BibitemShut {NoStop}%
\bibitem [{\citenamefont {{Godfrey}}\ \emph {et~al.}(2023)\citenamefont {{Godfrey}}, \citenamefont {{Edelman}},\ and\ \citenamefont {{Farr}}}]{Godfrey2023}%
  \BibitemOpen
  \bibfield  {author} {\bibinfo {author} {\bibfnamefont {J.}~\bibnamefont {{Godfrey}}}, \bibinfo {author} {\bibfnamefont {B.}~\bibnamefont {{Edelman}}},\ and\ \bibinfo {author} {\bibfnamefont {B.}~\bibnamefont {{Farr}}},\ }\href {https://doi.org/10.48550/arXiv.2304.01288} {\bibfield  {journal} {\bibinfo  {journal} {arXiv e-prints}\ ,\ \bibinfo {eid} {arXiv:2304.01288}} (\bibinfo {year} {2023})},\ \Eprint {https://arxiv.org/abs/2304.01288} {arXiv:2304.01288 [astro-ph.HE]} \BibitemShut {NoStop}%
\bibitem [{\citenamefont {{Miller}}\ \emph {et~al.}(2024)\citenamefont {{Miller}}, \citenamefont {{Ko}}, \citenamefont {{Callister}},\ and\ \citenamefont {{Chatziioannou}}}]{Miller2024}%
  \BibitemOpen
  \bibfield  {author} {\bibinfo {author} {\bibfnamefont {S.~J.}\ \bibnamefont {{Miller}}}, \bibinfo {author} {\bibfnamefont {Z.}~\bibnamefont {{Ko}}}, \bibinfo {author} {\bibfnamefont {T.}~\bibnamefont {{Callister}}},\ and\ \bibinfo {author} {\bibfnamefont {K.}~\bibnamefont {{Chatziioannou}}},\ }\href {https://doi.org/10.1103/PhysRevD.109.104036} {\bibfield  {journal} {\bibinfo  {journal} {\prd}\ }\textbf {\bibinfo {volume} {109}},\ \bibinfo {eid} {104036} (\bibinfo {year} {2024})},\ \Eprint {https://arxiv.org/abs/2401.05613} {arXiv:2401.05613 [gr-qc]} \BibitemShut {NoStop}%
\bibitem [{\citenamefont {{Romero-Shaw}}\ \emph {et~al.}(2021)\citenamefont {{Romero-Shaw}}, \citenamefont {{Kremer}}, \citenamefont {{Lasky}}, \citenamefont {{Thrane}},\ and\ \citenamefont {{Samsing}}}]{Romero-Shaw2021}%
  \BibitemOpen
  \bibfield  {author} {\bibinfo {author} {\bibfnamefont {I.~M.}\ \bibnamefont {{Romero-Shaw}}}, \bibinfo {author} {\bibfnamefont {K.}~\bibnamefont {{Kremer}}}, \bibinfo {author} {\bibfnamefont {P.~D.}\ \bibnamefont {{Lasky}}}, \bibinfo {author} {\bibfnamefont {E.}~\bibnamefont {{Thrane}}},\ and\ \bibinfo {author} {\bibfnamefont {J.}~\bibnamefont {{Samsing}}},\ }\href {https://doi.org/10.1093/mnras/stab1815} {\bibfield  {journal} {\bibinfo  {journal} {\mnras}\ }\textbf {\bibinfo {volume} {506}},\ \bibinfo {pages} {2362} (\bibinfo {year} {2021})},\ \Eprint {https://arxiv.org/abs/2011.14541} {arXiv:2011.14541 [astro-ph.HE]} \BibitemShut {NoStop}%
\bibitem [{\citenamefont {{Saini}}(2024)}]{Saini2024}%
  \BibitemOpen
  \bibfield  {author} {\bibinfo {author} {\bibfnamefont {P.}~\bibnamefont {{Saini}}},\ }\href {https://doi.org/10.1093/mnras/stae037} {\bibfield  {journal} {\bibinfo  {journal} {\mnras}\ }\textbf {\bibinfo {volume} {528}},\ \bibinfo {pages} {833} (\bibinfo {year} {2024})},\ \Eprint {https://arxiv.org/abs/2308.07565} {arXiv:2308.07565 [astro-ph.HE]} \BibitemShut {NoStop}%
\bibitem [{\citenamefont {{Nicholl}}\ \emph {et~al.}(2021)\citenamefont {{Nicholl}}, \citenamefont {{Margalit}}, \citenamefont {{Schmidt}}, \citenamefont {{Smith}}, \citenamefont {{Ridley}},\ and\ \citenamefont {{Nuttall}}}]{Nicholl2021}%
  \BibitemOpen
  \bibfield  {author} {\bibinfo {author} {\bibfnamefont {M.}~\bibnamefont {{Nicholl}}}, \bibinfo {author} {\bibfnamefont {B.}~\bibnamefont {{Margalit}}}, \bibinfo {author} {\bibfnamefont {P.}~\bibnamefont {{Schmidt}}}, \bibinfo {author} {\bibfnamefont {G.~P.}\ \bibnamefont {{Smith}}}, \bibinfo {author} {\bibfnamefont {E.~J.}\ \bibnamefont {{Ridley}}},\ and\ \bibinfo {author} {\bibfnamefont {J.}~\bibnamefont {{Nuttall}}},\ }\href {https://doi.org/10.1093/mnras/stab1523} {\bibfield  {journal} {\bibinfo  {journal} {\mnras}\ }\textbf {\bibinfo {volume} {505}},\ \bibinfo {pages} {3016} (\bibinfo {year} {2021})},\ \Eprint {https://arxiv.org/abs/2102.02229} {arXiv:2102.02229 [astro-ph.HE]} \BibitemShut {NoStop}%
\bibitem [{\citenamefont {{Gompertz}}\ \emph {et~al.}(2023)\citenamefont {{Gompertz}}, \citenamefont {{Nicholl}}, \citenamefont {{Smith}}, \citenamefont {{Harisankar}}, \citenamefont {{Pratten}}, \citenamefont {{Schmidt}},\ and\ \citenamefont {{Smith}}}]{Gompertz2023}%
  \BibitemOpen
  \bibfield  {author} {\bibinfo {author} {\bibfnamefont {B.~P.}\ \bibnamefont {{Gompertz}}}, \bibinfo {author} {\bibfnamefont {M.}~\bibnamefont {{Nicholl}}}, \bibinfo {author} {\bibfnamefont {J.~C.}\ \bibnamefont {{Smith}}}, \bibinfo {author} {\bibfnamefont {S.}~\bibnamefont {{Harisankar}}}, \bibinfo {author} {\bibfnamefont {G.}~\bibnamefont {{Pratten}}}, \bibinfo {author} {\bibfnamefont {P.}~\bibnamefont {{Schmidt}}},\ and\ \bibinfo {author} {\bibfnamefont {G.~P.}\ \bibnamefont {{Smith}}},\ }\href {https://doi.org/10.1093/mnras/stad2990} {\bibfield  {journal} {\bibinfo  {journal} {\mnras}\ }\textbf {\bibinfo {volume} {526}},\ \bibinfo {pages} {4585} (\bibinfo {year} {2023})},\ \Eprint {https://arxiv.org/abs/2305.07582} {arXiv:2305.07582 [astro-ph.HE]} \BibitemShut {NoStop}%
\bibitem [{\citenamefont {{Chattopadhyay}}\ \emph {et~al.}(2020)\citenamefont {{Chattopadhyay}}, \citenamefont {{Stevenson}}, \citenamefont {{Hurley}}, \citenamefont {{Rossi}},\ and\ \citenamefont {{Flynn}}}]{2020MNRAS.494.1587C}%
  \BibitemOpen
  \bibfield  {author} {\bibinfo {author} {\bibfnamefont {D.}~\bibnamefont {{Chattopadhyay}}}, \bibinfo {author} {\bibfnamefont {S.}~\bibnamefont {{Stevenson}}}, \bibinfo {author} {\bibfnamefont {J.~R.}\ \bibnamefont {{Hurley}}}, \bibinfo {author} {\bibfnamefont {L.~J.}\ \bibnamefont {{Rossi}}},\ and\ \bibinfo {author} {\bibfnamefont {C.}~\bibnamefont {{Flynn}}},\ }\href {https://doi.org/10.1093/mnras/staa756} {\bibfield  {journal} {\bibinfo  {journal} {\mnras}\ }\textbf {\bibinfo {volume} {494}},\ \bibinfo {pages} {1587} (\bibinfo {year} {2020})},\ \Eprint {https://arxiv.org/abs/1912.02415} {arXiv:1912.02415 [astro-ph.HE]} \BibitemShut {NoStop}%
\bibitem [{\citenamefont {{Barbieri}}\ \emph {et~al.}(2019)\citenamefont {{Barbieri}}, \citenamefont {{Salafia}}, \citenamefont {{Perego}}, \citenamefont {{Colpi}},\ and\ \citenamefont {{Ghirlanda}}}]{Barbieri2019}%
  \BibitemOpen
  \bibfield  {author} {\bibinfo {author} {\bibfnamefont {C.}~\bibnamefont {{Barbieri}}}, \bibinfo {author} {\bibfnamefont {O.~S.}\ \bibnamefont {{Salafia}}}, \bibinfo {author} {\bibfnamefont {A.}~\bibnamefont {{Perego}}}, \bibinfo {author} {\bibfnamefont {M.}~\bibnamefont {{Colpi}}},\ and\ \bibinfo {author} {\bibfnamefont {G.}~\bibnamefont {{Ghirlanda}}},\ }\href {https://doi.org/10.1051/0004-6361/201935443} {\bibfield  {journal} {\bibinfo  {journal} {\aap}\ }\textbf {\bibinfo {volume} {625}},\ \bibinfo {eid} {A152} (\bibinfo {year} {2019})},\ \Eprint {https://arxiv.org/abs/1903.04543} {arXiv:1903.04543 [astro-ph.HE]} \BibitemShut {NoStop}%
\bibitem [{\citenamefont {{Coughlin}}\ \emph {et~al.}(2019{\natexlab{a}})\citenamefont {{Coughlin}}, \citenamefont {{Dietrich}}, \citenamefont {{Margalit}},\ and\ \citenamefont {{Metzger}}}]{Coughlin2019}%
  \BibitemOpen
  \bibfield  {author} {\bibinfo {author} {\bibfnamefont {M.~W.}\ \bibnamefont {{Coughlin}}}, \bibinfo {author} {\bibfnamefont {T.}~\bibnamefont {{Dietrich}}}, \bibinfo {author} {\bibfnamefont {B.}~\bibnamefont {{Margalit}}},\ and\ \bibinfo {author} {\bibfnamefont {B.~D.}\ \bibnamefont {{Metzger}}},\ }\href {https://doi.org/10.1093/mnrasl/slz133} {\bibfield  {journal} {\bibinfo  {journal} {\mnras}\ }\textbf {\bibinfo {volume} {489}},\ \bibinfo {pages} {L91} (\bibinfo {year} {2019}{\natexlab{a}})},\ \Eprint {https://arxiv.org/abs/1812.04803} {arXiv:1812.04803 [astro-ph.HE]} \BibitemShut {NoStop}%
\bibitem [{\citenamefont {{Dietrich}}\ \emph {et~al.}(2020)\citenamefont {{Dietrich}}, \citenamefont {{Coughlin}}, \citenamefont {{Pang}}, \citenamefont {{Bulla}}, \citenamefont {{Heinzel}}, \citenamefont {{Issa}}, \citenamefont {{Tews}},\ and\ \citenamefont {{Antier}}}]{Dietrich2020}%
  \BibitemOpen
  \bibfield  {author} {\bibinfo {author} {\bibfnamefont {T.}~\bibnamefont {{Dietrich}}}, \bibinfo {author} {\bibfnamefont {M.~W.}\ \bibnamefont {{Coughlin}}}, \bibinfo {author} {\bibfnamefont {P.~T.~H.}\ \bibnamefont {{Pang}}}, \bibinfo {author} {\bibfnamefont {M.}~\bibnamefont {{Bulla}}}, \bibinfo {author} {\bibfnamefont {J.}~\bibnamefont {{Heinzel}}}, \bibinfo {author} {\bibfnamefont {L.}~\bibnamefont {{Issa}}}, \bibinfo {author} {\bibfnamefont {I.}~\bibnamefont {{Tews}}},\ and\ \bibinfo {author} {\bibfnamefont {S.}~\bibnamefont {{Antier}}},\ }\href {https://doi.org/10.1126/science.abb4317} {\bibfield  {journal} {\bibinfo  {journal} {Science}\ }\textbf {\bibinfo {volume} {370}},\ \bibinfo {pages} {1450} (\bibinfo {year} {2020})},\ \Eprint {https://arxiv.org/abs/2002.11355} {arXiv:2002.11355 [astro-ph.HE]} \BibitemShut {NoStop}%
\bibitem [{\citenamefont {{Breschi}}\ \emph {et~al.}(2021)\citenamefont {{Breschi}}, \citenamefont {{Perego}}, \citenamefont {{Bernuzzi}}, \citenamefont {{Del Pozzo}}, \citenamefont {{Nedora}}, \citenamefont {{Radice}},\ and\ \citenamefont {{Vescovi}}}]{Breschi2021}%
  \BibitemOpen
  \bibfield  {author} {\bibinfo {author} {\bibfnamefont {M.}~\bibnamefont {{Breschi}}}, \bibinfo {author} {\bibfnamefont {A.}~\bibnamefont {{Perego}}}, \bibinfo {author} {\bibfnamefont {S.}~\bibnamefont {{Bernuzzi}}}, \bibinfo {author} {\bibfnamefont {W.}~\bibnamefont {{Del Pozzo}}}, \bibinfo {author} {\bibfnamefont {V.}~\bibnamefont {{Nedora}}}, \bibinfo {author} {\bibfnamefont {D.}~\bibnamefont {{Radice}}},\ and\ \bibinfo {author} {\bibfnamefont {D.}~\bibnamefont {{Vescovi}}},\ }\href {https://doi.org/10.1093/mnras/stab1287} {\bibfield  {journal} {\bibinfo  {journal} {\mnras}\ }\textbf {\bibinfo {volume} {505}},\ \bibinfo {pages} {1661} (\bibinfo {year} {2021})},\ \Eprint {https://arxiv.org/abs/2101.01201} {arXiv:2101.01201 [astro-ph.HE]} \BibitemShut {NoStop}%
\bibitem [{\citenamefont {{Raaijmakers}}\ \emph {et~al.}(2021{\natexlab{a}})\citenamefont {{Raaijmakers}}, \citenamefont {{Nissanke}}, \citenamefont {{Foucart}} \emph {et~al.}}]{Raaijmakers2021}%
  \BibitemOpen
  \bibfield  {author} {\bibinfo {author} {\bibfnamefont {G.}~\bibnamefont {{Raaijmakers}}}, \bibinfo {author} {\bibfnamefont {S.}~\bibnamefont {{Nissanke}}}, \bibinfo {author} {\bibfnamefont {F.}~\bibnamefont {{Foucart}}}, \emph {et~al.},\ }\href {https://doi.org/10.3847/1538-4357/ac222d} {\bibfield  {journal} {\bibinfo  {journal} {\apj}\ }\textbf {\bibinfo {volume} {922}},\ \bibinfo {eid} {269} (\bibinfo {year} {2021}{\natexlab{a}})},\ \Eprint {https://arxiv.org/abs/2102.11569} {arXiv:2102.11569 [astro-ph.HE]} \BibitemShut {NoStop}%
\bibitem [{\citenamefont {{Ruiz}}\ \emph {et~al.}(2021)\citenamefont {{Ruiz}}, \citenamefont {{Shapiro}},\ and\ \citenamefont {{Tsokaros}}}]{Ruiz2021}%
  \BibitemOpen
  \bibfield  {author} {\bibinfo {author} {\bibfnamefont {M.}~\bibnamefont {{Ruiz}}}, \bibinfo {author} {\bibfnamefont {S.~L.}\ \bibnamefont {{Shapiro}}},\ and\ \bibinfo {author} {\bibfnamefont {A.}~\bibnamefont {{Tsokaros}}},\ }\href {https://doi.org/10.3389/fspas.2021.656907} {\bibfield  {journal} {\bibinfo  {journal} {Frontiers in Astronomy and Space Sciences}\ }\textbf {\bibinfo {volume} {8}},\ \bibinfo {eid} {39} (\bibinfo {year} {2021})},\ \Eprint {https://arxiv.org/abs/2102.03366} {arXiv:2102.03366 [astro-ph.HE]} \BibitemShut {NoStop}%
\bibitem [{\citenamefont {{Bavera}}\ \emph {et~al.}(2022)\citenamefont {{Bavera}}, \citenamefont {{Fragos}}, \citenamefont {{Zapartas}} \emph {et~al.}}]{Bavera2022}%
  \BibitemOpen
  \bibfield  {author} {\bibinfo {author} {\bibfnamefont {S.~S.}\ \bibnamefont {{Bavera}}}, \bibinfo {author} {\bibfnamefont {T.}~\bibnamefont {{Fragos}}}, \bibinfo {author} {\bibfnamefont {E.}~\bibnamefont {{Zapartas}}}, \emph {et~al.},\ }\href {https://doi.org/10.1051/0004-6361/202141979} {\bibfield  {journal} {\bibinfo  {journal} {\aap}\ }\textbf {\bibinfo {volume} {657}},\ \bibinfo {eid} {L8} (\bibinfo {year} {2022})},\ \Eprint {https://arxiv.org/abs/2106.15841} {arXiv:2106.15841 [astro-ph.HE]} \BibitemShut {NoStop}%
\bibitem [{\citenamefont {{D'Orazio}}\ \emph {et~al.}(2022)\citenamefont {{D'Orazio}}, \citenamefont {{Haiman}}, \citenamefont {{Levin}}, \citenamefont {{Samsing}},\ and\ \citenamefont {{Vigna-G{\'o}mez}}}]{DOrazio2022}%
  \BibitemOpen
  \bibfield  {author} {\bibinfo {author} {\bibfnamefont {D.~J.}\ \bibnamefont {{D'Orazio}}}, \bibinfo {author} {\bibfnamefont {Z.}~\bibnamefont {{Haiman}}}, \bibinfo {author} {\bibfnamefont {J.}~\bibnamefont {{Levin}}}, \bibinfo {author} {\bibfnamefont {J.}~\bibnamefont {{Samsing}}},\ and\ \bibinfo {author} {\bibfnamefont {A.}~\bibnamefont {{Vigna-G{\'o}mez}}},\ }\href {https://doi.org/10.3847/1538-4357/ac4bdb} {\bibfield  {journal} {\bibinfo  {journal} {\apj}\ }\textbf {\bibinfo {volume} {927}},\ \bibinfo {eid} {56} (\bibinfo {year} {2022})},\ \Eprint {https://arxiv.org/abs/2112.01979} {arXiv:2112.01979 [astro-ph.HE]} \BibitemShut {NoStop}%
\bibitem [{\citenamefont {{Colombo}}\ \emph {et~al.}(2024)\citenamefont {{Colombo}}, \citenamefont {{Duqu{\'e}}}, \citenamefont {{Salafia}} \emph {et~al.}}]{Colombo2024}%
  \BibitemOpen
  \bibfield  {author} {\bibinfo {author} {\bibfnamefont {A.}~\bibnamefont {{Colombo}}}, \bibinfo {author} {\bibfnamefont {R.}~\bibnamefont {{Duqu{\'e}}}}, \bibinfo {author} {\bibfnamefont {O.~S.}\ \bibnamefont {{Salafia}}}, \emph {et~al.},\ }\href {https://doi.org/10.1051/0004-6361/202348384} {\bibfield  {journal} {\bibinfo  {journal} {\aap}\ }\textbf {\bibinfo {volume} {686}},\ \bibinfo {eid} {A265} (\bibinfo {year} {2024})},\ \Eprint {https://arxiv.org/abs/2310.16894} {arXiv:2310.16894 [astro-ph.HE]} \BibitemShut {NoStop}%
\bibitem [{\citenamefont {{Bisero}}\ \emph {et~al.}(2025)\citenamefont {{Bisero}}, \citenamefont {{Vergani}}, \citenamefont {{Loffredo}}, \citenamefont {{Branchesi}}, \citenamefont {{Hazra}}, \citenamefont {{Dupletsa}},\ and\ \citenamefont {{Anderson}}}]{Bisero2025}%
  \BibitemOpen
  \bibfield  {author} {\bibinfo {author} {\bibfnamefont {S.}~\bibnamefont {{Bisero}}}, \bibinfo {author} {\bibfnamefont {S.~D.}\ \bibnamefont {{Vergani}}}, \bibinfo {author} {\bibfnamefont {E.}~\bibnamefont {{Loffredo}}}, \bibinfo {author} {\bibfnamefont {M.}~\bibnamefont {{Branchesi}}}, \bibinfo {author} {\bibfnamefont {N.}~\bibnamefont {{Hazra}}}, \bibinfo {author} {\bibfnamefont {U.}~\bibnamefont {{Dupletsa}}},\ and\ \bibinfo {author} {\bibfnamefont {R.~I.}\ \bibnamefont {{Anderson}}},\ }\href {https://doi.org/10.48550/arXiv.2507.02055} {\bibfield  {journal} {\bibinfo  {journal} {arXiv e-prints}\ ,\ \bibinfo {eid} {arXiv:2507.02055}} (\bibinfo {year} {2025})},\ \Eprint {https://arxiv.org/abs/2507.02055} {arXiv:2507.02055 [astro-ph.HE]} \BibitemShut {NoStop}%
\bibitem [{\citenamefont {{Loffredo}}\ \emph {et~al.}(2025)\citenamefont {{Loffredo}}, \citenamefont {{Hazra}}, \citenamefont {{Dupletsa}}, \citenamefont {{Branchesi}}, \citenamefont {{Ronchini}}, \citenamefont {{Santoliquido}}, \citenamefont {{Perego}}, \citenamefont {{Banerjee}}, \citenamefont {{Bisero}}, \citenamefont {{Ricigliano}}, \citenamefont {{Vergani}}, \citenamefont {{Andreoni}}, \citenamefont {{Cantiello}}, \citenamefont {{Harms}}, \citenamefont {{Mapelli}},\ and\ \citenamefont {{Oganesyan}}}]{2025A&A...697A..36L}%
  \BibitemOpen
  \bibfield  {author} {\bibinfo {author} {\bibfnamefont {E.}~\bibnamefont {{Loffredo}}}, \bibinfo {author} {\bibfnamefont {N.}~\bibnamefont {{Hazra}}}, \bibinfo {author} {\bibfnamefont {U.}~\bibnamefont {{Dupletsa}}}, \bibinfo {author} {\bibfnamefont {M.}~\bibnamefont {{Branchesi}}}, \bibinfo {author} {\bibfnamefont {S.}~\bibnamefont {{Ronchini}}}, \bibinfo {author} {\bibfnamefont {F.}~\bibnamefont {{Santoliquido}}}, \bibinfo {author} {\bibfnamefont {A.}~\bibnamefont {{Perego}}}, \bibinfo {author} {\bibfnamefont {B.}~\bibnamefont {{Banerjee}}}, \bibinfo {author} {\bibfnamefont {S.}~\bibnamefont {{Bisero}}}, \bibinfo {author} {\bibfnamefont {G.}~\bibnamefont {{Ricigliano}}}, \bibinfo {author} {\bibfnamefont {S.}~\bibnamefont {{Vergani}}}, \bibinfo {author} {\bibfnamefont {I.}~\bibnamefont {{Andreoni}}}, \bibinfo {author} {\bibfnamefont {M.}~\bibnamefont {{Cantiello}}}, \bibinfo {author} {\bibfnamefont {J.}~\bibnamefont {{Harms}}}, \bibinfo {author} {\bibfnamefont {M.}~\bibnamefont {{Mapelli}}},\ and\ \bibinfo
  {author} {\bibfnamefont {G.}~\bibnamefont {{Oganesyan}}},\ }\href {https://doi.org/10.1051/0004-6361/202452863} {\bibfield  {journal} {\bibinfo  {journal} {\aap}\ }\textbf {\bibinfo {volume} {697}},\ \bibinfo {eid} {A36} (\bibinfo {year} {2025})},\ \Eprint {https://arxiv.org/abs/2411.02342} {arXiv:2411.02342 [astro-ph.HE]} \BibitemShut {NoStop}%
\bibitem [{\citenamefont {{Risti{\'c}}}\ \emph {et~al.}(2025)\citenamefont {{Risti{\'c}}}, \citenamefont {{O'Shaughnessy}}, \citenamefont {{Wagner}}, \citenamefont {{Fontes}}, \citenamefont {{Fryer}}, \citenamefont {{Korobkin}}, \citenamefont {{Mumpower}},\ and\ \citenamefont {{Wollaeger}}}]{2025arXiv250312320R}%
  \BibitemOpen
  \bibfield  {author} {\bibinfo {author} {\bibfnamefont {M.}~\bibnamefont {{Risti{\'c}}}}, \bibinfo {author} {\bibfnamefont {R.}~\bibnamefont {{O'Shaughnessy}}}, \bibinfo {author} {\bibfnamefont {K.}~\bibnamefont {{Wagner}}}, \bibinfo {author} {\bibfnamefont {C.~J.}\ \bibnamefont {{Fontes}}}, \bibinfo {author} {\bibfnamefont {C.~L.}\ \bibnamefont {{Fryer}}}, \bibinfo {author} {\bibfnamefont {O.}~\bibnamefont {{Korobkin}}}, \bibinfo {author} {\bibfnamefont {M.~R.}\ \bibnamefont {{Mumpower}}},\ and\ \bibinfo {author} {\bibfnamefont {R.~T.}\ \bibnamefont {{Wollaeger}}},\ }\href {https://doi.org/10.48550/arXiv.2503.12320} {\bibfield  {journal} {\bibinfo  {journal} {arXiv e-prints}\ ,\ \bibinfo {eid} {arXiv:2503.12320}} (\bibinfo {year} {2025})},\ \Eprint {https://arxiv.org/abs/2503.12320} {arXiv:2503.12320 [astro-ph.HE]} \BibitemShut {NoStop}%
\bibitem [{\citenamefont {{Branchesi}}\ \emph {et~al.}(2023{\natexlab{a}})\citenamefont {{Branchesi}}, \citenamefont {{Maggiore}}, \citenamefont {{Alonso}} \emph {et~al.}}]{2023JCAP...07..068B}%
  \BibitemOpen
  \bibfield  {author} {\bibinfo {author} {\bibfnamefont {M.}~\bibnamefont {{Branchesi}}}, \bibinfo {author} {\bibfnamefont {M.}~\bibnamefont {{Maggiore}}}, \bibinfo {author} {\bibfnamefont {D.}~\bibnamefont {{Alonso}}}, \emph {et~al.},\ }\href {https://doi.org/10.1088/1475-7516/2023/07/068} {\bibfield  {journal} {\bibinfo  {journal} {\jcap}\ }\textbf {\bibinfo {volume} {2023}},\ \bibinfo {eid} {068} (\bibinfo {year} {2023}{\natexlab{a}})},\ \Eprint {https://arxiv.org/abs/2303.15923} {arXiv:2303.15923 [gr-qc]} \BibitemShut {NoStop}%
\bibitem [{\citenamefont {{Abac}}\ \emph {et~al.}(2025{\natexlab{a}})\citenamefont {{Abac}}, \citenamefont {{Abramo}}, \citenamefont {{Albanesi}} \emph {et~al.}}]{2025arXiv250312263A}%
  \BibitemOpen
  \bibfield  {author} {\bibinfo {author} {\bibfnamefont {A.}~\bibnamefont {{Abac}}}, \bibinfo {author} {\bibfnamefont {R.}~\bibnamefont {{Abramo}}}, \bibinfo {author} {\bibfnamefont {S.}~\bibnamefont {{Albanesi}}}, \emph {et~al.},\ }\href {https://doi.org/10.48550/arXiv.2503.12263} {\bibfield  {journal} {\bibinfo  {journal} {arXiv e-prints}\ ,\ \bibinfo {eid} {arXiv:2503.12263}} (\bibinfo {year} {2025}{\natexlab{a}})},\ \Eprint {https://arxiv.org/abs/2503.12263} {arXiv:2503.12263 [gr-qc]} \BibitemShut {NoStop}%
\bibitem [{\citenamefont {{Bauswein}}\ \emph {et~al.}(2013)\citenamefont {{Bauswein}}, \citenamefont {{Goriely}},\ and\ \citenamefont {{Janka}}}]{Bauswein2013}%
  \BibitemOpen
  \bibfield  {author} {\bibinfo {author} {\bibfnamefont {A.}~\bibnamefont {{Bauswein}}}, \bibinfo {author} {\bibfnamefont {S.}~\bibnamefont {{Goriely}}},\ and\ \bibinfo {author} {\bibfnamefont {H.~T.}\ \bibnamefont {{Janka}}},\ }\href {https://doi.org/10.1088/0004-637X/773/1/78} {\bibfield  {journal} {\bibinfo  {journal} {\apj}\ }\textbf {\bibinfo {volume} {773}},\ \bibinfo {eid} {78} (\bibinfo {year} {2013})},\ \Eprint {https://arxiv.org/abs/1302.6530} {arXiv:1302.6530 [astro-ph.SR]} \BibitemShut {NoStop}%
\bibitem [{\citenamefont {{Patricelli}}\ \emph {et~al.}(2022)\citenamefont {{Patricelli}}, \citenamefont {{Bernardini}}, \citenamefont {{Mapelli}}, \citenamefont {{D'Avanzo}}, \citenamefont {{Santoliquido}}, \citenamefont {{Cella}}, \citenamefont {{Razzano}},\ and\ \citenamefont {{Cuoco}}}]{2022MNRAS.513.4159P}%
  \BibitemOpen
  \bibfield  {author} {\bibinfo {author} {\bibfnamefont {B.}~\bibnamefont {{Patricelli}}}, \bibinfo {author} {\bibfnamefont {M.~G.}\ \bibnamefont {{Bernardini}}}, \bibinfo {author} {\bibfnamefont {M.}~\bibnamefont {{Mapelli}}}, \bibinfo {author} {\bibfnamefont {P.}~\bibnamefont {{D'Avanzo}}}, \bibinfo {author} {\bibfnamefont {F.}~\bibnamefont {{Santoliquido}}}, \bibinfo {author} {\bibfnamefont {G.}~\bibnamefont {{Cella}}}, \bibinfo {author} {\bibfnamefont {M.}~\bibnamefont {{Razzano}}},\ and\ \bibinfo {author} {\bibfnamefont {E.}~\bibnamefont {{Cuoco}}},\ }\href {https://doi.org/10.1093/mnras/stac1167} {\bibfield  {journal} {\bibinfo  {journal} {\mnras}\ }\textbf {\bibinfo {volume} {513}},\ \bibinfo {pages} {4159} (\bibinfo {year} {2022})},\ \Eprint {https://arxiv.org/abs/2204.12504} {arXiv:2204.12504 [astro-ph.HE]} \BibitemShut {NoStop}%
\bibitem [{\citenamefont {{\noopsort{Team COMPAS}}{Team COMPAS: Riley, J.}}\ \emph {et~al.}(2022)\citenamefont {{\noopsort{Team COMPAS}}{Team COMPAS: Riley, J.}}, \citenamefont {{Agrawal}}, \citenamefont {{Barrett}} \emph {et~al.}}]{COMPAS2022}%
  \BibitemOpen
  \bibfield  {author} {\bibinfo {author} {\bibnamefont {{\noopsort{Team COMPAS}}{Team COMPAS: Riley, J.}}}, \bibinfo {author} {\bibfnamefont {P.}~\bibnamefont {{Agrawal}}}, \bibinfo {author} {\bibfnamefont {J.~W.}\ \bibnamefont {{Barrett}}}, \emph {et~al.},\ }\href {https://doi.org/10.3847/1538-4365/ac416c} {\bibfield  {journal} {\bibinfo  {journal} {\apjs}\ }\textbf {\bibinfo {volume} {258}},\ \bibinfo {eid} {34} (\bibinfo {year} {2022})},\ \Eprint {https://arxiv.org/abs/2109.10352} {arXiv:2109.10352 [astro-ph.IM]} \BibitemShut {NoStop}%
\bibitem [{\citenamefont {{Iacovelli}}\ \emph {et~al.}(2022{\natexlab{a}})\citenamefont {{Iacovelli}}, \citenamefont {{Mancarella}}, \citenamefont {{Foffa}},\ and\ \citenamefont {{Maggiore}}}]{gwfast2022}%
  \BibitemOpen
  \bibfield  {author} {\bibinfo {author} {\bibfnamefont {F.}~\bibnamefont {{Iacovelli}}}, \bibinfo {author} {\bibfnamefont {M.}~\bibnamefont {{Mancarella}}}, \bibinfo {author} {\bibfnamefont {S.}~\bibnamefont {{Foffa}}},\ and\ \bibinfo {author} {\bibfnamefont {M.}~\bibnamefont {{Maggiore}}},\ }\href {https://doi.org/10.3847/1538-4365/ac9129} {\bibfield  {journal} {\bibinfo  {journal} {\apjs}\ }\textbf {\bibinfo {volume} {263}},\ \bibinfo {eid} {2} (\bibinfo {year} {2022}{\natexlab{a}})},\ \Eprint {https://arxiv.org/abs/2207.06910} {arXiv:2207.06910 [astro-ph.IM]} \BibitemShut {NoStop}%
\bibitem [{\citenamefont {{Guillochon}}\ \emph {et~al.}(2018)\citenamefont {{Guillochon}}, \citenamefont {{Nicholl}}, \citenamefont {{Villar}}, \citenamefont {{Mockler}}, \citenamefont {{Narayan}}, \citenamefont {{Mandel}}, \citenamefont {{Berger}},\ and\ \citenamefont {{Williams}}}]{MOSFiT2018}%
  \BibitemOpen
  \bibfield  {author} {\bibinfo {author} {\bibfnamefont {J.}~\bibnamefont {{Guillochon}}}, \bibinfo {author} {\bibfnamefont {M.}~\bibnamefont {{Nicholl}}}, \bibinfo {author} {\bibfnamefont {V.~A.}\ \bibnamefont {{Villar}}}, \bibinfo {author} {\bibfnamefont {B.}~\bibnamefont {{Mockler}}}, \bibinfo {author} {\bibfnamefont {G.}~\bibnamefont {{Narayan}}}, \bibinfo {author} {\bibfnamefont {K.~S.}\ \bibnamefont {{Mandel}}}, \bibinfo {author} {\bibfnamefont {E.}~\bibnamefont {{Berger}}},\ and\ \bibinfo {author} {\bibfnamefont {P.~K.~G.}\ \bibnamefont {{Williams}}},\ }\href {https://doi.org/10.3847/1538-4365/aab761} {\bibfield  {journal} {\bibinfo  {journal} {\apjs}\ }\textbf {\bibinfo {volume} {236}},\ \bibinfo {eid} {6} (\bibinfo {year} {2018})},\ \Eprint {https://arxiv.org/abs/1710.02145} {arXiv:1710.02145 [astro-ph.IM]} \BibitemShut {NoStop}%
\bibitem [{\citenamefont {{Abac}}\ \emph {et~al.}(2025{\natexlab{b}})\citenamefont {{Abac}}, \citenamefont {{Abramo}}, \citenamefont {{Albanesi}} \emph {et~al.}}]{ET2025}%
  \BibitemOpen
  \bibfield  {author} {\bibinfo {author} {\bibfnamefont {A.}~\bibnamefont {{Abac}}}, \bibinfo {author} {\bibfnamefont {R.}~\bibnamefont {{Abramo}}}, \bibinfo {author} {\bibfnamefont {S.}~\bibnamefont {{Albanesi}}}, \emph {et~al.},\ }\href {https://doi.org/10.48550/arXiv.2503.12263} {\bibfield  {journal} {\bibinfo  {journal} {arXiv e-prints}\ ,\ \bibinfo {eid} {arXiv:2503.12263}} (\bibinfo {year} {2025}{\natexlab{b}})},\ \Eprint {https://arxiv.org/abs/2503.12263} {arXiv:2503.12263 [gr-qc]} \BibitemShut {NoStop}%
\bibitem [{\citenamefont {{Evans}}\ \emph {et~al.}(2021)\citenamefont {{Evans}}, \citenamefont {{Adhikari}}, \citenamefont {{Afle}}, \citenamefont {{Ballmer}} \emph {et~al.}}]{CE2021}%
  \BibitemOpen
  \bibfield  {author} {\bibinfo {author} {\bibfnamefont {M.}~\bibnamefont {{Evans}}}, \bibinfo {author} {\bibfnamefont {R.~X.}\ \bibnamefont {{Adhikari}}}, \bibinfo {author} {\bibfnamefont {C.}~\bibnamefont {{Afle}}}, \bibinfo {author} {\bibfnamefont {S.~W.}\ \bibnamefont {{Ballmer}}}, \emph {et~al.},\ }\bibfield  {journal} {\bibinfo  {journal} {arXiv e-prints}\ }\href {https://doi.org/10.48550/arXiv.2109.09882} {10.48550/arXiv.2109.09882} (\bibinfo {year} {2021})\BibitemShut {NoStop}%
\bibitem [{\citenamefont {{Sari}}\ \emph {et~al.}(1998{\natexlab{a}})\citenamefont {{Sari}}, \citenamefont {{Piran}},\ and\ \citenamefont {{Narayan}}}]{Sari98}%
  \BibitemOpen
  \bibfield  {author} {\bibinfo {author} {\bibfnamefont {R.}~\bibnamefont {{Sari}}}, \bibinfo {author} {\bibfnamefont {T.}~\bibnamefont {{Piran}}},\ and\ \bibinfo {author} {\bibfnamefont {R.}~\bibnamefont {{Narayan}}},\ }\href {https://doi.org/10.1086/311269} {\bibfield  {journal} {\bibinfo  {journal} {\apjl}\ }\textbf {\bibinfo {volume} {497}},\ \bibinfo {pages} {L17} (\bibinfo {year} {1998}{\natexlab{a}})},\ \Eprint {https://arxiv.org/abs/astro-ph/9712005} {arXiv:astro-ph/9712005 [astro-ph]} \BibitemShut {NoStop}%
\bibitem [{\citenamefont {{Planck Collaboration}}\ \emph {et~al.}(2020)\citenamefont {{Planck Collaboration}}, \citenamefont {{Aghanim}} \emph {et~al.}}]{Planck2018}%
  \BibitemOpen
  \bibfield  {author} {\bibinfo {author} {\bibnamefont {{Planck Collaboration}}}, \bibinfo {author} {\bibfnamefont {N.}~\bibnamefont {{Aghanim}}}, \emph {et~al.},\ }\href {https://doi.org/10.1051/0004-6361/201833910} {\bibfield  {journal} {\bibinfo  {journal} {\aap}\ }\textbf {\bibinfo {volume} {641}},\ \bibinfo {eid} {A6} (\bibinfo {year} {2020})},\ \Eprint {https://arxiv.org/abs/1807.06209} {arXiv:1807.06209 [astro-ph.CO]} \BibitemShut {NoStop}%
\bibitem [{\citenamefont {{Stevenson}}\ \emph {et~al.}(2017{\natexlab{b}})\citenamefont {{Stevenson}}, \citenamefont {{Vigna-G{\'o}mez}}, \citenamefont {{Mandel}}, \citenamefont {{Barrett}}, \citenamefont {{Neijssel}}, \citenamefont {{Perkins}},\ and\ \citenamefont {{de Mink}}}]{Stevenson2017b}%
  \BibitemOpen
  \bibfield  {author} {\bibinfo {author} {\bibfnamefont {S.}~\bibnamefont {{Stevenson}}}, \bibinfo {author} {\bibfnamefont {A.}~\bibnamefont {{Vigna-G{\'o}mez}}}, \bibinfo {author} {\bibfnamefont {I.}~\bibnamefont {{Mandel}}}, \bibinfo {author} {\bibfnamefont {J.~W.}\ \bibnamefont {{Barrett}}}, \bibinfo {author} {\bibfnamefont {C.~J.}\ \bibnamefont {{Neijssel}}}, \bibinfo {author} {\bibfnamefont {D.}~\bibnamefont {{Perkins}}},\ and\ \bibinfo {author} {\bibfnamefont {S.~E.}\ \bibnamefont {{de Mink}}},\ }\href {https://doi.org/10.1038/ncomms14906} {\bibfield  {journal} {\bibinfo  {journal} {Nature Communications}\ }\textbf {\bibinfo {volume} {8}},\ \bibinfo {eid} {14906} (\bibinfo {year} {2017}{\natexlab{b}})},\ \Eprint {https://arxiv.org/abs/1704.01352} {arXiv:1704.01352 [astro-ph.HE]} \BibitemShut {NoStop}%
\bibitem [{\citenamefont {{Vigna-G{\'o}mez}}\ \emph {et~al.}(2018)\citenamefont {{Vigna-G{\'o}mez}}, \citenamefont {{Neijssel}}, \citenamefont {{Stevenson}} \emph {et~al.}}]{VignaGomez2018}%
  \BibitemOpen
  \bibfield  {author} {\bibinfo {author} {\bibfnamefont {A.}~\bibnamefont {{Vigna-G{\'o}mez}}}, \bibinfo {author} {\bibfnamefont {C.~J.}\ \bibnamefont {{Neijssel}}}, \bibinfo {author} {\bibfnamefont {S.}~\bibnamefont {{Stevenson}}}, \emph {et~al.},\ }\href {https://doi.org/10.1093/mnras/sty2463} {\bibfield  {journal} {\bibinfo  {journal} {\mnras}\ }\textbf {\bibinfo {volume} {481}},\ \bibinfo {pages} {4009} (\bibinfo {year} {2018})},\ \Eprint {https://arxiv.org/abs/1805.07974} {arXiv:1805.07974 [astro-ph.SR]} \BibitemShut {NoStop}%
\bibitem [{\citenamefont {{Vigna-G{\'o}mez}}\ \emph {et~al.}(2020)\citenamefont {{Vigna-G{\'o}mez}}, \citenamefont {{MacLeod}}, \citenamefont {{Neijssel}} \emph {et~al.}}]{VignaGomez2020}%
  \BibitemOpen
  \bibfield  {author} {\bibinfo {author} {\bibfnamefont {A.}~\bibnamefont {{Vigna-G{\'o}mez}}}, \bibinfo {author} {\bibfnamefont {M.}~\bibnamefont {{MacLeod}}}, \bibinfo {author} {\bibfnamefont {C.~J.}\ \bibnamefont {{Neijssel}}}, \emph {et~al.},\ }\href {https://doi.org/10.1017/pasa.2020.31} {\ \textbf {\bibinfo {volume} {37}},\ \bibinfo {eid} {e038} (\bibinfo {year} {2020})},\ \Eprint {https://arxiv.org/abs/2001.09829} {arXiv:2001.09829 [astro-ph.SR]} \BibitemShut {NoStop}%
\bibitem [{\citenamefont {{Kroupa}}(2001)}]{Kroupa2001}%
  \BibitemOpen
  \bibfield  {author} {\bibinfo {author} {\bibfnamefont {P.}~\bibnamefont {{Kroupa}}},\ }\href {https://doi.org/10.1046/j.1365-8711.2001.04022.x} {\bibfield  {journal} {\bibinfo  {journal} {\mnras}\ }\textbf {\bibinfo {volume} {322}},\ \bibinfo {pages} {231} (\bibinfo {year} {2001})},\ \Eprint {https://arxiv.org/abs/astro-ph/0009005} {arXiv:astro-ph/0009005 [astro-ph]} \BibitemShut {NoStop}%
\bibitem [{\citenamefont {{Sana}}\ \emph {et~al.}(2012)\citenamefont {{Sana}}, \citenamefont {{de Mink}}, \citenamefont {{de Koter}} \emph {et~al.}}]{Sana2012}%
  \BibitemOpen
  \bibfield  {author} {\bibinfo {author} {\bibfnamefont {H.}~\bibnamefont {{Sana}}}, \bibinfo {author} {\bibfnamefont {S.~E.}\ \bibnamefont {{de Mink}}}, \bibinfo {author} {\bibfnamefont {A.}~\bibnamefont {{de Koter}}}, \emph {et~al.},\ }\href {https://doi.org/10.1126/science.1223344} {\bibfield  {journal} {\bibinfo  {journal} {Science}\ }\textbf {\bibinfo {volume} {337}},\ \bibinfo {pages} {444} (\bibinfo {year} {2012})},\ \Eprint {https://arxiv.org/abs/1207.6397} {arXiv:1207.6397 [astro-ph.SR]} \BibitemShut {NoStop}%
\bibitem [{\citenamefont {{{\"O}pik}}(1924)}]{Opik1924}%
  \BibitemOpen
  \bibfield  {author} {\bibinfo {author} {\bibfnamefont {E.}~\bibnamefont {{{\"O}pik}}},\ }\href@noop {} {\bibfield  {journal} {\bibinfo  {journal} {Publications of the Tartu Astrofizica Observatory}\ }\textbf {\bibinfo {volume} {25}},\ \bibinfo {pages} {1} (\bibinfo {year} {1924})}\BibitemShut {NoStop}%
\bibitem [{\citenamefont {{Neijssel}}\ \emph {et~al.}(2019)\citenamefont {{Neijssel}}, \citenamefont {{Vigna-G{\'o}mez}}, \citenamefont {{Stevenson}} \emph {et~al.}}]{Neijssel2019}%
  \BibitemOpen
  \bibfield  {author} {\bibinfo {author} {\bibfnamefont {C.~J.}\ \bibnamefont {{Neijssel}}}, \bibinfo {author} {\bibfnamefont {A.}~\bibnamefont {{Vigna-G{\'o}mez}}}, \bibinfo {author} {\bibfnamefont {S.}~\bibnamefont {{Stevenson}}}, \emph {et~al.},\ }\href {https://doi.org/10.1093/mnras/stz2840} {\bibfield  {journal} {\bibinfo  {journal} {\mnras}\ }\textbf {\bibinfo {volume} {490}},\ \bibinfo {pages} {3740} (\bibinfo {year} {2019})},\ \Eprint {https://arxiv.org/abs/1906.08136} {arXiv:1906.08136 [astro-ph.SR]} \BibitemShut {NoStop}%
\bibitem [{\citenamefont {{Peters}}(1964)}]{Peters1964}%
  \BibitemOpen
  \bibfield  {author} {\bibinfo {author} {\bibfnamefont {P.~C.}\ \bibnamefont {{Peters}}},\ }\href {https://doi.org/10.1103/PhysRev.136.B1224} {\bibfield  {journal} {\bibinfo  {journal} {Physical Review}\ }\textbf {\bibinfo {volume} {136}},\ \bibinfo {pages} {1224} (\bibinfo {year} {1964})}\BibitemShut {NoStop}%
\bibitem [{\citenamefont {{Andreoni}}\ \emph {et~al.}(2024{\natexlab{a}})\citenamefont {{Andreoni}}, \citenamefont {{Margutti}}, \citenamefont {{Banovetz}} \emph {et~al.}}]{Andreoni2024a}%
  \BibitemOpen
  \bibfield  {author} {\bibinfo {author} {\bibfnamefont {I.}~\bibnamefont {{Andreoni}}}, \bibinfo {author} {\bibfnamefont {R.}~\bibnamefont {{Margutti}}}, \bibinfo {author} {\bibfnamefont {J.}~\bibnamefont {{Banovetz}}}, \emph {et~al.},\ }\href {https://doi.org/10.48550/arXiv.2411.04793} {\bibfield  {journal} {\bibinfo  {journal} {arXiv e-prints}\ ,\ \bibinfo {eid} {arXiv:2411.04793}} (\bibinfo {year} {2024}{\natexlab{a}})},\ \Eprint {https://arxiv.org/abs/2411.04793} {arXiv:2411.04793 [astro-ph.IM]} \BibitemShut {NoStop}%
\bibitem [{\citenamefont {{Andreoni}}\ \emph {et~al.}(2024{\natexlab{b}})\citenamefont {{Andreoni}}, \citenamefont {{Coughlin}}, \citenamefont {{Criswell}} \emph {et~al.}}]{Andreoni2024b}%
  \BibitemOpen
  \bibfield  {author} {\bibinfo {author} {\bibfnamefont {I.}~\bibnamefont {{Andreoni}}}, \bibinfo {author} {\bibfnamefont {M.~W.}\ \bibnamefont {{Coughlin}}}, \bibinfo {author} {\bibfnamefont {A.~W.}\ \bibnamefont {{Criswell}}}, \emph {et~al.},\ }\href {https://doi.org/10.1016/j.astropartphys.2023.102904} {\bibfield  {journal} {\bibinfo  {journal} {Astroparticle Physics}\ }\textbf {\bibinfo {volume} {155}},\ \bibinfo {eid} {102904} (\bibinfo {year} {2024}{\natexlab{b}})},\ \Eprint {https://arxiv.org/abs/2307.09511} {arXiv:2307.09511 [astro-ph.HE]} \BibitemShut {NoStop}%
\bibitem [{\citenamefont {{Soberman}}\ \emph {et~al.}(1997)\citenamefont {{Soberman}}, \citenamefont {{Phinney}},\ and\ \citenamefont {{van den Heuvel}}}]{Soberman1997}%
  \BibitemOpen
  \bibfield  {author} {\bibinfo {author} {\bibfnamefont {G.~E.}\ \bibnamefont {{Soberman}}}, \bibinfo {author} {\bibfnamefont {E.~S.}\ \bibnamefont {{Phinney}}},\ and\ \bibinfo {author} {\bibfnamefont {E.~P.~J.}\ \bibnamefont {{van den Heuvel}}},\ }\href {https://doi.org/10.48550/arXiv.astro-ph/9703016} {\bibfield  {journal} {\bibinfo  {journal} {\aap}\ }\textbf {\bibinfo {volume} {327}},\ \bibinfo {pages} {620} (\bibinfo {year} {1997})},\ \Eprint {https://arxiv.org/abs/astro-ph/9703016} {arXiv:astro-ph/9703016 [astro-ph]} \BibitemShut {NoStop}%
\bibitem [{\citenamefont {{Farrow}}\ \emph {et~al.}(2019)\citenamefont {{Farrow}}, \citenamefont {{Zhu}},\ and\ \citenamefont {{Thrane}}}]{2019ApJ...876...18F}%
  \BibitemOpen
  \bibfield  {author} {\bibinfo {author} {\bibfnamefont {N.}~\bibnamefont {{Farrow}}}, \bibinfo {author} {\bibfnamefont {X.-J.}\ \bibnamefont {{Zhu}}},\ and\ \bibinfo {author} {\bibfnamefont {E.}~\bibnamefont {{Thrane}}},\ }\href {https://doi.org/10.3847/1538-4357/ab12e3} {\bibfield  {journal} {\bibinfo  {journal} {\apj}\ }\textbf {\bibinfo {volume} {876}},\ \bibinfo {eid} {18} (\bibinfo {year} {2019})},\ \Eprint {https://arxiv.org/abs/1902.03300} {arXiv:1902.03300 [astro-ph.HE]} \BibitemShut {NoStop}%
\bibitem [{\citenamefont {{{\"O}zel}}\ and\ \citenamefont {{Freire}}(2016)}]{2016ARA&A..54..401O}%
  \BibitemOpen
  \bibfield  {author} {\bibinfo {author} {\bibfnamefont {F.}~\bibnamefont {{{\"O}zel}}}\ and\ \bibinfo {author} {\bibfnamefont {P.}~\bibnamefont {{Freire}}},\ }\href {https://doi.org/10.1146/annurev-astro-081915-023322} {\bibfield  {journal} {\bibinfo  {journal} {\araa}\ }\textbf {\bibinfo {volume} {54}},\ \bibinfo {pages} {401} (\bibinfo {year} {2016})},\ \Eprint {https://arxiv.org/abs/1603.02698} {arXiv:1603.02698 [astro-ph.HE]} \BibitemShut {NoStop}%
\bibitem [{\citenamefont {{Abbott}}\ \emph {et~al.}(2020{\natexlab{a}})\citenamefont {{Abbott}} \emph {et~al.}}]{GW190425discovery}%
  \BibitemOpen
  \bibfield  {author} {\bibinfo {author} {\bibfnamefont {B.~P.}\ \bibnamefont {{Abbott}}} \emph {et~al.},\ }\href {https://doi.org/10.3847/2041-8213/ab75f5} {\bibfield  {journal} {\bibinfo  {journal} {\apjl}\ }\textbf {\bibinfo {volume} {892}},\ \bibinfo {eid} {L3} (\bibinfo {year} {2020}{\natexlab{a}})},\ \Eprint {https://arxiv.org/abs/2001.01761} {arXiv:2001.01761 [astro-ph.HE]} \BibitemShut {NoStop}%
\bibitem [{\citenamefont {{Abbott}}\ \emph {et~al.}(2019{\natexlab{b}})\citenamefont {{Abbott}} \emph {et~al.}}]{GWTC1}%
  \BibitemOpen
  \bibfield  {author} {\bibinfo {author} {\bibfnamefont {B.~P.}\ \bibnamefont {{Abbott}}} \emph {et~al.},\ }\href {https://doi.org/10.1103/PhysRevX.9.031040} {\bibfield  {journal} {\bibinfo  {journal} {Physical Review X}\ }\textbf {\bibinfo {volume} {9}},\ \bibinfo {eid} {031040} (\bibinfo {year} {2019}{\natexlab{b}})},\ \Eprint {https://arxiv.org/abs/1811.12907} {arXiv:1811.12907 [astro-ph.HE]} \BibitemShut {NoStop}%
\bibitem [{\citenamefont {{Abbott}}\ \emph {et~al.}(2021)\citenamefont {{Abbott}} \emph {et~al.}}]{GWTC2}%
  \BibitemOpen
  \bibfield  {author} {\bibinfo {author} {\bibfnamefont {R.}~\bibnamefont {{Abbott}}} \emph {et~al.},\ }\href {https://doi.org/10.1103/PhysRevX.11.021053} {\bibfield  {journal} {\bibinfo  {journal} {Physical Review X}\ }\textbf {\bibinfo {volume} {11}},\ \bibinfo {eid} {021053} (\bibinfo {year} {2021})},\ \Eprint {https://arxiv.org/abs/2010.14527} {arXiv:2010.14527 [gr-qc]} \BibitemShut {NoStop}%
\bibitem [{\citenamefont {{Abbott}}\ \emph {et~al.}(2023)\citenamefont {{Abbott}} \emph {et~al.}}]{GWTC3}%
  \BibitemOpen
  \bibfield  {author} {\bibinfo {author} {\bibfnamefont {R.}~\bibnamefont {{Abbott}}} \emph {et~al.},\ }\href {https://doi.org/10.1103/PhysRevX.13.041039} {\bibfield  {journal} {\bibinfo  {journal} {Physical Review X}\ }\textbf {\bibinfo {volume} {13}},\ \bibinfo {eid} {041039} (\bibinfo {year} {2023})},\ \Eprint {https://arxiv.org/abs/2111.03606} {arXiv:2111.03606 [gr-qc]} \BibitemShut {NoStop}%
\bibitem [{\citenamefont {{The LIGO Scientific Collaboration}}\ \emph {et~al.}(2025)\citenamefont {{The LIGO Scientific Collaboration}}, \citenamefont {{the Virgo Collaboration}}, \citenamefont {{the KAGRA Collaboration}} \emph {et~al.}}]{GWTC4}%
  \BibitemOpen
  \bibfield  {author} {\bibinfo {author} {\bibnamefont {{The LIGO Scientific Collaboration}}}, \bibinfo {author} {\bibnamefont {{the Virgo Collaboration}}}, \bibinfo {author} {\bibnamefont {{the KAGRA Collaboration}}}, \emph {et~al.},\ }\href {https://doi.org/10.48550/arXiv.2508.18082} {\bibfield  {journal} {\bibinfo  {journal} {arXiv e-prints}\ ,\ \bibinfo {eid} {arXiv:2508.18082}} (\bibinfo {year} {2025})},\ \Eprint {https://arxiv.org/abs/2508.18082} {arXiv:2508.18082 [gr-qc]} \BibitemShut {NoStop}%
\bibitem [{\citenamefont {{Tauris}}\ \emph {et~al.}(2017)\citenamefont {{Tauris}}, \citenamefont {{Kramer}}, \citenamefont {{Freire}} \emph {et~al.}}]{2017ApJ...846..170T}%
  \BibitemOpen
  \bibfield  {author} {\bibinfo {author} {\bibfnamefont {T.~M.}\ \bibnamefont {{Tauris}}}, \bibinfo {author} {\bibfnamefont {M.}~\bibnamefont {{Kramer}}}, \bibinfo {author} {\bibfnamefont {P.~C.~C.}\ \bibnamefont {{Freire}}}, \emph {et~al.},\ }\href {https://doi.org/10.3847/1538-4357/aa7e89} {\bibfield  {journal} {\bibinfo  {journal} {\apj}\ }\textbf {\bibinfo {volume} {846}},\ \bibinfo {eid} {170} (\bibinfo {year} {2017})},\ \Eprint {https://arxiv.org/abs/1706.09438} {arXiv:1706.09438 [astro-ph.HE]} \BibitemShut {NoStop}%
\bibitem [{\citenamefont {{Cutler}}\ and\ \citenamefont {{Flanagan}}(1994)}]{Cutler1994}%
  \BibitemOpen
  \bibfield  {author} {\bibinfo {author} {\bibfnamefont {C.}~\bibnamefont {{Cutler}}}\ and\ \bibinfo {author} {\bibfnamefont {{\'E}.~E.}\ \bibnamefont {{Flanagan}}},\ }\href {https://doi.org/10.1103/PhysRevD.49.2658} {\bibfield  {journal} {\bibinfo  {journal} {\prd}\ }\textbf {\bibinfo {volume} {49}},\ \bibinfo {pages} {2658} (\bibinfo {year} {1994})},\ \Eprint {https://arxiv.org/abs/gr-qc/9402014} {arXiv:gr-qc/9402014 [gr-qc]} \BibitemShut {NoStop}%
\bibitem [{\citenamefont {{Kidder}}(1995)}]{Kidder1995}%
  \BibitemOpen
  \bibfield  {author} {\bibinfo {author} {\bibfnamefont {L.~E.}\ \bibnamefont {{Kidder}}},\ }\href {https://doi.org/10.1103/PhysRevD.52.821} {\bibfield  {journal} {\bibinfo  {journal} {\prd}\ }\textbf {\bibinfo {volume} {52}},\ \bibinfo {pages} {821} (\bibinfo {year} {1995})},\ \Eprint {https://arxiv.org/abs/gr-qc/9506022} {arXiv:gr-qc/9506022 [gr-qc]} \BibitemShut {NoStop}%
\bibitem [{\citenamefont {{Dhani}}\ \emph {et~al.}(2025)\citenamefont {{Dhani}}, \citenamefont {{V{\"o}lkel}}, \citenamefont {{Buonanno}}, \citenamefont {{Estelles}}, \citenamefont {{Gair}}, \citenamefont {{Pfeiffer}}, \citenamefont {{Pompili}},\ and\ \citenamefont {{Toubiana}}}]{2025PhRvX..15c1036D}%
  \BibitemOpen
  \bibfield  {author} {\bibinfo {author} {\bibfnamefont {A.}~\bibnamefont {{Dhani}}}, \bibinfo {author} {\bibfnamefont {S.~H.}\ \bibnamefont {{V{\"o}lkel}}}, \bibinfo {author} {\bibfnamefont {A.}~\bibnamefont {{Buonanno}}}, \bibinfo {author} {\bibfnamefont {H.}~\bibnamefont {{Estelles}}}, \bibinfo {author} {\bibfnamefont {J.}~\bibnamefont {{Gair}}}, \bibinfo {author} {\bibfnamefont {H.~P.}\ \bibnamefont {{Pfeiffer}}}, \bibinfo {author} {\bibfnamefont {L.}~\bibnamefont {{Pompili}}},\ and\ \bibinfo {author} {\bibfnamefont {A.}~\bibnamefont {{Toubiana}}},\ }\href {https://doi.org/10.1103/5pks-qz6b} {\bibfield  {journal} {\bibinfo  {journal} {Physical Review X}\ }\textbf {\bibinfo {volume} {15}},\ \bibinfo {eid} {031036} (\bibinfo {year} {2025})},\ \Eprint {https://arxiv.org/abs/2404.05811} {arXiv:2404.05811 [gr-qc]} \BibitemShut {NoStop}%
\bibitem [{\citenamefont {{Abbott}}\ \emph {et~al.}(2020{\natexlab{b}})\citenamefont {{Abbott}}, \citenamefont {{Abbott}}, \citenamefont {{Abbott}} \emph {et~al.}}]{Abbott2020}%
  \BibitemOpen
  \bibfield  {author} {\bibinfo {author} {\bibfnamefont {B.~P.}\ \bibnamefont {{Abbott}}}, \bibinfo {author} {\bibfnamefont {R.}~\bibnamefont {{Abbott}}}, \bibinfo {author} {\bibfnamefont {T.~D.}\ \bibnamefont {{Abbott}}}, \emph {et~al.},\ }\href {https://doi.org/10.1088/1361-6382/ab685e} {\bibfield  {journal} {\bibinfo  {journal} {Classical and Quantum Gravity}\ }\textbf {\bibinfo {volume} {37}},\ \bibinfo {eid} {055002} (\bibinfo {year} {2020}{\natexlab{b}})},\ \Eprint {https://arxiv.org/abs/1908.11170} {arXiv:1908.11170 [gr-qc]} \BibitemShut {NoStop}%
\bibitem [{\citenamefont {{Dietrich}}\ \emph {et~al.}(2019)\citenamefont {{Dietrich}}, \citenamefont {{Samajdar}}, \citenamefont {{Khan}}, \citenamefont {{Johnson-McDaniel}}, \citenamefont {{Dudi}},\ and\ \citenamefont {{Tichy}}}]{Dietrich2019}%
  \BibitemOpen
  \bibfield  {author} {\bibinfo {author} {\bibfnamefont {T.}~\bibnamefont {{Dietrich}}}, \bibinfo {author} {\bibfnamefont {A.}~\bibnamefont {{Samajdar}}}, \bibinfo {author} {\bibfnamefont {S.}~\bibnamefont {{Khan}}}, \bibinfo {author} {\bibfnamefont {N.~K.}\ \bibnamefont {{Johnson-McDaniel}}}, \bibinfo {author} {\bibfnamefont {R.}~\bibnamefont {{Dudi}}},\ and\ \bibinfo {author} {\bibfnamefont {W.}~\bibnamefont {{Tichy}}},\ }\href {https://doi.org/10.1103/PhysRevD.100.044003} {\bibfield  {journal} {\bibinfo  {journal} {\prd}\ }\textbf {\bibinfo {volume} {100}},\ \bibinfo {eid} {044003} (\bibinfo {year} {2019})},\ \Eprint {https://arxiv.org/abs/1905.06011} {arXiv:1905.06011 [gr-qc]} \BibitemShut {NoStop}%
\bibitem [{\citenamefont {{Barack}}\ and\ \citenamefont {{Cutler}}(2004)}]{Barack2004}%
  \BibitemOpen
  \bibfield  {author} {\bibinfo {author} {\bibfnamefont {L.}~\bibnamefont {{Barack}}}\ and\ \bibinfo {author} {\bibfnamefont {C.}~\bibnamefont {{Cutler}}},\ }\href {https://doi.org/10.1103/PhysRevD.69.082005} {\bibfield  {journal} {\bibinfo  {journal} {\prd}\ }\textbf {\bibinfo {volume} {69}},\ \bibinfo {eid} {082005} (\bibinfo {year} {2004})},\ \Eprint {https://arxiv.org/abs/gr-qc/0310125} {arXiv:gr-qc/0310125 [gr-qc]} \BibitemShut {NoStop}%
\bibitem [{\citenamefont {{Wen}}\ and\ \citenamefont {{Chen}}(2010)}]{Wen2010}%
  \BibitemOpen
  \bibfield  {author} {\bibinfo {author} {\bibfnamefont {L.}~\bibnamefont {{Wen}}}\ and\ \bibinfo {author} {\bibfnamefont {Y.}~\bibnamefont {{Chen}}},\ }\href {https://doi.org/10.1103/PhysRevD.81.082001} {\bibfield  {journal} {\bibinfo  {journal} {\prd}\ }\textbf {\bibinfo {volume} {81}},\ \bibinfo {eid} {082001} (\bibinfo {year} {2010})},\ \Eprint {https://arxiv.org/abs/1003.2504} {arXiv:1003.2504 [astro-ph.CO]} \BibitemShut {NoStop}%
\bibitem [{\citenamefont {{The PyCBC team}}(2018)}]{pycbc}%
  \BibitemOpen
  \bibfield  {author} {\bibinfo {author} {\bibnamefont {{The PyCBC team}}},\ }\href@noop {} {\bibinfo {title} {{PyCBC: Gravitational-wave data analysis toolkit}}},\ \bibinfo {howpublished} {Astrophysics Source Code Library, record ascl:1805.030} (\bibinfo {year} {2018})\BibitemShut {NoStop}%
\bibitem [{\citenamefont {{Srivastava}}\ \emph {et~al.}(2022)\citenamefont {{Srivastava}}, \citenamefont {{Davis}}, \citenamefont {{Kuns}}, \citenamefont {{Landry}}, \citenamefont {{Ballmer}}, \citenamefont {{Evans}}, \citenamefont {{Hall}}, \citenamefont {{Read}},\ and\ \citenamefont {{Sathyaprakash}}}]{CEpsd}%
  \BibitemOpen
  \bibfield  {author} {\bibinfo {author} {\bibfnamefont {V.}~\bibnamefont {{Srivastava}}}, \bibinfo {author} {\bibfnamefont {D.}~\bibnamefont {{Davis}}}, \bibinfo {author} {\bibfnamefont {K.}~\bibnamefont {{Kuns}}}, \bibinfo {author} {\bibfnamefont {P.}~\bibnamefont {{Landry}}}, \bibinfo {author} {\bibfnamefont {S.}~\bibnamefont {{Ballmer}}}, \bibinfo {author} {\bibfnamefont {M.}~\bibnamefont {{Evans}}}, \bibinfo {author} {\bibfnamefont {E.~D.}\ \bibnamefont {{Hall}}}, \bibinfo {author} {\bibfnamefont {J.}~\bibnamefont {{Read}}},\ and\ \bibinfo {author} {\bibfnamefont {B.~S.}\ \bibnamefont {{Sathyaprakash}}},\ }\href {https://doi.org/10.3847/1538-4357/ac5f04} {\bibfield  {journal} {\bibinfo  {journal} {\apj}\ }\textbf {\bibinfo {volume} {931}},\ \bibinfo {eid} {22} (\bibinfo {year} {2022})},\ \Eprint {https://arxiv.org/abs/2201.10668} {arXiv:2201.10668 [gr-qc]} \BibitemShut {NoStop}%
\bibitem [{\citenamefont {{Branchesi}}\ \emph {et~al.}(2023{\natexlab{b}})\citenamefont {{Branchesi}}, \citenamefont {{Maggiore}}, \citenamefont {{Alonso}} \emph {et~al.}}]{ETpsd}%
  \BibitemOpen
  \bibfield  {author} {\bibinfo {author} {\bibfnamefont {M.}~\bibnamefont {{Branchesi}}}, \bibinfo {author} {\bibfnamefont {M.}~\bibnamefont {{Maggiore}}}, \bibinfo {author} {\bibfnamefont {D.}~\bibnamefont {{Alonso}}}, \emph {et~al.},\ }\href {https://doi.org/10.1088/1475-7516/2023/07/068} {\bibfield  {journal} {\bibinfo  {journal} {\jcap}\ }\textbf {\bibinfo {volume} {2023}},\ \bibinfo {eid} {068} (\bibinfo {year} {2023}{\natexlab{b}})},\ \Eprint {https://arxiv.org/abs/2303.15923} {arXiv:2303.15923 [gr-qc]} \BibitemShut {NoStop}%
\bibitem [{\citenamefont {{Iacovelli}}\ \emph {et~al.}(2022{\natexlab{b}})\citenamefont {{Iacovelli}}, \citenamefont {{Mancarella}}, \citenamefont {{Foffa}},\ and\ \citenamefont {{Maggiore}}}]{Iacovelli2022}%
  \BibitemOpen
  \bibfield  {author} {\bibinfo {author} {\bibfnamefont {F.}~\bibnamefont {{Iacovelli}}}, \bibinfo {author} {\bibfnamefont {M.}~\bibnamefont {{Mancarella}}}, \bibinfo {author} {\bibfnamefont {S.}~\bibnamefont {{Foffa}}},\ and\ \bibinfo {author} {\bibfnamefont {M.}~\bibnamefont {{Maggiore}}},\ }\href {https://doi.org/10.3847/1538-4357/ac9cd4} {\bibfield  {journal} {\bibinfo  {journal} {\apj}\ }\textbf {\bibinfo {volume} {941}},\ \bibinfo {eid} {208} (\bibinfo {year} {2022}{\natexlab{b}})},\ \Eprint {https://arxiv.org/abs/2207.02771} {arXiv:2207.02771 [gr-qc]} \BibitemShut {NoStop}%
\bibitem [{\citenamefont {{Maggiore}}\ \emph {et~al.}(2024)\citenamefont {{Maggiore}}, \citenamefont {{Iacovelli}}, \citenamefont {{Belgacem}}, \citenamefont {{Mancarella}},\ and\ \citenamefont {{Muttoni}}}]{Maggiore2024}%
  \BibitemOpen
  \bibfield  {author} {\bibinfo {author} {\bibfnamefont {M.}~\bibnamefont {{Maggiore}}}, \bibinfo {author} {\bibfnamefont {F.}~\bibnamefont {{Iacovelli}}}, \bibinfo {author} {\bibfnamefont {E.}~\bibnamefont {{Belgacem}}}, \bibinfo {author} {\bibfnamefont {M.}~\bibnamefont {{Mancarella}}},\ and\ \bibinfo {author} {\bibfnamefont {N.}~\bibnamefont {{Muttoni}}},\ }\href {https://doi.org/10.48550/arXiv.2411.05754} {\bibfield  {journal} {\bibinfo  {journal} {arXiv e-prints}\ ,\ \bibinfo {eid} {arXiv:2411.05754}} (\bibinfo {year} {2024})},\ \Eprint {https://arxiv.org/abs/2411.05754} {arXiv:2411.05754 [gr-qc]} \BibitemShut {NoStop}%
\bibitem [{\citenamefont {{Radice}}\ \emph {et~al.}(2020)\citenamefont {{Radice}}, \citenamefont {{Bernuzzi}},\ and\ \citenamefont {{Perego}}}]{Radice2020}%
  \BibitemOpen
  \bibfield  {author} {\bibinfo {author} {\bibfnamefont {D.}~\bibnamefont {{Radice}}}, \bibinfo {author} {\bibfnamefont {S.}~\bibnamefont {{Bernuzzi}}},\ and\ \bibinfo {author} {\bibfnamefont {A.}~\bibnamefont {{Perego}}},\ }\href {https://doi.org/10.1146/annurev-nucl-013120-114541} {\bibfield  {journal} {\bibinfo  {journal} {Annual Review of Nuclear and Particle Science}\ }\textbf {\bibinfo {volume} {70}},\ \bibinfo {pages} {95} (\bibinfo {year} {2020})},\ \Eprint {https://arxiv.org/abs/2002.03863} {arXiv:2002.03863 [astro-ph.HE]} \BibitemShut {NoStop}%
\bibitem [{\citenamefont {{Nakar}}(2020)}]{Nakar2020}%
  \BibitemOpen
  \bibfield  {author} {\bibinfo {author} {\bibfnamefont {E.}~\bibnamefont {{Nakar}}},\ }\href {https://doi.org/10.1016/j.physrep.2020.08.008} {\bibfield  {journal} {\bibinfo  {journal} {\physrep}\ }\textbf {\bibinfo {volume} {886}},\ \bibinfo {pages} {1} (\bibinfo {year} {2020})},\ \Eprint {https://arxiv.org/abs/1912.05659} {arXiv:1912.05659 [astro-ph.HE]} \BibitemShut {NoStop}%
\bibitem [{\citenamefont {{Coughlin}}\ \emph {et~al.}(2019{\natexlab{b}})\citenamefont {{Coughlin}}, \citenamefont {{Antier}}, \citenamefont {{Corre}}, \citenamefont {{Alqassimi}}, \citenamefont {{Anand}}, \citenamefont {{Christensen}}, \citenamefont {{Coulter}}, \citenamefont {{Foley}}, \citenamefont {{Guessoum}}, \citenamefont {{Mikulski}}, \citenamefont {{Mualla}}, \citenamefont {{Reed}},\ and\ \citenamefont {{Tao}}}]{Coughlin2019b}%
  \BibitemOpen
  \bibfield  {author} {\bibinfo {author} {\bibfnamefont {M.~W.}\ \bibnamefont {{Coughlin}}}, \bibinfo {author} {\bibfnamefont {S.}~\bibnamefont {{Antier}}}, \bibinfo {author} {\bibfnamefont {D.}~\bibnamefont {{Corre}}}, \bibinfo {author} {\bibfnamefont {K.}~\bibnamefont {{Alqassimi}}}, \bibinfo {author} {\bibfnamefont {S.}~\bibnamefont {{Anand}}}, \bibinfo {author} {\bibfnamefont {N.}~\bibnamefont {{Christensen}}}, \bibinfo {author} {\bibfnamefont {D.~A.}\ \bibnamefont {{Coulter}}}, \bibinfo {author} {\bibfnamefont {R.~J.}\ \bibnamefont {{Foley}}}, \bibinfo {author} {\bibfnamefont {N.}~\bibnamefont {{Guessoum}}}, \bibinfo {author} {\bibfnamefont {T.~M.}\ \bibnamefont {{Mikulski}}}, \bibinfo {author} {\bibfnamefont {M.~A.}\ \bibnamefont {{Mualla}}}, \bibinfo {author} {\bibfnamefont {D.}~\bibnamefont {{Reed}}},\ and\ \bibinfo {author} {\bibfnamefont {D.}~\bibnamefont {{Tao}}},\ }\href {https://doi.org/10.1093/mnras/stz2485} {\bibfield  {journal} {\bibinfo  {journal} {\mnras}\ }\textbf {\bibinfo {volume}
  {489}},\ \bibinfo {pages} {5775} (\bibinfo {year} {2019}{\natexlab{b}})},\ \Eprint {https://arxiv.org/abs/1909.01244} {arXiv:1909.01244 [astro-ph.IM]} \BibitemShut {NoStop}%
\bibitem [{\citenamefont {{Raaijmakers}}\ \emph {et~al.}(2021{\natexlab{b}})\citenamefont {{Raaijmakers}}, \citenamefont {{Greif}}, \citenamefont {{Hebeler}}, \citenamefont {{Hinderer}}, \citenamefont {{Nissanke}}, \citenamefont {{Schwenk}}, \citenamefont {{Riley}}, \citenamefont {{Watts}}, \citenamefont {{Lattimer}},\ and\ \citenamefont {{Ho}}}]{Raaijmakers2021b}%
  \BibitemOpen
  \bibfield  {author} {\bibinfo {author} {\bibfnamefont {G.}~\bibnamefont {{Raaijmakers}}}, \bibinfo {author} {\bibfnamefont {S.~K.}\ \bibnamefont {{Greif}}}, \bibinfo {author} {\bibfnamefont {K.}~\bibnamefont {{Hebeler}}}, \bibinfo {author} {\bibfnamefont {T.}~\bibnamefont {{Hinderer}}}, \bibinfo {author} {\bibfnamefont {S.}~\bibnamefont {{Nissanke}}}, \bibinfo {author} {\bibfnamefont {A.}~\bibnamefont {{Schwenk}}}, \bibinfo {author} {\bibfnamefont {T.~E.}\ \bibnamefont {{Riley}}}, \bibinfo {author} {\bibfnamefont {A.~L.}\ \bibnamefont {{Watts}}}, \bibinfo {author} {\bibfnamefont {J.~M.}\ \bibnamefont {{Lattimer}}},\ and\ \bibinfo {author} {\bibfnamefont {W.~C.~G.}\ \bibnamefont {{Ho}}},\ }\href {https://doi.org/10.3847/2041-8213/ac089a} {\bibfield  {journal} {\bibinfo  {journal} {\apjl}\ }\textbf {\bibinfo {volume} {918}},\ \bibinfo {eid} {L29} (\bibinfo {year} {2021}{\natexlab{b}})},\ \Eprint {https://arxiv.org/abs/2105.06981} {arXiv:2105.06981 [astro-ph.HE]} \BibitemShut {NoStop}%
\bibitem [{\citenamefont {{Villar}}\ \emph {et~al.}(2017)\citenamefont {{Villar}}, \citenamefont {{Guillochon}}, \citenamefont {{Berger}}, \citenamefont {{Metzger}}, \citenamefont {{Cowperthwaite}}, \citenamefont {{Nicholl}}, \citenamefont {{Alexander}}, \citenamefont {{Blanchard}}, \citenamefont {{Chornock}}, \citenamefont {{Eftekhari}}, \citenamefont {{Fong}}, \citenamefont {{Margutti}},\ and\ \citenamefont {{Williams}}}]{Villar2017}%
  \BibitemOpen
  \bibfield  {author} {\bibinfo {author} {\bibfnamefont {V.~A.}\ \bibnamefont {{Villar}}}, \bibinfo {author} {\bibfnamefont {J.}~\bibnamefont {{Guillochon}}}, \bibinfo {author} {\bibfnamefont {E.}~\bibnamefont {{Berger}}}, \bibinfo {author} {\bibfnamefont {B.~D.}\ \bibnamefont {{Metzger}}}, \bibinfo {author} {\bibfnamefont {P.~S.}\ \bibnamefont {{Cowperthwaite}}}, \bibinfo {author} {\bibfnamefont {M.}~\bibnamefont {{Nicholl}}}, \bibinfo {author} {\bibfnamefont {K.~D.}\ \bibnamefont {{Alexander}}}, \bibinfo {author} {\bibfnamefont {P.~K.}\ \bibnamefont {{Blanchard}}}, \bibinfo {author} {\bibfnamefont {R.}~\bibnamefont {{Chornock}}}, \bibinfo {author} {\bibfnamefont {T.}~\bibnamefont {{Eftekhari}}}, \bibinfo {author} {\bibfnamefont {W.}~\bibnamefont {{Fong}}}, \bibinfo {author} {\bibfnamefont {R.}~\bibnamefont {{Margutti}}},\ and\ \bibinfo {author} {\bibfnamefont {P.~K.~G.}\ \bibnamefont {{Williams}}},\ }\href {https://doi.org/10.3847/2041-8213/aa9c84} {\bibfield  {journal} {\bibinfo  {journal} {\apjl}\
  }\textbf {\bibinfo {volume} {851}},\ \bibinfo {eid} {L21} (\bibinfo {year} {2017})},\ \Eprint {https://arxiv.org/abs/1710.11576} {arXiv:1710.11576 [astro-ph.HE]} \BibitemShut {NoStop}%
\bibitem [{\citenamefont {{Metzger}}(2019)}]{Metzger2019}%
  \BibitemOpen
  \bibfield  {author} {\bibinfo {author} {\bibfnamefont {B.~D.}\ \bibnamefont {{Metzger}}},\ }\href {https://doi.org/10.1007/s41114-019-0024-0} {\bibfield  {journal} {\bibinfo  {journal} {Living Reviews in Relativity}\ }\textbf {\bibinfo {volume} {23}},\ \bibinfo {eid} {1} (\bibinfo {year} {2019})},\ \Eprint {https://arxiv.org/abs/1910.01617} {arXiv:1910.01617 [astro-ph.HE]} \BibitemShut {NoStop}%
\bibitem [{\citenamefont {{Sari}}\ \emph {et~al.}(1998{\natexlab{b}})\citenamefont {{Sari}}, \citenamefont {{Piran}},\ and\ \citenamefont {{Narayan}}}]{afterglow_theory}%
  \BibitemOpen
  \bibfield  {author} {\bibinfo {author} {\bibfnamefont {R.}~\bibnamefont {{Sari}}}, \bibinfo {author} {\bibfnamefont {T.}~\bibnamefont {{Piran}}},\ and\ \bibinfo {author} {\bibfnamefont {R.}~\bibnamefont {{Narayan}}},\ }\href {https://doi.org/10.1086/311269} {\bibfield  {journal} {\bibinfo  {journal} {\apj}\ }\textbf {\bibinfo {volume} {497}},\ \bibinfo {pages} {L17} (\bibinfo {year} {1998}{\natexlab{b}})},\ \Eprint {https://arxiv.org/abs/9712005} {arXiv:9712005 [astro-ph.HE]} \BibitemShut {NoStop}%
\bibitem [{\citenamefont {{Blandford}}\ and\ \citenamefont {{McKee}}(1976)}]{synchrotron_eqns}%
  \BibitemOpen
  \bibfield  {author} {\bibinfo {author} {\bibfnamefont {R.}~\bibnamefont {{Blandford}}}\ and\ \bibinfo {author} {\bibfnamefont {C.}~\bibnamefont {{McKee}}},\ }\href {https://doi.org/10.1063/1.861619} {\bibfield  {journal} {\bibinfo  {journal} {Phys. Fluids}\ }\textbf {\bibinfo {volume} {19}},\ \bibinfo {pages} {1130} (\bibinfo {year} {1976})}\BibitemShut {NoStop}%
\bibitem [{\citenamefont {{van der Horst}}(2007)}]{Afterglow_eqns}%
  \BibitemOpen
  \bibfield  {author} {\bibinfo {author} {\bibfnamefont {A.}~\bibnamefont {{van der Horst}}},\ }\emph {\bibinfo {title} {{Broadband View of Blast Wave Physics: A Study of Gamma-Ray Burst Afterglows}}},\ \href@noop {} {Ph.D. thesis},\ \bibinfo  {school} {Universiteit van Amsterdam} (\bibinfo {year} {2007})\BibitemShut {NoStop}%
\bibitem [{\citenamefont {{Safi-Harb}}\ \emph {et~al.}(2019{\natexlab{b}})\citenamefont {{Safi-Harb}}, \citenamefont {{Doerksen}}, \citenamefont {{Rogers}},\ and\ \citenamefont {{Fryer}}}]{SafiHarb2019}%
  \BibitemOpen
  \bibfield  {author} {\bibinfo {author} {\bibfnamefont {S.}~\bibnamefont {{Safi-Harb}}}, \bibinfo {author} {\bibfnamefont {N.}~\bibnamefont {{Doerksen}}}, \bibinfo {author} {\bibfnamefont {A.}~\bibnamefont {{Rogers}}},\ and\ \bibinfo {author} {\bibfnamefont {C.~L.}\ \bibnamefont {{Fryer}}},\ }\href {https://doi.org/10.48550/arXiv.1812.11320} {\bibfield  {journal} {\bibinfo  {journal} {\jrasc}\ }\textbf {\bibinfo {volume} {113}},\ \bibinfo {pages} {7} (\bibinfo {year} {2019}{\natexlab{b}})},\ \Eprint {https://arxiv.org/abs/1812.11320} {arXiv:1812.11320 [astro-ph.HE]} \BibitemShut {NoStop}%
\bibitem [{\citenamefont {{Magee}}\ \emph {et~al.}(2019)\citenamefont {{Magee}}, \citenamefont {{Fong}}, \citenamefont {{Caudill}}, \citenamefont {{Messick}}, \citenamefont {{Cannon}}, \citenamefont {{Godwin}}, \citenamefont {{Hanna}}, \citenamefont {{Kapadia}}, \citenamefont {{Meacher}}, \citenamefont {{Mohite}}, \citenamefont {{Mukherjee}}, \citenamefont {{Pace}}, \citenamefont {{Sachdev}}, \citenamefont {{Shikauchi}},\ and\ \citenamefont {{Singer}}}]{Magee2019}%
  \BibitemOpen
  \bibfield  {author} {\bibinfo {author} {\bibfnamefont {R.}~\bibnamefont {{Magee}}}, \bibinfo {author} {\bibfnamefont {H.}~\bibnamefont {{Fong}}}, \bibinfo {author} {\bibfnamefont {S.}~\bibnamefont {{Caudill}}}, \bibinfo {author} {\bibfnamefont {C.}~\bibnamefont {{Messick}}}, \bibinfo {author} {\bibfnamefont {K.}~\bibnamefont {{Cannon}}}, \bibinfo {author} {\bibfnamefont {P.}~\bibnamefont {{Godwin}}}, \bibinfo {author} {\bibfnamefont {C.}~\bibnamefont {{Hanna}}}, \bibinfo {author} {\bibfnamefont {S.}~\bibnamefont {{Kapadia}}}, \bibinfo {author} {\bibfnamefont {D.}~\bibnamefont {{Meacher}}}, \bibinfo {author} {\bibfnamefont {S.~R.}\ \bibnamefont {{Mohite}}}, \bibinfo {author} {\bibfnamefont {D.}~\bibnamefont {{Mukherjee}}}, \bibinfo {author} {\bibfnamefont {A.}~\bibnamefont {{Pace}}}, \bibinfo {author} {\bibfnamefont {S.}~\bibnamefont {{Sachdev}}}, \bibinfo {author} {\bibfnamefont {M.}~\bibnamefont {{Shikauchi}}},\ and\ \bibinfo {author} {\bibfnamefont {L.}~\bibnamefont {{Singer}}},\ }\href
  {https://doi.org/10.3847/2041-8213/ab20cf} {\bibfield  {journal} {\bibinfo  {journal} {\apjl}\ }\textbf {\bibinfo {volume} {878}},\ \bibinfo {eid} {L17} (\bibinfo {year} {2019})},\ \Eprint {https://arxiv.org/abs/1901.09884} {arXiv:1901.09884 [gr-qc]} \BibitemShut {NoStop}%
\bibitem [{\citenamefont {{Astropy Collaboration}}\ \emph {et~al.}(2022)\citenamefont {{Astropy Collaboration}}, \citenamefont {{Price-Whelan}}, \citenamefont {{Lim}}, \citenamefont {{Earl}} \emph {et~al.}}]{Astropy}%
  \BibitemOpen
  \bibfield  {author} {\bibinfo {author} {\bibnamefont {{Astropy Collaboration}}}, \bibinfo {author} {\bibfnamefont {A.~M.}\ \bibnamefont {{Price-Whelan}}}, \bibinfo {author} {\bibfnamefont {P.~L.}\ \bibnamefont {{Lim}}}, \bibinfo {author} {\bibfnamefont {N.}~\bibnamefont {{Earl}}}, \emph {et~al.},\ }\href {https://doi.org/10.3847/1538-4357/ac7c74} {\bibfield  {journal} {\bibinfo  {journal} {\apj}\ }\textbf {\bibinfo {volume} {935}},\ \bibinfo {eid} {167} (\bibinfo {year} {2022})},\ \Eprint {https://arxiv.org/abs/2206.14220} {arXiv:2206.14220 [astro-ph.IM]} \BibitemShut {NoStop}%
\bibitem [{\citenamefont {{Broekgaarden}}\ \emph {et~al.}(2022)\citenamefont {{Broekgaarden}}, \citenamefont {{Berger}}, \citenamefont {{Stevenson}} \emph {et~al.}}]{Broekgaarden2022}%
  \BibitemOpen
  \bibfield  {author} {\bibinfo {author} {\bibfnamefont {F.~S.}\ \bibnamefont {{Broekgaarden}}}, \bibinfo {author} {\bibfnamefont {E.}~\bibnamefont {{Berger}}}, \bibinfo {author} {\bibfnamefont {S.}~\bibnamefont {{Stevenson}}}, \emph {et~al.},\ }\href {https://doi.org/10.1093/mnras/stac1677} {\bibfield  {journal} {\bibinfo  {journal} {\mnras}\ }\textbf {\bibinfo {volume} {516}},\ \bibinfo {pages} {5737} (\bibinfo {year} {2022})},\ \Eprint {https://arxiv.org/abs/2112.05763} {arXiv:2112.05763 [astro-ph.HE]} \BibitemShut {NoStop}%
\bibitem [{\citenamefont {{Mandel}}\ and\ \citenamefont {{Broekgaarden}}(2022)}]{Mandel2022}%
  \BibitemOpen
  \bibfield  {author} {\bibinfo {author} {\bibfnamefont {I.}~\bibnamefont {{Mandel}}}\ and\ \bibinfo {author} {\bibfnamefont {F.~S.}\ \bibnamefont {{Broekgaarden}}},\ }\href {https://doi.org/10.1007/s41114-021-00034-3} {\bibfield  {journal} {\bibinfo  {journal} {Living Reviews in Relativity}\ }\textbf {\bibinfo {volume} {25}},\ \bibinfo {eid} {1} (\bibinfo {year} {2022})},\ \Eprint {https://arxiv.org/abs/2107.14239} {arXiv:2107.14239 [astro-ph.HE]} \BibitemShut {NoStop}%
\bibitem [{\citenamefont {{van Son}}\ \emph {et~al.}(2022)\citenamefont {{van Son}}, \citenamefont {{de Mink}}, \citenamefont {{Callister}}, \citenamefont {{Justham}}, \citenamefont {{Renzo}}, \citenamefont {{Wagg}}, \citenamefont {{Broekgaarden}}, \citenamefont {{Kummer}}, \citenamefont {{Pakmor}},\ and\ \citenamefont {{Mandel}}}]{vanSon2022}%
  \BibitemOpen
  \bibfield  {author} {\bibinfo {author} {\bibfnamefont {L.~A.~C.}\ \bibnamefont {{van Son}}}, \bibinfo {author} {\bibfnamefont {S.~E.}\ \bibnamefont {{de Mink}}}, \bibinfo {author} {\bibfnamefont {T.}~\bibnamefont {{Callister}}}, \bibinfo {author} {\bibfnamefont {S.}~\bibnamefont {{Justham}}}, \bibinfo {author} {\bibfnamefont {M.}~\bibnamefont {{Renzo}}}, \bibinfo {author} {\bibfnamefont {T.}~\bibnamefont {{Wagg}}}, \bibinfo {author} {\bibfnamefont {F.~S.}\ \bibnamefont {{Broekgaarden}}}, \bibinfo {author} {\bibfnamefont {F.}~\bibnamefont {{Kummer}}}, \bibinfo {author} {\bibfnamefont {R.}~\bibnamefont {{Pakmor}}},\ and\ \bibinfo {author} {\bibfnamefont {I.}~\bibnamefont {{Mandel}}},\ }\href {https://doi.org/10.3847/1538-4357/ac64a3} {\bibfield  {journal} {\bibinfo  {journal} {\apj}\ }\textbf {\bibinfo {volume} {931}},\ \bibinfo {eid} {17} (\bibinfo {year} {2022})},\ \Eprint {https://arxiv.org/abs/2110.01634} {arXiv:2110.01634 [astro-ph.HE]} \BibitemShut {NoStop}%
\bibitem [{\citenamefont {{Fishbach}}\ and\ \citenamefont {{van Son}}(2023)}]{Fishbach2023}%
  \BibitemOpen
  \bibfield  {author} {\bibinfo {author} {\bibfnamefont {M.}~\bibnamefont {{Fishbach}}}\ and\ \bibinfo {author} {\bibfnamefont {L.}~\bibnamefont {{van Son}}},\ }\href {https://doi.org/10.3847/2041-8213/ad0560} {\bibfield  {journal} {\bibinfo  {journal} {\apjl}\ }\textbf {\bibinfo {volume} {957}},\ \bibinfo {eid} {L31} (\bibinfo {year} {2023})},\ \Eprint {https://arxiv.org/abs/2307.15824} {arXiv:2307.15824 [astro-ph.GA]} \BibitemShut {NoStop}%
\bibitem [{\citenamefont {{Schiebelbein-Zwack}}\ and\ \citenamefont {{Fishbach}}(2024)}]{SchiebelbeinZwack2024}%
  \BibitemOpen
  \bibfield  {author} {\bibinfo {author} {\bibfnamefont {A.}~\bibnamefont {{Schiebelbein-Zwack}}}\ and\ \bibinfo {author} {\bibfnamefont {M.}~\bibnamefont {{Fishbach}}},\ }\href {https://doi.org/10.3847/1538-4357/ad5353} {\bibfield  {journal} {\bibinfo  {journal} {\apj}\ }\textbf {\bibinfo {volume} {970}},\ \bibinfo {eid} {128} (\bibinfo {year} {2024})},\ \Eprint {https://arxiv.org/abs/2403.17156} {arXiv:2403.17156 [astro-ph.HE]} \BibitemShut {NoStop}%
\bibitem [{\citenamefont {{de S{\'a}}}\ \emph {et~al.}(2024)\citenamefont {{de S{\'a}}}, \citenamefont {{Rocha}}, \citenamefont {{Bernardo}}, \citenamefont {{Bachega}},\ and\ \citenamefont {{Horvath}}}]{deSa2024}%
  \BibitemOpen
  \bibfield  {author} {\bibinfo {author} {\bibfnamefont {L.~M.}\ \bibnamefont {{de S{\'a}}}}, \bibinfo {author} {\bibfnamefont {L.~S.}\ \bibnamefont {{Rocha}}}, \bibinfo {author} {\bibfnamefont {A.}~\bibnamefont {{Bernardo}}}, \bibinfo {author} {\bibfnamefont {R.~R.~A.}\ \bibnamefont {{Bachega}}},\ and\ \bibinfo {author} {\bibfnamefont {J.~E.}\ \bibnamefont {{Horvath}}},\ }\href {https://doi.org/10.1093/mnras/stae2281} {\bibfield  {journal} {\bibinfo  {journal} {\mnras}\ }\textbf {\bibinfo {volume} {535}},\ \bibinfo {pages} {2041} (\bibinfo {year} {2024})},\ \Eprint {https://arxiv.org/abs/2410.01451} {arXiv:2410.01451 [astro-ph.HE]} \BibitemShut {NoStop}%
\bibitem [{\citenamefont {{Fishbach}}(2025)}]{Fishbach2025}%
  \BibitemOpen
  \bibfield  {author} {\bibinfo {author} {\bibfnamefont {M.}~\bibnamefont {{Fishbach}}},\ }\href {https://doi.org/10.1088/1361-6382/adaf70} {\bibfield  {journal} {\bibinfo  {journal} {Classical and Quantum Gravity}\ }\textbf {\bibinfo {volume} {42}},\ \bibinfo {eid} {055009} (\bibinfo {year} {2025})},\ \Eprint {https://arxiv.org/abs/2411.08658} {arXiv:2411.08658 [astro-ph.HE]} \BibitemShut {NoStop}%
\bibitem [{\citenamefont {{Sgalletta}}\ \emph {et~al.}(2025)\citenamefont {{Sgalletta}}, \citenamefont {{Mapelli}}, \citenamefont {{Boco}}, \citenamefont {{Santoliquido}}, \citenamefont {{Artale}}, \citenamefont {{Iorio}}, \citenamefont {{Lapi}},\ and\ \citenamefont {{Spera}}}]{Sgalletta2025}%
  \BibitemOpen
  \bibfield  {author} {\bibinfo {author} {\bibfnamefont {C.}~\bibnamefont {{Sgalletta}}}, \bibinfo {author} {\bibfnamefont {M.}~\bibnamefont {{Mapelli}}}, \bibinfo {author} {\bibfnamefont {L.}~\bibnamefont {{Boco}}}, \bibinfo {author} {\bibfnamefont {F.}~\bibnamefont {{Santoliquido}}}, \bibinfo {author} {\bibfnamefont {M.~C.}\ \bibnamefont {{Artale}}}, \bibinfo {author} {\bibfnamefont {G.}~\bibnamefont {{Iorio}}}, \bibinfo {author} {\bibfnamefont {A.}~\bibnamefont {{Lapi}}},\ and\ \bibinfo {author} {\bibfnamefont {M.}~\bibnamefont {{Spera}}},\ }\href {https://doi.org/10.1051/0004-6361/202452757} {\bibfield  {journal} {\bibinfo  {journal} {\aap}\ }\textbf {\bibinfo {volume} {698}},\ \bibinfo {eid} {A144} (\bibinfo {year} {2025})},\ \Eprint {https://arxiv.org/abs/2410.21401} {arXiv:2410.21401 [astro-ph.HE]} \BibitemShut {NoStop}%
\bibitem [{\citenamefont {{Roy}}\ \emph {et~al.}(2025)\citenamefont {{Roy}}, \citenamefont {{van Son}}, \citenamefont {{Ray}},\ and\ \citenamefont {{Farr}}}]{2025ApJ...985L..33R}%
  \BibitemOpen
  \bibfield  {author} {\bibinfo {author} {\bibfnamefont {S.~K.}\ \bibnamefont {{Roy}}}, \bibinfo {author} {\bibfnamefont {L.~A.~C.}\ \bibnamefont {{van Son}}}, \bibinfo {author} {\bibfnamefont {A.}~\bibnamefont {{Ray}}},\ and\ \bibinfo {author} {\bibfnamefont {W.~M.}\ \bibnamefont {{Farr}}},\ }\href {https://doi.org/10.3847/2041-8213/add34a} {\bibfield  {journal} {\bibinfo  {journal} {\apjl}\ }\textbf {\bibinfo {volume} {985}},\ \bibinfo {eid} {L33} (\bibinfo {year} {2025})},\ \Eprint {https://arxiv.org/abs/2411.02494} {arXiv:2411.02494 [astro-ph.CO]} \BibitemShut {NoStop}%
\bibitem [{\citenamefont {{Breivik}}\ \emph {et~al.}(2019)\citenamefont {{Breivik}}, \citenamefont {{Price-Whelan}}, \citenamefont {{D'Orazio}}, \citenamefont {{Hogg}}, \citenamefont {{Johnson}}, \citenamefont {{Moe}}, \citenamefont {{Morton}},\ and\ \citenamefont {{Tayar}}}]{Breivik2019}%
  \BibitemOpen
  \bibfield  {author} {\bibinfo {author} {\bibfnamefont {K.}~\bibnamefont {{Breivik}}}, \bibinfo {author} {\bibfnamefont {A.~M.}\ \bibnamefont {{Price-Whelan}}}, \bibinfo {author} {\bibfnamefont {D.~J.}\ \bibnamefont {{D'Orazio}}}, \bibinfo {author} {\bibfnamefont {D.~W.}\ \bibnamefont {{Hogg}}}, \bibinfo {author} {\bibfnamefont {L.~C.}\ \bibnamefont {{Johnson}}}, \bibinfo {author} {\bibfnamefont {M.}~\bibnamefont {{Moe}}}, \bibinfo {author} {\bibfnamefont {T.~D.}\ \bibnamefont {{Morton}}},\ and\ \bibinfo {author} {\bibfnamefont {J.}~\bibnamefont {{Tayar}}},\ }\href {https://doi.org/10.48550/arXiv.1903.05094} {\bibfield  {journal} {\bibinfo  {journal} {arXiv e-prints}\ ,\ \bibinfo {eid} {arXiv:1903.05094}} (\bibinfo {year} {2019})},\ \Eprint {https://arxiv.org/abs/1903.05094} {arXiv:1903.05094 [astro-ph.SR]} \BibitemShut {NoStop}%
\bibitem [{\citenamefont {{Littenberg}}\ \emph {et~al.}(2013)\citenamefont {{Littenberg}}, \citenamefont {{Baker}}, \citenamefont {{Buonanno}},\ and\ \citenamefont {{Kelly}}}]{Littenberg2013}%
  \BibitemOpen
  \bibfield  {author} {\bibinfo {author} {\bibfnamefont {T.~B.}\ \bibnamefont {{Littenberg}}}, \bibinfo {author} {\bibfnamefont {J.~G.}\ \bibnamefont {{Baker}}}, \bibinfo {author} {\bibfnamefont {A.}~\bibnamefont {{Buonanno}}},\ and\ \bibinfo {author} {\bibfnamefont {B.~J.}\ \bibnamefont {{Kelly}}},\ }\href {https://doi.org/10.1103/PhysRevD.87.104003} {\bibfield  {journal} {\bibinfo  {journal} {\prd}\ }\textbf {\bibinfo {volume} {87}},\ \bibinfo {eid} {104003} (\bibinfo {year} {2013})},\ \Eprint {https://arxiv.org/abs/1210.0893} {arXiv:1210.0893 [gr-qc]} \BibitemShut {NoStop}%
\bibitem [{\citenamefont {{Kapil}}\ \emph {et~al.}(2024)\citenamefont {{Kapil}}, \citenamefont {{Reali}}, \citenamefont {{Cotesta}},\ and\ \citenamefont {{Berti}}}]{Kapil2024}%
  \BibitemOpen
  \bibfield  {author} {\bibinfo {author} {\bibfnamefont {V.}~\bibnamefont {{Kapil}}}, \bibinfo {author} {\bibfnamefont {L.}~\bibnamefont {{Reali}}}, \bibinfo {author} {\bibfnamefont {R.}~\bibnamefont {{Cotesta}}},\ and\ \bibinfo {author} {\bibfnamefont {E.}~\bibnamefont {{Berti}}},\ }\href {https://doi.org/10.1103/PhysRevD.109.104043} {\bibfield  {journal} {\bibinfo  {journal} {\prd}\ }\textbf {\bibinfo {volume} {109}},\ \bibinfo {eid} {104043} (\bibinfo {year} {2024})},\ \Eprint {https://arxiv.org/abs/2404.00090} {arXiv:2404.00090 [gr-qc]} \BibitemShut {NoStop}%
\bibitem [{\citenamefont {{Gaspari}}\ \emph {et~al.}(2024)\citenamefont {{Gaspari}}, \citenamefont {{Stevance}}, \citenamefont {{Levan}}, \citenamefont {{Chrimes}},\ and\ \citenamefont {{Lyman}}}]{Gaspari2024}%
  \BibitemOpen
  \bibfield  {author} {\bibinfo {author} {\bibfnamefont {N.}~\bibnamefont {{Gaspari}}}, \bibinfo {author} {\bibfnamefont {H.~F.}\ \bibnamefont {{Stevance}}}, \bibinfo {author} {\bibfnamefont {A.~J.}\ \bibnamefont {{Levan}}}, \bibinfo {author} {\bibfnamefont {A.~A.}\ \bibnamefont {{Chrimes}}},\ and\ \bibinfo {author} {\bibfnamefont {J.~D.}\ \bibnamefont {{Lyman}}},\ }\href {https://doi.org/10.1051/0004-6361/202450908} {\bibfield  {journal} {\bibinfo  {journal} {\aap}\ }\textbf {\bibinfo {volume} {692}},\ \bibinfo {eid} {A21} (\bibinfo {year} {2024})},\ \Eprint {https://arxiv.org/abs/2410.19480} {arXiv:2410.19480 [astro-ph.HE]} \BibitemShut {NoStop}%
\bibitem [{\citenamefont {{Boesky}}\ \emph {et~al.}(2024)\citenamefont {{Boesky}}, \citenamefont {{Broekgaarden}},\ and\ \citenamefont {{Berger}}}]{2024ApJ...976...24B}%
  \BibitemOpen
  \bibfield  {author} {\bibinfo {author} {\bibfnamefont {A.~P.}\ \bibnamefont {{Boesky}}}, \bibinfo {author} {\bibfnamefont {F.~S.}\ \bibnamefont {{Broekgaarden}}},\ and\ \bibinfo {author} {\bibfnamefont {E.}~\bibnamefont {{Berger}}},\ }\href {https://doi.org/10.3847/1538-4357/ad7fe3} {\bibfield  {journal} {\bibinfo  {journal} {\apj}\ }\textbf {\bibinfo {volume} {976}},\ \bibinfo {eid} {24} (\bibinfo {year} {2024})},\ \Eprint {https://arxiv.org/abs/2405.01630} {arXiv:2405.01630 [astro-ph.HE]} \BibitemShut {NoStop}%
\bibitem [{\citenamefont {{Dominik}}\ \emph {et~al.}(2012)\citenamefont {{Dominik}}, \citenamefont {{Belczynski}}, \citenamefont {{Fryer}}, \citenamefont {{Holz}}, \citenamefont {{Berti}}, \citenamefont {{Bulik}}, \citenamefont {{Mandel}},\ and\ \citenamefont {{O'Shaughnessy}}}]{2012ApJ...759...52D}%
  \BibitemOpen
  \bibfield  {author} {\bibinfo {author} {\bibfnamefont {M.}~\bibnamefont {{Dominik}}}, \bibinfo {author} {\bibfnamefont {K.}~\bibnamefont {{Belczynski}}}, \bibinfo {author} {\bibfnamefont {C.}~\bibnamefont {{Fryer}}}, \bibinfo {author} {\bibfnamefont {D.~E.}\ \bibnamefont {{Holz}}}, \bibinfo {author} {\bibfnamefont {E.}~\bibnamefont {{Berti}}}, \bibinfo {author} {\bibfnamefont {T.}~\bibnamefont {{Bulik}}}, \bibinfo {author} {\bibfnamefont {I.}~\bibnamefont {{Mandel}}},\ and\ \bibinfo {author} {\bibfnamefont {R.}~\bibnamefont {{O'Shaughnessy}}},\ }\href {https://doi.org/10.1088/0004-637X/759/1/52} {\bibfield  {journal} {\bibinfo  {journal} {\apj}\ }\textbf {\bibinfo {volume} {759}},\ \bibinfo {eid} {52} (\bibinfo {year} {2012})},\ \Eprint {https://arxiv.org/abs/1202.4901} {arXiv:1202.4901 [astro-ph.HE]} \BibitemShut {NoStop}%
\bibitem [{\citenamefont {{Giacobbo}}\ and\ \citenamefont {{Mapelli}}(2018)}]{2018MNRAS.480.2011G}%
  \BibitemOpen
  \bibfield  {author} {\bibinfo {author} {\bibfnamefont {N.}~\bibnamefont {{Giacobbo}}}\ and\ \bibinfo {author} {\bibfnamefont {M.}~\bibnamefont {{Mapelli}}},\ }\href {https://doi.org/10.1093/mnras/sty1999} {\bibfield  {journal} {\bibinfo  {journal} {\mnras}\ }\textbf {\bibinfo {volume} {480}},\ \bibinfo {pages} {2011} (\bibinfo {year} {2018})},\ \Eprint {https://arxiv.org/abs/1806.00001} {arXiv:1806.00001 [astro-ph.HE]} \BibitemShut {NoStop}%
\bibitem [{\citenamefont {{Broekgaarden}}\ \emph {et~al.}(2021)\citenamefont {{Broekgaarden}}, \citenamefont {{Berger}}, \citenamefont {{Neijssel}}, \citenamefont {{Vigna-G{\'o}mez}}, \citenamefont {{Chattopadhyay}}, \citenamefont {{Stevenson}}, \citenamefont {{Chruslinska}}, \citenamefont {{Justham}}, \citenamefont {{de Mink}},\ and\ \citenamefont {{Mandel}}}]{2021MNRAS.508.5028B}%
  \BibitemOpen
  \bibfield  {author} {\bibinfo {author} {\bibfnamefont {F.~S.}\ \bibnamefont {{Broekgaarden}}}, \bibinfo {author} {\bibfnamefont {E.}~\bibnamefont {{Berger}}}, \bibinfo {author} {\bibfnamefont {C.~J.}\ \bibnamefont {{Neijssel}}}, \bibinfo {author} {\bibfnamefont {A.}~\bibnamefont {{Vigna-G{\'o}mez}}}, \bibinfo {author} {\bibfnamefont {D.}~\bibnamefont {{Chattopadhyay}}}, \bibinfo {author} {\bibfnamefont {S.}~\bibnamefont {{Stevenson}}}, \bibinfo {author} {\bibfnamefont {M.}~\bibnamefont {{Chruslinska}}}, \bibinfo {author} {\bibfnamefont {S.}~\bibnamefont {{Justham}}}, \bibinfo {author} {\bibfnamefont {S.~E.}\ \bibnamefont {{de Mink}}},\ and\ \bibinfo {author} {\bibfnamefont {I.}~\bibnamefont {{Mandel}}},\ }\href {https://doi.org/10.1093/mnras/stab2716} {\bibfield  {journal} {\bibinfo  {journal} {\mnras}\ }\textbf {\bibinfo {volume} {508}},\ \bibinfo {pages} {5028} (\bibinfo {year} {2021})},\ \Eprint {https://arxiv.org/abs/2103.02608} {arXiv:2103.02608 [astro-ph.HE]} \BibitemShut {NoStop}%
\bibitem [{\citenamefont {{Dorozsmai}}\ and\ \citenamefont {{Toonen}}(2024)}]{2024MNRAS.530.3706D}%
  \BibitemOpen
  \bibfield  {author} {\bibinfo {author} {\bibfnamefont {A.}~\bibnamefont {{Dorozsmai}}}\ and\ \bibinfo {author} {\bibfnamefont {S.}~\bibnamefont {{Toonen}}},\ }\href {https://doi.org/10.1093/mnras/stae152} {\bibfield  {journal} {\bibinfo  {journal} {\mnras}\ }\textbf {\bibinfo {volume} {530}},\ \bibinfo {pages} {3706} (\bibinfo {year} {2024})},\ \Eprint {https://arxiv.org/abs/2207.08837} {arXiv:2207.08837 [astro-ph.SR]} \BibitemShut {NoStop}%
\bibitem [{\citenamefont {{Chruslinska}}\ \emph {et~al.}(2018)\citenamefont {{Chruslinska}}, \citenamefont {{Belczynski}}, \citenamefont {{Klencki}},\ and\ \citenamefont {{Benacquista}}}]{2018MNRAS.474.2937C}%
  \BibitemOpen
  \bibfield  {author} {\bibinfo {author} {\bibfnamefont {M.}~\bibnamefont {{Chruslinska}}}, \bibinfo {author} {\bibfnamefont {K.}~\bibnamefont {{Belczynski}}}, \bibinfo {author} {\bibfnamefont {J.}~\bibnamefont {{Klencki}}},\ and\ \bibinfo {author} {\bibfnamefont {M.}~\bibnamefont {{Benacquista}}},\ }\href {https://doi.org/10.1093/mnras/stx2923} {\bibfield  {journal} {\bibinfo  {journal} {\mnras}\ }\textbf {\bibinfo {volume} {474}},\ \bibinfo {pages} {2937} (\bibinfo {year} {2018})},\ \Eprint {https://arxiv.org/abs/1708.07885} {arXiv:1708.07885 [astro-ph.HE]} \BibitemShut {NoStop}%
\bibitem [{\citenamefont {{Mandel}}\ and\ \citenamefont {{M{\"u}ller}}(2020)}]{2020MNRAS.499.3214M}%
  \BibitemOpen
  \bibfield  {author} {\bibinfo {author} {\bibfnamefont {I.}~\bibnamefont {{Mandel}}}\ and\ \bibinfo {author} {\bibfnamefont {B.}~\bibnamefont {{M{\"u}ller}}},\ }\href {https://doi.org/10.1093/mnras/staa3043} {\bibfield  {journal} {\bibinfo  {journal} {\mnras}\ }\textbf {\bibinfo {volume} {499}},\ \bibinfo {pages} {3214} (\bibinfo {year} {2020})},\ \Eprint {https://arxiv.org/abs/2006.08360} {arXiv:2006.08360 [astro-ph.HE]} \BibitemShut {NoStop}%
\bibitem [{\citenamefont {{Liu}}\ \emph {et~al.}(2024)\citenamefont {{Liu}}, \citenamefont {{Zhang}},\ and\ \citenamefont {{Wang}}}]{Liu2024}%
  \BibitemOpen
  \bibfield  {author} {\bibinfo {author} {\bibfnamefont {M.}~\bibnamefont {{Liu}}}, \bibinfo {author} {\bibfnamefont {J.}~\bibnamefont {{Zhang}}},\ and\ \bibinfo {author} {\bibfnamefont {C.}~\bibnamefont {{Wang}}},\ }\href {https://doi.org/10.48550/arXiv.2403.08223} {\bibfield  {journal} {\bibinfo  {journal} {arXiv e-prints}\ ,\ \bibinfo {eid} {arXiv:2403.08223}} (\bibinfo {year} {2024})},\ \Eprint {https://arxiv.org/abs/2403.08223} {arXiv:2403.08223 [astro-ph.HE]} \BibitemShut {NoStop}%
\bibitem [{\citenamefont {{Steinle}}\ \emph {et~al.}(2025)\citenamefont {{Steinle}}, \citenamefont {{Safi-Harb}},\ and\ \citenamefont {{Fryer}}}]{Steinle2025}%
  \BibitemOpen
  \bibfield  {author} {\bibinfo {author} {\bibfnamefont {N.}~\bibnamefont {{Steinle}}}, \bibinfo {author} {\bibfnamefont {S.}~\bibnamefont {{Safi-Harb}}},\ and\ \bibinfo {author} {\bibfnamefont {C.}~\bibnamefont {{Fryer}}},\ }\href@noop {} {\  (\bibinfo {year} {2025})},\ \bibinfo {note} {in preparation}\BibitemShut {NoStop}%
\bibitem [{\citenamefont {{Renzo}}\ \emph {et~al.}(2017)\citenamefont {{Renzo}}, \citenamefont {{Ott}}, \citenamefont {{Shore}},\ and\ \citenamefont {{de Mink}}}]{Renzo2017}%
  \BibitemOpen
  \bibfield  {author} {\bibinfo {author} {\bibfnamefont {M.}~\bibnamefont {{Renzo}}}, \bibinfo {author} {\bibfnamefont {C.~D.}\ \bibnamefont {{Ott}}}, \bibinfo {author} {\bibfnamefont {S.~N.}\ \bibnamefont {{Shore}}},\ and\ \bibinfo {author} {\bibfnamefont {S.~E.}\ \bibnamefont {{de Mink}}},\ }\href {https://doi.org/10.1051/0004-6361/201730698} {\bibfield  {journal} {\bibinfo  {journal} {\aap}\ }\textbf {\bibinfo {volume} {603}},\ \bibinfo {eid} {A118} (\bibinfo {year} {2017})},\ \Eprint {https://arxiv.org/abs/1703.09705} {arXiv:1703.09705 [astro-ph.SR]} \BibitemShut {NoStop}%
\bibitem [{\citenamefont {{van Son}}\ \emph {et~al.}(2025)\citenamefont {{van Son}}, \citenamefont {{Roy}}, \citenamefont {{Mandel}}, \citenamefont {{Farr}}, \citenamefont {{Lam}}, \citenamefont {{Merritt}}, \citenamefont {{Broekgaarden}}, \citenamefont {{Sander}},\ and\ \citenamefont {{Andrews}}}]{vanSon2025}%
  \BibitemOpen
  \bibfield  {author} {\bibinfo {author} {\bibfnamefont {L.~A.~C.}\ \bibnamefont {{van Son}}}, \bibinfo {author} {\bibfnamefont {S.~K.}\ \bibnamefont {{Roy}}}, \bibinfo {author} {\bibfnamefont {I.}~\bibnamefont {{Mandel}}}, \bibinfo {author} {\bibfnamefont {W.~M.}\ \bibnamefont {{Farr}}}, \bibinfo {author} {\bibfnamefont {A.}~\bibnamefont {{Lam}}}, \bibinfo {author} {\bibfnamefont {J.}~\bibnamefont {{Merritt}}}, \bibinfo {author} {\bibfnamefont {F.~S.}\ \bibnamefont {{Broekgaarden}}}, \bibinfo {author} {\bibfnamefont {A.~A.~C.}\ \bibnamefont {{Sander}}},\ and\ \bibinfo {author} {\bibfnamefont {J.~J.}\ \bibnamefont {{Andrews}}},\ }\href {https://doi.org/10.3847/1538-4357/ada14a} {\bibfield  {journal} {\bibinfo  {journal} {\apj}\ }\textbf {\bibinfo {volume} {979}},\ \bibinfo {eid} {209} (\bibinfo {year} {2025})},\ \Eprint {https://arxiv.org/abs/2411.02484} {arXiv:2411.02484 [astro-ph.HE]} \BibitemShut {NoStop}%
\bibitem [{\citenamefont {{Richards}}\ \emph {et~al.}(2023)\citenamefont {{Richards}}, \citenamefont {{Eldridge}}, \citenamefont {{Briel}}, \citenamefont {{Stevance}},\ and\ \citenamefont {{Willcox}}}]{Richards2023}%
  \BibitemOpen
  \bibfield  {author} {\bibinfo {author} {\bibfnamefont {S.~M.}\ \bibnamefont {{Richards}}}, \bibinfo {author} {\bibfnamefont {J.~J.}\ \bibnamefont {{Eldridge}}}, \bibinfo {author} {\bibfnamefont {M.~M.}\ \bibnamefont {{Briel}}}, \bibinfo {author} {\bibfnamefont {H.~F.}\ \bibnamefont {{Stevance}}},\ and\ \bibinfo {author} {\bibfnamefont {R.}~\bibnamefont {{Willcox}}},\ }\href {https://doi.org/10.1093/mnras/stad977} {\bibfield  {journal} {\bibinfo  {journal} {\mnras}\ }\textbf {\bibinfo {volume} {522}},\ \bibinfo {pages} {3972} (\bibinfo {year} {2023})},\ \Eprint {https://arxiv.org/abs/2208.02407} {arXiv:2208.02407 [astro-ph.HE]} \BibitemShut {NoStop}%
\bibitem [{\citenamefont {{Fumagalli}}\ \emph {et~al.}(2024)\citenamefont {{Fumagalli}}, \citenamefont {{Romero-Shaw}}, \citenamefont {{Gerosa}}, \citenamefont {{De Renzis}}, \citenamefont {{Kritos}},\ and\ \citenamefont {{Olejak}}}]{Fumagalli2024}%
  \BibitemOpen
  \bibfield  {author} {\bibinfo {author} {\bibfnamefont {G.}~\bibnamefont {{Fumagalli}}}, \bibinfo {author} {\bibfnamefont {I.}~\bibnamefont {{Romero-Shaw}}}, \bibinfo {author} {\bibfnamefont {D.}~\bibnamefont {{Gerosa}}}, \bibinfo {author} {\bibfnamefont {V.}~\bibnamefont {{De Renzis}}}, \bibinfo {author} {\bibfnamefont {K.}~\bibnamefont {{Kritos}}},\ and\ \bibinfo {author} {\bibfnamefont {A.}~\bibnamefont {{Olejak}}},\ }\href {https://doi.org/10.1103/PhysRevD.110.063012} {\bibfield  {journal} {\bibinfo  {journal} {\prd}\ }\textbf {\bibinfo {volume} {110}},\ \bibinfo {eid} {063012} (\bibinfo {year} {2024})},\ \Eprint {https://arxiv.org/abs/2405.14945} {arXiv:2405.14945 [astro-ph.HE]} \BibitemShut {NoStop}%
\bibitem [{\citenamefont {{Vitale}}\ \emph {et~al.}(2022)\citenamefont {{Vitale}}, \citenamefont {{Gerosa}}, \citenamefont {{Farr}},\ and\ \citenamefont {{Taylor}}}]{Vitale2022}%
  \BibitemOpen
  \bibfield  {author} {\bibinfo {author} {\bibfnamefont {S.}~\bibnamefont {{Vitale}}}, \bibinfo {author} {\bibfnamefont {D.}~\bibnamefont {{Gerosa}}}, \bibinfo {author} {\bibfnamefont {W.~M.}\ \bibnamefont {{Farr}}},\ and\ \bibinfo {author} {\bibfnamefont {S.~R.}\ \bibnamefont {{Taylor}}},\ }in\ \href {https://doi.org/10.1007/978-981-15-4702-7_45-1} {\emph {\bibinfo {booktitle} {Handbook of Gravitational Wave Astronomy}}},\ \bibinfo {editor} {edited by\ \bibinfo {editor} {\bibfnamefont {C.}~\bibnamefont {{Bambi}}}, \bibinfo {editor} {\bibfnamefont {S.}~\bibnamefont {{Katsanevas}}},\ and\ \bibinfo {editor} {\bibfnamefont {K.~D.}\ \bibnamefont {{Kokkotas}}}}\ (\bibinfo {year} {2022})\ p.~\bibinfo {pages} {45}\BibitemShut {NoStop}%
\bibitem [{\citenamefont {{Fragos}}\ \emph {et~al.}(2023)\citenamefont {{Fragos}}, \citenamefont {{Andrews}}, \citenamefont {{Bavera}} \emph {et~al.}}]{Fragos2023}%
  \BibitemOpen
  \bibfield  {author} {\bibinfo {author} {\bibfnamefont {T.}~\bibnamefont {{Fragos}}}, \bibinfo {author} {\bibfnamefont {J.~J.}\ \bibnamefont {{Andrews}}}, \bibinfo {author} {\bibfnamefont {S.~S.}\ \bibnamefont {{Bavera}}}, \emph {et~al.},\ }\href {https://doi.org/10.3847/1538-4365/ac90c1} {\bibfield  {journal} {\bibinfo  {journal} {\apjs}\ }\textbf {\bibinfo {volume} {264}},\ \bibinfo {eid} {45} (\bibinfo {year} {2023})},\ \Eprint {https://arxiv.org/abs/2202.05892} {arXiv:2202.05892 [astro-ph.SR]} \BibitemShut {NoStop}%
\bibitem [{\citenamefont {{Andrews}}\ \emph {et~al.}(2024)\citenamefont {{Andrews}}, \citenamefont {{Bavera}}, \citenamefont {{Briel}} \emph {et~al.}}]{Andrews2024}%
  \BibitemOpen
  \bibfield  {author} {\bibinfo {author} {\bibfnamefont {J.~J.}\ \bibnamefont {{Andrews}}}, \bibinfo {author} {\bibfnamefont {S.~S.}\ \bibnamefont {{Bavera}}}, \bibinfo {author} {\bibfnamefont {M.}~\bibnamefont {{Briel}}}, \emph {et~al.},\ }\href {https://doi.org/10.48550/arXiv.2411.02376} {\bibfield  {journal} {\bibinfo  {journal} {arXiv e-prints}\ ,\ \bibinfo {eid} {arXiv:2411.02376}} (\bibinfo {year} {2024})},\ \Eprint {https://arxiv.org/abs/2411.02376} {arXiv:2411.02376 [astro-ph.GA]} \BibitemShut {NoStop}%
\bibitem [{\citenamefont {{Lattimer}}(2010)}]{Lattimer2010}%
  \BibitemOpen
  \bibfield  {author} {\bibinfo {author} {\bibfnamefont {J.~M.}\ \bibnamefont {{Lattimer}}},\ }\href {https://doi.org/10.1016/j.newar.2010.09.013} {\bibfield  {journal} {\bibinfo  {journal} {\nar}\ }\textbf {\bibinfo {volume} {54}},\ \bibinfo {pages} {101} (\bibinfo {year} {2010})}\BibitemShut {NoStop}%
\bibitem [{\citenamefont {{Lattimer}}(2021)}]{Lattimer2021}%
  \BibitemOpen
  \bibfield  {author} {\bibinfo {author} {\bibfnamefont {J.~M.}\ \bibnamefont {{Lattimer}}},\ }\href {https://doi.org/10.1146/annurev-nucl-102419-124827} {\bibfield  {journal} {\bibinfo  {journal} {Annual Review of Nuclear and Particle Science}\ }\textbf {\bibinfo {volume} {71}},\ \bibinfo {pages} {433} (\bibinfo {year} {2021})}\BibitemShut {NoStop}%
\bibitem [{\citenamefont {{Radice}}\ \emph {et~al.}(2018)\citenamefont {{Radice}}, \citenamefont {{Perego}}, \citenamefont {{Zappa}},\ and\ \citenamefont {{Bernuzzi}}}]{Radice2018}%
  \BibitemOpen
  \bibfield  {author} {\bibinfo {author} {\bibfnamefont {D.}~\bibnamefont {{Radice}}}, \bibinfo {author} {\bibfnamefont {A.}~\bibnamefont {{Perego}}}, \bibinfo {author} {\bibfnamefont {F.}~\bibnamefont {{Zappa}}},\ and\ \bibinfo {author} {\bibfnamefont {S.}~\bibnamefont {{Bernuzzi}}},\ }\href {https://doi.org/10.3847/2041-8213/aaa402} {\bibfield  {journal} {\bibinfo  {journal} {\apjl}\ }\textbf {\bibinfo {volume} {852}},\ \bibinfo {eid} {L29} (\bibinfo {year} {2018})},\ \Eprint {https://arxiv.org/abs/1711.03647} {arXiv:1711.03647 [astro-ph.HE]} \BibitemShut {NoStop}%
\bibitem [{\citenamefont {{Bauswein}}(2019)}]{Bauswein2019}%
  \BibitemOpen
  \bibfield  {author} {\bibinfo {author} {\bibfnamefont {A.}~\bibnamefont {{Bauswein}}},\ }\href {https://doi.org/10.1016/j.aop.2019.167958} {\bibfield  {journal} {\bibinfo  {journal} {Annals of Physics}\ }\textbf {\bibinfo {volume} {411}},\ \bibinfo {eid} {167958} (\bibinfo {year} {2019})},\ \Eprint {https://arxiv.org/abs/2103.16371} {arXiv:2103.16371 [astro-ph.HE]} \BibitemShut {NoStop}%
\bibitem [{\citenamefont {{Margalit}}\ and\ \citenamefont {{Metzger}}(2019)}]{Margalit2019a}%
  \BibitemOpen
  \bibfield  {author} {\bibinfo {author} {\bibfnamefont {B.}~\bibnamefont {{Margalit}}}\ and\ \bibinfo {author} {\bibfnamefont {B.~D.}\ \bibnamefont {{Metzger}}},\ }\href {https://doi.org/10.3847/2041-8213/ab2ae2} {\bibfield  {journal} {\bibinfo  {journal} {\apjl}\ }\textbf {\bibinfo {volume} {880}},\ \bibinfo {eid} {L15} (\bibinfo {year} {2019})},\ \Eprint {https://arxiv.org/abs/1904.11995} {arXiv:1904.11995 [astro-ph.HE]} \BibitemShut {NoStop}%
\bibitem [{\citenamefont {{Margalit}}(2019)}]{Margalit2019b}%
  \BibitemOpen
  \bibfield  {author} {\bibinfo {author} {\bibfnamefont {B.}~\bibnamefont {{Margalit}}},\ }\href {https://doi.org/10.1016/j.aop.2019.167925} {\bibfield  {journal} {\bibinfo  {journal} {Annals of Physics}\ }\textbf {\bibinfo {volume} {410}},\ \bibinfo {eid} {167925} (\bibinfo {year} {2019})}\BibitemShut {NoStop}%
\bibitem [{\citenamefont {{Marczenko}}\ \emph {et~al.}(2020)\citenamefont {{Marczenko}}, \citenamefont {{Blaschke}}, \citenamefont {{Redlich}},\ and\ \citenamefont {{Sasaki}}}]{Marczenko2020}%
  \BibitemOpen
  \bibfield  {author} {\bibinfo {author} {\bibfnamefont {M.}~\bibnamefont {{Marczenko}}}, \bibinfo {author} {\bibfnamefont {D.}~\bibnamefont {{Blaschke}}}, \bibinfo {author} {\bibfnamefont {K.}~\bibnamefont {{Redlich}}},\ and\ \bibinfo {author} {\bibfnamefont {C.}~\bibnamefont {{Sasaki}}},\ }\href {https://doi.org/10.1051/0004-6361/202038211} {\bibfield  {journal} {\bibinfo  {journal} {\aap}\ }\textbf {\bibinfo {volume} {643}},\ \bibinfo {eid} {A82} (\bibinfo {year} {2020})},\ \Eprint {https://arxiv.org/abs/2004.09566} {arXiv:2004.09566 [astro-ph.HE]} \BibitemShut {NoStop}%
\bibitem [{\citenamefont {{G{\"u}ven}}\ \emph {et~al.}(2020)\citenamefont {{G{\"u}ven}}, \citenamefont {{Bozkurt}}, \citenamefont {{Khan}},\ and\ \citenamefont {{Margueron}}}]{Guven2020}%
  \BibitemOpen
  \bibfield  {author} {\bibinfo {author} {\bibfnamefont {H.}~\bibnamefont {{G{\"u}ven}}}, \bibinfo {author} {\bibfnamefont {K.}~\bibnamefont {{Bozkurt}}}, \bibinfo {author} {\bibfnamefont {E.}~\bibnamefont {{Khan}}},\ and\ \bibinfo {author} {\bibfnamefont {J.}~\bibnamefont {{Margueron}}},\ }\href {https://doi.org/10.1103/PhysRevC.102.015805} {\bibfield  {journal} {\bibinfo  {journal} {\prc}\ }\textbf {\bibinfo {volume} {102}},\ \bibinfo {eid} {015805} (\bibinfo {year} {2020})},\ \Eprint {https://arxiv.org/abs/2001.10259} {arXiv:2001.10259 [nucl-th]} \BibitemShut {NoStop}%
\bibitem [{\citenamefont {{Lund}}\ \emph {et~al.}(2024)\citenamefont {{Lund}}, \citenamefont {{Somasundaram}}, \citenamefont {{McLaughlin}}, \citenamefont {{Miller}}, \citenamefont {{Mumpower}},\ and\ \citenamefont {{Tews}}}]{Lund2024}%
  \BibitemOpen
  \bibfield  {author} {\bibinfo {author} {\bibfnamefont {K.~A.}\ \bibnamefont {{Lund}}}, \bibinfo {author} {\bibfnamefont {R.}~\bibnamefont {{Somasundaram}}}, \bibinfo {author} {\bibfnamefont {G.~C.}\ \bibnamefont {{McLaughlin}}}, \bibinfo {author} {\bibfnamefont {J.~M.}\ \bibnamefont {{Miller}}}, \bibinfo {author} {\bibfnamefont {M.~R.}\ \bibnamefont {{Mumpower}}},\ and\ \bibinfo {author} {\bibfnamefont {I.}~\bibnamefont {{Tews}}},\ }\href {https://doi.org/10.48550/arXiv.2408.07686} {\bibfield  {journal} {\bibinfo  {journal} {arXiv e-prints}\ ,\ \bibinfo {eid} {arXiv:2408.07686}} (\bibinfo {year} {2024})},\ \Eprint {https://arxiv.org/abs/2408.07686} {arXiv:2408.07686 [astro-ph.HE]} \BibitemShut {NoStop}%
\bibitem [{\citenamefont {{Breschi}}\ \emph {et~al.}(2024)\citenamefont {{Breschi}}, \citenamefont {{Gamba}}, \citenamefont {{Carullo}}, \citenamefont {{Godzieba}}, \citenamefont {{Bernuzzi}}, \citenamefont {{Perego}},\ and\ \citenamefont {{Radice}}}]{Breschi2024}%
  \BibitemOpen
  \bibfield  {author} {\bibinfo {author} {\bibfnamefont {M.}~\bibnamefont {{Breschi}}}, \bibinfo {author} {\bibfnamefont {R.}~\bibnamefont {{Gamba}}}, \bibinfo {author} {\bibfnamefont {G.}~\bibnamefont {{Carullo}}}, \bibinfo {author} {\bibfnamefont {D.}~\bibnamefont {{Godzieba}}}, \bibinfo {author} {\bibfnamefont {S.}~\bibnamefont {{Bernuzzi}}}, \bibinfo {author} {\bibfnamefont {A.}~\bibnamefont {{Perego}}},\ and\ \bibinfo {author} {\bibfnamefont {D.}~\bibnamefont {{Radice}}},\ }\href {https://doi.org/10.1051/0004-6361/202449173} {\bibfield  {journal} {\bibinfo  {journal} {\aap}\ }\textbf {\bibinfo {volume} {689}},\ \bibinfo {eid} {A51} (\bibinfo {year} {2024})},\ \Eprint {https://arxiv.org/abs/2401.03750} {arXiv:2401.03750 [astro-ph.HE]} \BibitemShut {NoStop}%
\bibitem [{\citenamefont {{Guo}}\ and\ \citenamefont {{Niu}}(2024)}]{Guo2024}%
  \BibitemOpen
  \bibfield  {author} {\bibinfo {author} {\bibfnamefont {L.}~\bibnamefont {{Guo}}}\ and\ \bibinfo {author} {\bibfnamefont {Y.~F.}\ \bibnamefont {{Niu}}},\ }\href {https://doi.org/10.1103/PhysRevC.110.L012801} {\bibfield  {journal} {\bibinfo  {journal} {\prc}\ }\textbf {\bibinfo {volume} {110}},\ \bibinfo {eid} {L012801} (\bibinfo {year} {2024})},\ \Eprint {https://arxiv.org/abs/2311.09792} {arXiv:2311.09792 [astro-ph.HE]} \BibitemShut {NoStop}%
\bibitem [{\citenamefont {{Ecker}}\ \emph {et~al.}(2025)\citenamefont {{Ecker}}, \citenamefont {{Gorda}}, \citenamefont {{Kurkela}},\ and\ \citenamefont {{Rezzolla}}}]{Ecker2025}%
  \BibitemOpen
  \bibfield  {author} {\bibinfo {author} {\bibfnamefont {C.}~\bibnamefont {{Ecker}}}, \bibinfo {author} {\bibfnamefont {T.}~\bibnamefont {{Gorda}}}, \bibinfo {author} {\bibfnamefont {A.}~\bibnamefont {{Kurkela}}},\ and\ \bibinfo {author} {\bibfnamefont {L.}~\bibnamefont {{Rezzolla}}},\ }\href {https://doi.org/10.1038/s41467-025-56500-x} {\bibfield  {journal} {\bibinfo  {journal} {Nature Communications}\ }\textbf {\bibinfo {volume} {16}},\ \bibinfo {eid} {1320} (\bibinfo {year} {2025})},\ \Eprint {https://arxiv.org/abs/2403.03246} {arXiv:2403.03246 [astro-ph.HE]} \BibitemShut {NoStop}%
\bibitem [{\citenamefont {{Perna}}\ \emph {et~al.}(2025)\citenamefont {{Perna}}, \citenamefont {{Gottlieb}}, \citenamefont {{Shukla}},\ and\ \citenamefont {{Radice}}}]{Perna2025}%
  \BibitemOpen
  \bibfield  {author} {\bibinfo {author} {\bibfnamefont {R.}~\bibnamefont {{Perna}}}, \bibinfo {author} {\bibfnamefont {O.}~\bibnamefont {{Gottlieb}}}, \bibinfo {author} {\bibfnamefont {E.}~\bibnamefont {{Shukla}}},\ and\ \bibinfo {author} {\bibfnamefont {D.}~\bibnamefont {{Radice}}},\ }\href {https://doi.org/10.1103/PhysRevD.111.063015} {\bibfield  {journal} {\bibinfo  {journal} {\prd}\ }\textbf {\bibinfo {volume} {111}},\ \bibinfo {eid} {063015} (\bibinfo {year} {2025})},\ \Eprint {https://arxiv.org/abs/2412.07846} {arXiv:2412.07846 [astro-ph.HE]} \BibitemShut {NoStop}%
\bibitem [{\citenamefont {{Legred}}\ \emph {et~al.}(2024)\citenamefont {{Legred}}, \citenamefont {{Sy-Garcia}}, \citenamefont {{Chatziioannou}},\ and\ \citenamefont {{Essick}}}]{Legred2024}%
  \BibitemOpen
  \bibfield  {author} {\bibinfo {author} {\bibfnamefont {I.}~\bibnamefont {{Legred}}}, \bibinfo {author} {\bibfnamefont {B.~O.}\ \bibnamefont {{Sy-Garcia}}}, \bibinfo {author} {\bibfnamefont {K.}~\bibnamefont {{Chatziioannou}}},\ and\ \bibinfo {author} {\bibfnamefont {R.}~\bibnamefont {{Essick}}},\ }\href {https://doi.org/10.1103/PhysRevD.109.023020} {\bibfield  {journal} {\bibinfo  {journal} {\prd}\ }\textbf {\bibinfo {volume} {109}},\ \bibinfo {eid} {023020} (\bibinfo {year} {2024})},\ \Eprint {https://arxiv.org/abs/2310.10854} {arXiv:2310.10854 [astro-ph.HE]} \BibitemShut {NoStop}%
\bibitem [{\citenamefont {{Babiuc Hamilton}}\ and\ \citenamefont {{Messman}}(2025)}]{BabiucHamilton2025}%
  \BibitemOpen
  \bibfield  {author} {\bibinfo {author} {\bibfnamefont {M.~C.}\ \bibnamefont {{Babiuc Hamilton}}}\ and\ \bibinfo {author} {\bibfnamefont {W.~A.}\ \bibnamefont {{Messman}}},\ }\href {https://doi.org/10.1088/1361-6382/adccb1} {\bibfield  {journal} {\bibinfo  {journal} {Classical and Quantum Gravity}\ }\textbf {\bibinfo {volume} {42}},\ \bibinfo {eid} {095012} (\bibinfo {year} {2025})},\ \Eprint {https://arxiv.org/abs/2411.10552} {arXiv:2411.10552 [gr-qc]} \BibitemShut {NoStop}%
\bibitem [{\citenamefont {{Burgio}}\ \emph {et~al.}(2021)\citenamefont {{Burgio}}, \citenamefont {{Schulze}}, \citenamefont {{Vida{\~n}a}},\ and\ \citenamefont {{Wei}}}]{Burgio2021}%
  \BibitemOpen
  \bibfield  {author} {\bibinfo {author} {\bibfnamefont {G.~F.}\ \bibnamefont {{Burgio}}}, \bibinfo {author} {\bibfnamefont {H.~J.}\ \bibnamefont {{Schulze}}}, \bibinfo {author} {\bibfnamefont {I.}~\bibnamefont {{Vida{\~n}a}}},\ and\ \bibinfo {author} {\bibfnamefont {J.~B.}\ \bibnamefont {{Wei}}},\ }\href {https://doi.org/10.1016/j.ppnp.2021.103879} {\bibfield  {journal} {\bibinfo  {journal} {Progress in Particle and Nuclear Physics}\ }\textbf {\bibinfo {volume} {120}},\ \bibinfo {eid} {103879} (\bibinfo {year} {2021})},\ \Eprint {https://arxiv.org/abs/2105.03747} {arXiv:2105.03747 [nucl-th]} \BibitemShut {NoStop}%
\bibitem [{\citenamefont {{Ellis}}\ \emph {et~al.}(2018)\citenamefont {{Ellis}}, \citenamefont {{H{\"u}tsi}}, \citenamefont {{Kannike}}, \citenamefont {{Marzola}}, \citenamefont {{Raidal}},\ and\ \citenamefont {{Vaskonen}}}]{Ellis2018}%
  \BibitemOpen
  \bibfield  {author} {\bibinfo {author} {\bibfnamefont {J.}~\bibnamefont {{Ellis}}}, \bibinfo {author} {\bibfnamefont {G.}~\bibnamefont {{H{\"u}tsi}}}, \bibinfo {author} {\bibfnamefont {K.}~\bibnamefont {{Kannike}}}, \bibinfo {author} {\bibfnamefont {L.}~\bibnamefont {{Marzola}}}, \bibinfo {author} {\bibfnamefont {M.}~\bibnamefont {{Raidal}}},\ and\ \bibinfo {author} {\bibfnamefont {V.}~\bibnamefont {{Vaskonen}}},\ }\href {https://doi.org/10.1103/PhysRevD.97.123007} {\bibfield  {journal} {\bibinfo  {journal} {\prd}\ }\textbf {\bibinfo {volume} {97}},\ \bibinfo {eid} {123007} (\bibinfo {year} {2018})},\ \Eprint {https://arxiv.org/abs/1804.01418} {arXiv:1804.01418 [astro-ph.CO]} \BibitemShut {NoStop}%
\bibitem [{\citenamefont {{Alsing}}\ \emph {et~al.}(2018)\citenamefont {{Alsing}}, \citenamefont {{Silva}},\ and\ \citenamefont {{Berti}}}]{Alsing2018}%
  \BibitemOpen
  \bibfield  {author} {\bibinfo {author} {\bibfnamefont {J.}~\bibnamefont {{Alsing}}}, \bibinfo {author} {\bibfnamefont {H.~O.}\ \bibnamefont {{Silva}}},\ and\ \bibinfo {author} {\bibfnamefont {E.}~\bibnamefont {{Berti}}},\ }\href {https://doi.org/10.1093/mnras/sty1065} {\bibfield  {journal} {\bibinfo  {journal} {\mnras}\ }\textbf {\bibinfo {volume} {478}},\ \bibinfo {pages} {1377} (\bibinfo {year} {2018})},\ \Eprint {https://arxiv.org/abs/1709.07889} {arXiv:1709.07889 [astro-ph.HE]} \BibitemShut {NoStop}%
\bibitem [{\citenamefont {{Ai}}\ \emph {et~al.}(2020)\citenamefont {{Ai}}, \citenamefont {{Gao}},\ and\ \citenamefont {{Zhang}}}]{Ai2020}%
  \BibitemOpen
  \bibfield  {author} {\bibinfo {author} {\bibfnamefont {S.}~\bibnamefont {{Ai}}}, \bibinfo {author} {\bibfnamefont {H.}~\bibnamefont {{Gao}}},\ and\ \bibinfo {author} {\bibfnamefont {B.}~\bibnamefont {{Zhang}}},\ }\href {https://doi.org/10.3847/1538-4357/ab80bd} {\bibfield  {journal} {\bibinfo  {journal} {\apj}\ }\textbf {\bibinfo {volume} {893}},\ \bibinfo {eid} {146} (\bibinfo {year} {2020})},\ \Eprint {https://arxiv.org/abs/1912.06369} {arXiv:1912.06369 [astro-ph.HE]} \BibitemShut {NoStop}%
\bibitem [{\citenamefont {{Rocha}}\ \emph {et~al.}(2021)\citenamefont {{Rocha}}, \citenamefont {{Bachega}}, \citenamefont {{Horvath}},\ and\ \citenamefont {{Moraes}}}]{Rocha2021}%
  \BibitemOpen
  \bibfield  {author} {\bibinfo {author} {\bibfnamefont {L.~S.}\ \bibnamefont {{Rocha}}}, \bibinfo {author} {\bibfnamefont {R.~R.~A.}\ \bibnamefont {{Bachega}}}, \bibinfo {author} {\bibfnamefont {J.~E.}\ \bibnamefont {{Horvath}}},\ and\ \bibinfo {author} {\bibfnamefont {P.~H.~R.~S.}\ \bibnamefont {{Moraes}}},\ }\href {https://doi.org/10.48550/arXiv.2107.08822} {\bibfield  {journal} {\bibinfo  {journal} {arXiv e-prints}\ ,\ \bibinfo {eid} {arXiv:2107.08822}} (\bibinfo {year} {2021})},\ \Eprint {https://arxiv.org/abs/2107.08822} {arXiv:2107.08822 [astro-ph.HE]} \BibitemShut {NoStop}%
\bibitem [{\citenamefont {{Kalogera}}\ and\ \citenamefont {{Baym}}(1996)}]{Kalogera1996}%
  \BibitemOpen
  \bibfield  {author} {\bibinfo {author} {\bibfnamefont {V.}~\bibnamefont {{Kalogera}}}\ and\ \bibinfo {author} {\bibfnamefont {G.}~\bibnamefont {{Baym}}},\ }\href {https://doi.org/10.1086/310296} {\bibfield  {journal} {\bibinfo  {journal} {\apjl}\ }\textbf {\bibinfo {volume} {470}},\ \bibinfo {pages} {L61} (\bibinfo {year} {1996})},\ \Eprint {https://arxiv.org/abs/astro-ph/9608059} {arXiv:astro-ph/9608059 [astro-ph]} \BibitemShut {NoStop}%
\bibitem [{\citenamefont {{Lattimer}}\ and\ \citenamefont {{Prakash}}(2004)}]{Lattimer2004}%
  \BibitemOpen
  \bibfield  {author} {\bibinfo {author} {\bibfnamefont {J.~M.}\ \bibnamefont {{Lattimer}}}\ and\ \bibinfo {author} {\bibfnamefont {M.}~\bibnamefont {{Prakash}}},\ }\href {https://doi.org/10.1126/science.1090720} {\bibfield  {journal} {\bibinfo  {journal} {Science}\ }\textbf {\bibinfo {volume} {304}},\ \bibinfo {pages} {536} (\bibinfo {year} {2004})},\ \Eprint {https://arxiv.org/abs/astro-ph/0405262} {arXiv:astro-ph/0405262 [astro-ph]} \BibitemShut {NoStop}%
\bibitem [{\citenamefont {{Chamel}}\ \emph {et~al.}(2013)\citenamefont {{Chamel}}, \citenamefont {{Haensel}}, \citenamefont {{Zdunik}},\ and\ \citenamefont {{Fantina}}}]{Chamel2013}%
  \BibitemOpen
  \bibfield  {author} {\bibinfo {author} {\bibfnamefont {N.}~\bibnamefont {{Chamel}}}, \bibinfo {author} {\bibfnamefont {P.}~\bibnamefont {{Haensel}}}, \bibinfo {author} {\bibfnamefont {J.~L.}\ \bibnamefont {{Zdunik}}},\ and\ \bibinfo {author} {\bibfnamefont {A.~F.}\ \bibnamefont {{Fantina}}},\ }\href {https://doi.org/10.1142/S021830131330018X} {\bibfield  {journal} {\bibinfo  {journal} {International Journal of Modern Physics E}\ }\textbf {\bibinfo {volume} {22}},\ \bibinfo {eid} {1330018} (\bibinfo {year} {2013})},\ \Eprint {https://arxiv.org/abs/1307.3995} {arXiv:1307.3995 [astro-ph.HE]} \BibitemShut {NoStop}%
\bibitem [{\citenamefont {{Williams}}\ \emph {et~al.}(2024)\citenamefont {{Williams}}, \citenamefont {{Schmidt}},\ and\ \citenamefont {{Pratten}}}]{Williams2024}%
  \BibitemOpen
  \bibfield  {author} {\bibinfo {author} {\bibfnamefont {N.}~\bibnamefont {{Williams}}}, \bibinfo {author} {\bibfnamefont {P.}~\bibnamefont {{Schmidt}}},\ and\ \bibinfo {author} {\bibfnamefont {G.}~\bibnamefont {{Pratten}}},\ }\href {https://doi.org/10.1103/PhysRevD.110.104013} {\bibfield  {journal} {\bibinfo  {journal} {\prd}\ }\textbf {\bibinfo {volume} {110}},\ \bibinfo {eid} {104013} (\bibinfo {year} {2024})},\ \Eprint {https://arxiv.org/abs/2407.08538} {arXiv:2407.08538 [gr-qc]} \BibitemShut {NoStop}%
\bibitem [{\citenamefont {{Du}}\ \emph {et~al.}(2024)\citenamefont {{Du}}, \citenamefont {{Yorgancioglu}}, \citenamefont {{Rao}}, \citenamefont {{Kumar}}, \citenamefont {{Yi}}, \citenamefont {{Zhang}},\ and\ \citenamefont {{Zhang}}}]{Du2024}%
  \BibitemOpen
  \bibfield  {author} {\bibinfo {author} {\bibfnamefont {Y.~F.}\ \bibnamefont {{Du}}}, \bibinfo {author} {\bibfnamefont {E.~S.}\ \bibnamefont {{Yorgancioglu}}}, \bibinfo {author} {\bibfnamefont {J.~H.}\ \bibnamefont {{Rao}}}, \bibinfo {author} {\bibfnamefont {A.}~\bibnamefont {{Kumar}}}, \bibinfo {author} {\bibfnamefont {S.~X.}\ \bibnamefont {{Yi}}}, \bibinfo {author} {\bibfnamefont {S.~N.}\ \bibnamefont {{Zhang}}},\ and\ \bibinfo {author} {\bibfnamefont {S.}~\bibnamefont {{Zhang}}},\ }\href {https://doi.org/10.1093/mnras/stae2261} {\bibfield  {journal} {\bibinfo  {journal} {\mnras}\ }\textbf {\bibinfo {volume} {534}},\ \bibinfo {pages} {2715} (\bibinfo {year} {2024})},\ \Eprint {https://arxiv.org/abs/2409.19295} {arXiv:2409.19295 [astro-ph.HE]} \BibitemShut {NoStop}%
\bibitem [{\citenamefont {{Fong}}\ \emph {et~al.}(2015)\citenamefont {{Fong}}, \citenamefont {{Berger}}, \citenamefont {{Margutti}},\ and\ \citenamefont {{Zauderer}}}]{2015ApJ...815..102F}%
  \BibitemOpen
  \bibfield  {author} {\bibinfo {author} {\bibfnamefont {W.}~\bibnamefont {{Fong}}}, \bibinfo {author} {\bibfnamefont {E.}~\bibnamefont {{Berger}}}, \bibinfo {author} {\bibfnamefont {R.}~\bibnamefont {{Margutti}}},\ and\ \bibinfo {author} {\bibfnamefont {B.~A.}\ \bibnamefont {{Zauderer}}},\ }\href {https://doi.org/10.1088/0004-637X/815/2/102} {\bibfield  {journal} {\bibinfo  {journal} {\apj}\ }\textbf {\bibinfo {volume} {815}},\ \bibinfo {eid} {102} (\bibinfo {year} {2015})},\ \Eprint {https://arxiv.org/abs/1509.02922} {arXiv:1509.02922 [astro-ph.HE]} \BibitemShut {NoStop}%
\bibitem [{\citenamefont {{Ghirlanda}}\ \emph {et~al.}(2019)\citenamefont {{Ghirlanda}}, \citenamefont {{Salafia}}, \citenamefont {{Paragi}} \emph {et~al.}}]{2019Sci...363..968G}%
  \BibitemOpen
  \bibfield  {author} {\bibinfo {author} {\bibfnamefont {G.}~\bibnamefont {{Ghirlanda}}}, \bibinfo {author} {\bibfnamefont {O.~S.}\ \bibnamefont {{Salafia}}}, \bibinfo {author} {\bibfnamefont {Z.}~\bibnamefont {{Paragi}}}, \emph {et~al.},\ }\href {https://doi.org/10.1126/science.aau8815} {\bibfield  {journal} {\bibinfo  {journal} {Science}\ }\textbf {\bibinfo {volume} {363}},\ \bibinfo {pages} {968} (\bibinfo {year} {2019})},\ \Eprint {https://arxiv.org/abs/1808.00469} {arXiv:1808.00469 [astro-ph.HE]} \BibitemShut {NoStop}%
\bibitem [{\citenamefont {{Beniamini}}\ \emph {et~al.}(2020)\citenamefont {{Beniamini}}, \citenamefont {{Granot}},\ and\ \citenamefont {{Gill}}}]{2020MNRAS.493.3521B}%
  \BibitemOpen
  \bibfield  {author} {\bibinfo {author} {\bibfnamefont {P.}~\bibnamefont {{Beniamini}}}, \bibinfo {author} {\bibfnamefont {J.}~\bibnamefont {{Granot}}},\ and\ \bibinfo {author} {\bibfnamefont {R.}~\bibnamefont {{Gill}}},\ }\href {https://doi.org/10.1093/mnras/staa538} {\bibfield  {journal} {\bibinfo  {journal} {\mnras}\ }\textbf {\bibinfo {volume} {493}},\ \bibinfo {pages} {3521} (\bibinfo {year} {2020})},\ \Eprint {https://arxiv.org/abs/2001.02239} {arXiv:2001.02239 [astro-ph.HE]} \BibitemShut {NoStop}%
\bibitem [{\citenamefont {{Salafia}}\ \emph {et~al.}(2023)\citenamefont {{Salafia}}, \citenamefont {{Ravasio}}, \citenamefont {{Ghirlanda}},\ and\ \citenamefont {{Mandel}}}]{2023A&A...680A..45S}%
  \BibitemOpen
  \bibfield  {author} {\bibinfo {author} {\bibfnamefont {O.~S.}\ \bibnamefont {{Salafia}}}, \bibinfo {author} {\bibfnamefont {M.~E.}\ \bibnamefont {{Ravasio}}}, \bibinfo {author} {\bibfnamefont {G.}~\bibnamefont {{Ghirlanda}}},\ and\ \bibinfo {author} {\bibfnamefont {I.}~\bibnamefont {{Mandel}}},\ }\href {https://doi.org/10.1051/0004-6361/202347298} {\bibfield  {journal} {\bibinfo  {journal} {\aap}\ }\textbf {\bibinfo {volume} {680}},\ \bibinfo {eid} {A45} (\bibinfo {year} {2023})},\ \Eprint {https://arxiv.org/abs/2306.15488} {arXiv:2306.15488 [astro-ph.HE]} \BibitemShut {NoStop}%
\bibitem [{\citenamefont {{Gittins}}\ \emph {et~al.}(2025)\citenamefont {{Gittins}}, \citenamefont {{Matur}}, \citenamefont {{Andersson}},\ and\ \citenamefont {{Hawke}}}]{2025PhRvD.111b3049G}%
  \BibitemOpen
  \bibfield  {author} {\bibinfo {author} {\bibfnamefont {F.}~\bibnamefont {{Gittins}}}, \bibinfo {author} {\bibfnamefont {R.}~\bibnamefont {{Matur}}}, \bibinfo {author} {\bibfnamefont {N.}~\bibnamefont {{Andersson}}},\ and\ \bibinfo {author} {\bibfnamefont {I.}~\bibnamefont {{Hawke}}},\ }\href {https://doi.org/10.1103/PhysRevD.111.023049} {\bibfield  {journal} {\bibinfo  {journal} {\prd}\ }\textbf {\bibinfo {volume} {111}},\ \bibinfo {eid} {023049} (\bibinfo {year} {2025})},\ \Eprint {https://arxiv.org/abs/2409.13468} {arXiv:2409.13468 [gr-qc]} \BibitemShut {NoStop}%
\bibitem [{\citenamefont {{Dax}}\ \emph {et~al.}(2025)\citenamefont {{Dax}}, \citenamefont {{Green}}, \citenamefont {{Gair}}, \citenamefont {{Gupte}}, \citenamefont {{P{\"u}rrer}}, \citenamefont {{Raymond}}, \citenamefont {{Wildberger}}, \citenamefont {{Macke}}, \citenamefont {{Buonanno}},\ and\ \citenamefont {{Sch{\"o}lkopf}}}]{Dax2025}%
  \BibitemOpen
  \bibfield  {author} {\bibinfo {author} {\bibfnamefont {M.}~\bibnamefont {{Dax}}}, \bibinfo {author} {\bibfnamefont {S.~R.}\ \bibnamefont {{Green}}}, \bibinfo {author} {\bibfnamefont {J.}~\bibnamefont {{Gair}}}, \bibinfo {author} {\bibfnamefont {N.}~\bibnamefont {{Gupte}}}, \bibinfo {author} {\bibfnamefont {M.}~\bibnamefont {{P{\"u}rrer}}}, \bibinfo {author} {\bibfnamefont {V.}~\bibnamefont {{Raymond}}}, \bibinfo {author} {\bibfnamefont {J.}~\bibnamefont {{Wildberger}}}, \bibinfo {author} {\bibfnamefont {J.~H.}\ \bibnamefont {{Macke}}}, \bibinfo {author} {\bibfnamefont {A.}~\bibnamefont {{Buonanno}}},\ and\ \bibinfo {author} {\bibfnamefont {B.}~\bibnamefont {{Sch{\"o}lkopf}}},\ }\href {https://doi.org/10.1038/s41586-025-08593-z} {\bibfield  {journal} {\bibinfo  {journal} {\nat}\ }\textbf {\bibinfo {volume} {639}},\ \bibinfo {pages} {49} (\bibinfo {year} {2025})},\ \Eprint {https://arxiv.org/abs/2407.09602} {arXiv:2407.09602 [gr-qc]} \BibitemShut {NoStop}%
\bibitem [{\citenamefont {{Duque}}\ \emph {et~al.}(2020)\citenamefont {{Duque}}, \citenamefont {{Beniamini}}, \citenamefont {{Daigne}},\ and\ \citenamefont {{Mochkovitch}}}]{Duque2020}%
  \BibitemOpen
  \bibfield  {author} {\bibinfo {author} {\bibfnamefont {R.}~\bibnamefont {{Duque}}}, \bibinfo {author} {\bibfnamefont {P.}~\bibnamefont {{Beniamini}}}, \bibinfo {author} {\bibfnamefont {F.}~\bibnamefont {{Daigne}}},\ and\ \bibinfo {author} {\bibfnamefont {R.}~\bibnamefont {{Mochkovitch}}},\ }\href {https://doi.org/10.1051/0004-6361/201937115} {\bibfield  {journal} {\bibinfo  {journal} {\aap}\ }\textbf {\bibinfo {volume} {639}},\ \bibinfo {eid} {A15} (\bibinfo {year} {2020})},\ \Eprint {https://arxiv.org/abs/1911.03302} {arXiv:1911.03302 [astro-ph.HE]} \BibitemShut {NoStop}%
\bibitem [{\citenamefont {{Kumar Acharya}}\ \emph {et~al.}(2025)\citenamefont {{Kumar Acharya}}, \citenamefont {{Beniamini}},\ and\ \citenamefont {{Hotokezaka}}}]{KumarAcharya2025}%
  \BibitemOpen
  \bibfield  {author} {\bibinfo {author} {\bibfnamefont {S.}~\bibnamefont {{Kumar Acharya}}}, \bibinfo {author} {\bibfnamefont {P.}~\bibnamefont {{Beniamini}}},\ and\ \bibinfo {author} {\bibfnamefont {K.}~\bibnamefont {{Hotokezaka}}},\ }\href {https://doi.org/10.1051/0004-6361/202452290} {\bibfield  {journal} {\bibinfo  {journal} {\aap}\ }\textbf {\bibinfo {volume} {693}},\ \bibinfo {eid} {A108} (\bibinfo {year} {2025})},\ \Eprint {https://arxiv.org/abs/2409.11291} {arXiv:2409.11291 [astro-ph.HE]} \BibitemShut {NoStop}%
\bibitem [{\citenamefont {{Liu}}\ and\ \citenamefont {{Zou}}(2022)}]{Liu2022}%
  \BibitemOpen
  \bibfield  {author} {\bibinfo {author} {\bibfnamefont {Y.}~\bibnamefont {{Liu}}}\ and\ \bibinfo {author} {\bibfnamefont {Y.-C.}\ \bibnamefont {{Zou}}},\ }\href {https://doi.org/10.1103/PhysRevD.106.123024} {\bibfield  {journal} {\bibinfo  {journal} {\prd}\ }\textbf {\bibinfo {volume} {106}},\ \bibinfo {eid} {123024} (\bibinfo {year} {2022})},\ \Eprint {https://arxiv.org/abs/2211.02855} {arXiv:2211.02855 [astro-ph.HE]} \BibitemShut {NoStop}%
\bibitem [{\citenamefont {{Collette}}\ \emph {et~al.}(2022)\citenamefont {{Collette}}, \citenamefont {{Kluyver}}, \citenamefont {{Caswell}} \emph {et~al.}}]{h5py}%
  \BibitemOpen
  \bibfield  {author} {\bibinfo {author} {\bibfnamefont {A.}~\bibnamefont {{Collette}}}, \bibinfo {author} {\bibfnamefont {T.}~\bibnamefont {{Kluyver}}}, \bibinfo {author} {\bibfnamefont {T.~A.}\ \bibnamefont {{Caswell}}}, \emph {et~al.},\ }\href {https://doi.org/10.5281/zenodo.6575970} {\bibinfo {title} {{h5py/h5py: 3.7.0}}} (\bibinfo {year} {2022})\BibitemShut {NoStop}%
\bibitem [{\citenamefont {Foreman-Mackey}(2016)}]{corner}%
  \BibitemOpen
  \bibfield  {author} {\bibinfo {author} {\bibfnamefont {D.}~\bibnamefont {Foreman-Mackey}},\ }\href {https://doi.org/10.21105/joss.00024} {\bibfield  {journal} {\bibinfo  {journal} {The Journal of Open Source Software}\ }\textbf {\bibinfo {volume} {1}},\ \bibinfo {pages} {24} (\bibinfo {year} {2016})}\BibitemShut {NoStop}%
\bibitem [{\citenamefont {Hunter}(2007)}]{matplotlib}%
  \BibitemOpen
  \bibfield  {author} {\bibinfo {author} {\bibfnamefont {J.~D.}\ \bibnamefont {Hunter}},\ }\href {https://doi.org/10.1109/MCSE.2007.55} {\bibfield  {journal} {\bibinfo  {journal} {Computing in Science \& Engineering}\ }\textbf {\bibinfo {volume} {9}},\ \bibinfo {pages} {90} (\bibinfo {year} {2007})}\BibitemShut {NoStop}%
\end{thebibliography}%
\end{document}